\documentclass{article}

\PassOptionsToPackage{square,comma,numbers,sort&compress}{natbib}

\usepackage[final]{neurips_data_2022}

\usepackage[utf8]{inputenc} 
\usepackage[T1]{fontenc}    
\usepackage{url}            
\usepackage{booktabs}       
\usepackage{amsfonts}       
\usepackage{nicefrac}       
\usepackage{microtype}      
\usepackage{xcolor}         

\usepackage{tikz}
\usepackage{graphicx}
\usepackage{comment}
\usepackage{amsmath}
\usepackage{amssymb} 
\usepackage{tabularx}
\usepackage{threeparttable}
\usepackage{multirow}
\usepackage{subcaption}
\usepackage{array}
\usepackage{wrapfig}
\usepackage{placeins}
\usepackage[export]{adjustbox}
\newcolumntype{x}[1]{>{\raggedright}p{#1}}
\usepackage[inline]{enumitem}

\usepackage{makecell}

\usepackage[most]{tcolorbox}

\usepackage[pagebackref,breaklinks,colorlinks]{hyperref}
\hypersetup{
    linkcolor=cyan,
    urlcolor=cyan,
}
\usepackage[capitalize]{cleveref}

\crefname{section}{Sec.}{Secs.}
\Crefname{section}{Section}{Sections}
\Crefname{table}{Table}{Tables}
\crefname{table}{Tab.}{Tabs.}

\newlength{\twosubht}
\newsavebox{\twosubbox}
\newcommand{\ra}[1]{\renewcommand{\arraystretch}{#1}}

\title{ConfLab: A Data Collection Concept, Dataset, and Benchmark for Machine Analysis of Free-Standing Social Interactions in the Wild}

\author{%
    Chirag Raman$^1$\thanks{Equal contribution} \And 
    Jose Vargas-Quiros$^1$\footnotemark[1] \And 
    Stephanie Tan$^1$\footnotemark[1] \And  Ashraful Islam$^2$\\
    \AND Ekin Gedik$^1$ \qquad Hayley Hung$^1$\vspace{10pt}\\
    $^1$Delft University of Technology, Delft, The Netherlands\\
    {\fontfamily{cmtt}\selectfont \{c.a.raman, j.d.vargasquiros, s.tan-1, e.gedik, h.hung\}@tudelft.nl}\vspace{3pt}\\
    $^2$Rensselaer Polytechnic Institute, New York, USA\\
    {\fontfamily{cmtt}\selectfont islama6@rpi.edu}\\
}

\begin{document}

\maketitle

\begin{abstract}
Recording the dynamics of unscripted human interactions in the wild is challenging due to the delicate trade-offs between several factors: participant privacy, ecological validity, data fidelity, and logistical overheads. To address these, following a \textit{datasets for the community by the community} ethos, we propose the Conference Living Lab (ConfLab): a new concept for multimodal multisensor data collection of in-the-wild free-standing social conversations. For the first instantiation of ConfLab described here, we organized a real-life professional networking event at a major international conference. Involving $48$ conference attendees, the dataset captures a diverse mix of status, acquaintance, and networking motivations. Our capture setup improves upon the data fidelity of prior in-the-wild datasets while retaining privacy sensitivity: $8$ videos ($1920\times1080, 60$~fps) from a non-invasive overhead view, and custom wearable sensors with onboard recording of body motion (full $9$-axis IMU), privacy-preserving low-frequency audio ($1250$~Hz), and Bluetooth-based proximity. Additionally, we developed custom solutions for distributed hardware synchronization at acquisition, and time-efficient continuous annotation of body keypoints and actions at high sampling rates. Our benchmarks showcase some of the open research tasks related to in-the-wild privacy-preserving social data analysis: keypoints detection from overhead camera views, skeleton-based no-audio speaker detection, and F-formation detection. 
\end{abstract}

\section{Introduction}
A crucial challenge towards developing artificial socially intelligent systems is understanding how \textit{real-life} situational contexts affect social human behavior \cite{dudzik2021recognizing}. Social-science findings indeed show that the dynamics of how we conduct daily interactions vary significantly depending on the social situation \cite{Fleeson2007,GuardiaRyan2007,Hall2019}. Unfortunately, such dynamics are not adequately captured by many data collection setups where role-played or scripted scenarios are typical \cite{osborne2020social}.

\begin{figure*}[t!]
\makebox[\textwidth][c]{\includegraphics[width=\textwidth]{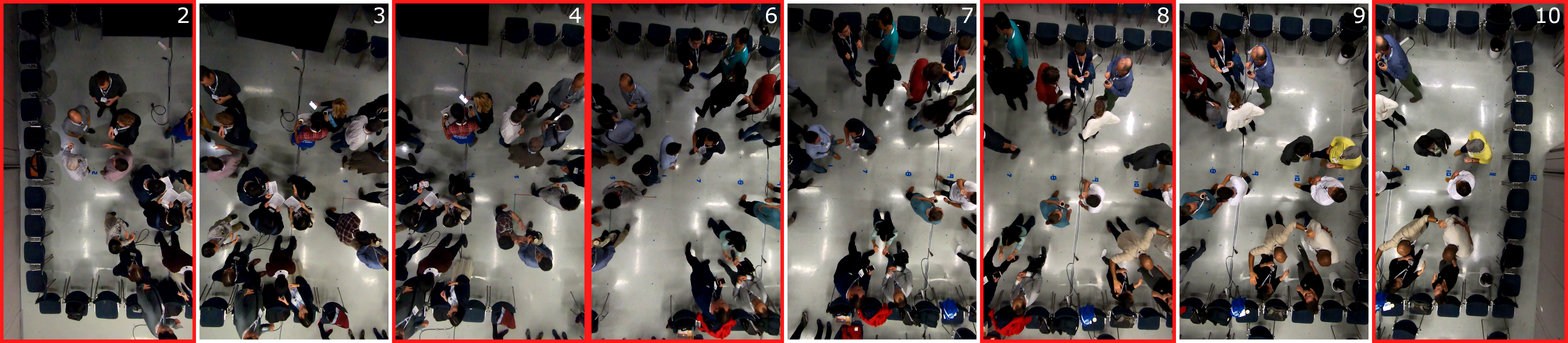}}
\caption{Snapshot of the interaction area from our cameras. We annotated only cameras highlighted with red borders (high scene overlap). For a clearer visual impression of the scene, we omit cameras 1 (few people recorded) and 5 (failed early in the event). Faces blurred to preserve privacy. 
}
\label{fig:scene}
\vspace{-10pt}
\end{figure*}

In this paper we address the problem of collecting a privacy-sensitive dataset of unscripted social dynamics of real-life relationships where encounters can influence someone's daily life. We argue that doing so requires recording these exchanges in the natural ecology, requiring an approach different from the typical setup of locally-organized studies. Specifically, we focus on free-standing interactions within the setting of an international conference (see Figure \ref{fig:scene}). 

\begin{figure}[b]
\begin{minipage}{0.6\textwidth}
    \centering
    \includegraphics[width=\linewidth]{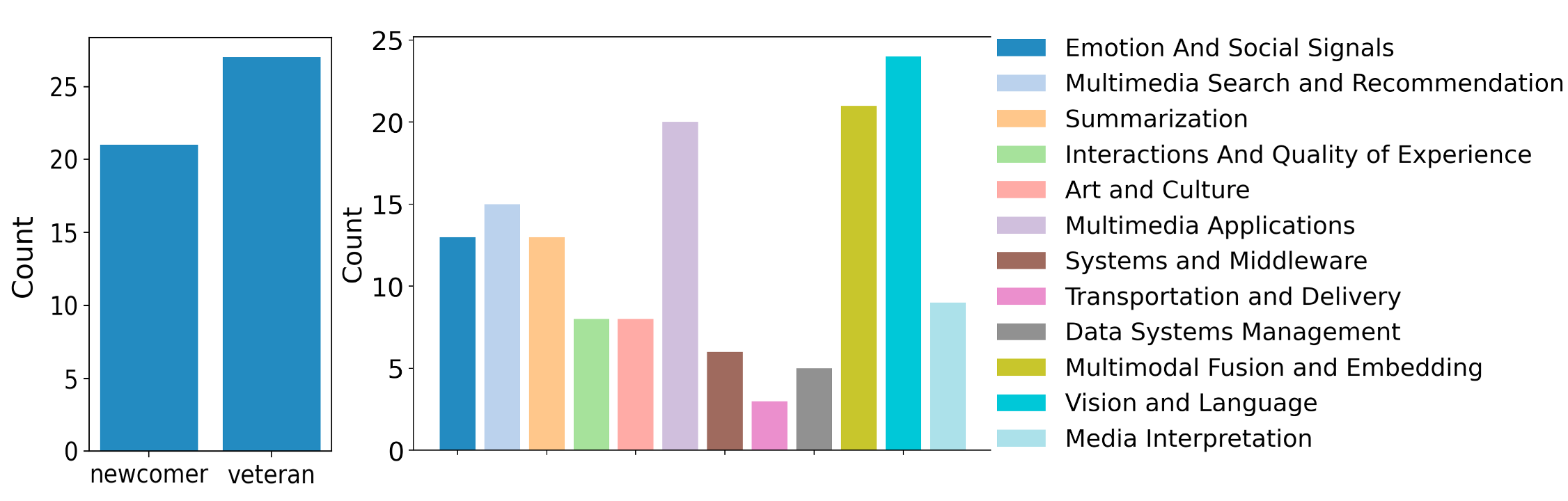}
    \caption{Frequency of newcomer/veteran participants (left) and reported research interests (right).}
    \label{fig:participant stats}
\end{minipage}\hfill
\begin{minipage}{0.38\textwidth}  
    \centering
    \includegraphics[width=\linewidth]{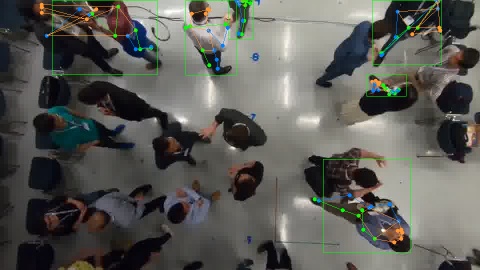}
    \caption{Keypoint detection using pretrained RSN \cite{cai2020learningRSN}. Additional SOTA results are in Appendix~\ref{app:kp_results}}
    \label{fig:pretrained-rsn}
\end{minipage}
\end{figure}

Recording an international community in its natural habitat is characterized by several intersecting challenges: an intrinsic trade-off exists between data fidelity, ecological validity, and privacy preservation. For ecological validity, a non-invasive capture setup is essential for mitigating any influence on behavior naturalness \cite{andrade2018internal, labonte2018wild, hung2019complex}.The most common solution involves mounting cameras from aerial perspectives such as top-down \cite{MnM2021, hung2011detecting} and elevated-side views \cite{alameda2015salsa, BMVC.25.23, ZenEtAl2010}. Now elevated-side views make it easy to capture sensitive personal information such as faces, which leads to several ethical concerns. For instance, capturing faces has been related to harmful downstream surveillance applications \cite{ft2019whosusing}. Besides, state-of-the-art (SOTA) body-keypoint estimation techniques perform poorly on aerial perspectives \cite{carissimi2018filling, MnM2021}, making the extraction of automatic pose annotations challenging (\figurename~\ref{fig:pretrained-rsn}). To avoid such issues, some researchers have turned to more privacy-preserving wearable sensors shown to benefit many behavior analysis tasks \cite{hung2019complex, gedikhungIMWUT2018, Gedik2017a}. 

In all, the closest related datasets (see Table~\ref{tab:dataset-comparison}) suffer from several technical limitations precluding the analysis and modeling of fine-grained social behavior: (i) lack of articulated pose annotations; (ii) a limited number of people in the scene, preventing complex interactions such as group splitting/merging behaviors, and (iii) an inadequate data sampling-rate and synchronization-latency to study time-sensitive social phenomena \cite[Sec.~3.3]{raman2020modular}. 
To address all these limitations, we propose the Conference Living Lab (ConfLab): a new concept for multimodal multisensor data collection
of ecologically-valid social settings. From the first instantiation of ConfLab, we provide a high-fidelity dataset of $48$ participants at a professional networking event.
\paragraph{Methodological Contributions:} We describe a data collection design that captures a diverse mix of real levels of seniority, acquaintance, affiliation, and motivation to network (see Figure \ref{fig:participant stats}). This was achieved by organizing ConfLab as part of a major international scientific conference. ConfLab had these goals: (i) a data collection effort follwing a \textit{by the community for the community} ethos: the more volunteers, the more data, (ii) volunteers who potentially use the data can experience first-hand potential privacy and ethical considerations related to sharing their own data, (iii) in light of recent data sourcing issues \cite{BirhaneP21, ft2019whosusing}, we incorporated privacy and invasiveness considerations directly into the decision-making process regarding sensor type, positioning, and sample-rates. 
\paragraph {Technical Contributions:}  \textbf{(i) aerial-view articulated pose}: our annotations of $17$ full-body keypoints enable improvements in (a) pose estimation and tracking, (b) pose-based recognition of social actions (under-explored in the top-down perspective), (c) pose-based F-formation
estimation (has not been possible 
from prior work \cite{hung2011detecting, SettiEtAl2015, SwoffordEtAl2020, VasconEtAl2014}), and (d) the direct study of interaction dynamics using full body poses (previously limited to lab settings \cite{Joo_2017_TPAMI}). \textbf{(ii) subtle body dynamics}: we are the first to use a full 9-axis Inertial Measurement Unit (IMU) enabling a richer representation of behaviour at higher sample rates; previous rates were found to be insufficient for downstream tasks \cite{Gedik2017a}. 
\textbf{(iii) enabling finer temporal-scale research questions}: a sub-second crossmodal latency of $\sim13$~ms along with higher
sampling rate of features ($60$~fps video, $56$~Hz IMU) opens the gateway for the in-the-wild study of nuanced time-sensitive social behaviors like mimicry and synchrony.

\begin{table*}[t!]
\caption{Comparison of ConfLab with prior datasets of free-standing conversation groups in in-the-wild social interaction settings. Conflab is the first and only social interaction dataset that offers skeletal keypoints and speaking status at high annotation resolution, as well as hardware synchronized camera and  multimodal wearable signals at high resolution.}
\label{tab:dataset-comparison}

\scriptsize \setlength{\tabcolsep}{3pt}
\begin{tabular*}{\textwidth}{@{}@{\extracolsep{\fill}}lcclll@{}}
\toprule
Dataset & \makecell[c]{People/\\Scene} & Video & Manual Annotations & Wearable Signals & Synchronization\\
\midrule
Cocktail \cite{ZenEtAl2010}$\dagger$  & $7$ & $512\times384$ &  \makecell[l]{F-formations\\($20$ and $30$~min, $1/5$~Hz)} &None&Unknown \\
\addlinespace[0.2cm]
CoffeeBreak \cite{BMVC.25.23} & $14$ & $1440\times1080$ & \makecell[l]{F-formations\\(130 frames in two sequences)} & None & None\\
\addlinespace[0.2cm]
IDIAP \cite{hung2011detecting} & $>50$ & \makecell[c]{$180$~min;\\$654\times439$\\$20$~fps} & \makecell[l]{F-formations\\($82$ independent frames)} & None & None\\
\addlinespace[0.2cm]
SALSA \cite{alameda2015salsa}$\dagger$ & $18$ & \makecell[c]{$60$~min;\\$1024\times768$\\$15$~fps} & \makecell[l]{Bounding boxes ($30$~min)\\Head \& body ori. ($30$~min)\\F-formations ($60$~min)\\(all $1/3$~Hz)} & \makecell[l]{Audio MFCCs ($30$~Hz)\\Acceleration ($20$~Hz)\\IR proximity ($1$~Hz)} & \makecell[l]{Post-hoc infra-red\\event-based (no-drift\\assumption)}\\
\addlinespace[0.2cm]
MnM \cite{MnM2021}$\dagger$   & $32$ & \makecell[c]{$30$~min;\\$1920\times1080$\\$30$~fps} & \makecell[l]{Bounding boxes ($30$~min, $1$~Hz $\ddagger$ )\\F-formations ($10$~min, $1$~Hz )\\Actions ($45$~min, $1$~Hz$\ddagger$)} & \makecell[l]{Accelerometer ($20$~Hz)\\{Radio proximity ($1$~Hz)}} & \makecell[l]{Intra-wearable sync via\\ gossiping protocol;\\Inter-modal sync using\\ manual inspection @$1$~Hz}\\
\addlinespace[0.2cm]
ConfLab & $48$ & \makecell[c]{$\sim45$~min;\\$1920\times1080$\\$\mathbf{60}$~\textbf{fps}} & \makecell[l]{\textbf{$\mathbf{17}$ keypoints ($\mathbf{16}$~min, $\mathbf{60}$~Hz)}\\F-formations ($16$~min, $1$~Hz)\\ \textbf{Speaking status ($\mathbf{16}$~min, $\mathbf{60}$~Hz)}} & \makecell[l]{
\textbf{Low-freq. audio ($\mathbf{1250}$~Hz)}\\
\textbf{BT proximity ($\mathbf{5}$~Hz)}\\\textbf{$\mathbf{9}$-axis IMU ($\mathbf{56}$~Hz)}} & \makecell[l]{\textbf{Wireless hardware sync at}\\\textbf{acquisition, max latency}\\\textbf{of $\sim\mathbf{13}$~ms} \cite{raman2020modular}}\\
\bottomrule

\end{tabular*}
\begin{tablenotes}[para]
	\item
	$\dagger$ Includes self-assessed personality ratings 
	\item
	$\ddagger$ Upsampled to $20$~Hz using Vatic \cite{Vondrick2013} 
	\item BT: Bluetooth  \hspace{3mm} IMU: Inertial Measurement Unit
\end{tablenotes}
\end{table*}

\section{Related Work}
\label{sec:related}

Early datasets of in-the-wild social events either spanned only a few minutes (e.g. Coffee Break \cite{BMVC.25.23}), or were recorded at such a large distance from the participants that performing robust, automated person detection or tracking with SOTA approaches was non-trivial (e.g. Idiap Poster Data \cite{hung2011detecting}). More recently, two different strategies have emerged to circumvent such issues. 

One approach involves fully instrumented labs with a high resolution multi-camera setup for video and audio data. Here automatic detectors \cite{7410886, bazzani2013social, Joo_2017_TPAMI} could be applied to obtain poses. This circumvents the cost- and labor-intensive process of manually labeling head poses, at the cost of less portable sensing setups. Notable examples of such in-the-lab studies include seated scenarios, such as the AMI meeting corpus \cite{amicorpus}, and more recently standing scenarios like the Panoptic Dataset \cite{Joo_2017_TPAMI}. Both enable the learning of multimodal behavioral dynamics. However, the dynamics of seated, scripted, or role-playing scenarios are different from that of an unconstrained social setting such as ours. In contrast, ConfLab moves out of the lab with a more modular and portable multimodal, multisensor solution that scales easily in the wild.


Another approach exploited wearable sensor data to allow for multimodal processing\textemdash sensors included 3 or 6 DOF inertial measurement units (IMU); infrared, bluetooth, or radio sensors to measure proximity; or microphones for speech behavior \cite{MnM2021,alameda2015salsa}. While proximity has been used as a proxy of face-to-face interaction \cite{alameda2015salsa, Cattuto2010DynamicsOP,hoffman_block_elmer_stadtfeld_2020, Atzmuellerwww18,olguin2008sensible}, recent findings highlight significant problems with such an assumption \cite{chaffin2017promise}. Such errors can have a significant impact on the machine-perceived experience of an individual, precluding the development of personalized technology. Chalcedony badges used by \citep{MnM2021} show more promising results with a radio-based proximity sensor and accelerometer \cite{8925179}, but such data remains insufficient for more downstream tasks due to the relatively low sample (20Hz) and annotation (1Hz) frequency \cite{Gedik2017a}. 
In light of these challenges in wearable sensing, ConfLab features custom-developed Midge sensors that enable more flexible and fine-grained on-device recording. At the same time, ConfLab enables researchers in the wearable and ubiquitous computing communities to investigate the benefit of exploiting wearable and multimodal data.

Furthermore, while both SALSA \cite{alameda2015salsa} and MatchNMingle \cite{MnM2021} capture a multimodal dataset of a large group of individuals involved in mingling behavior, the inter-modal synchronization is only guaranteed at 1/3~Hz and 1~Hz, respectively. Prior works coped with lower tolerances by computing summary statistics over input windows \cite{Gedik2017a, Cabrera-Quiros2018b, Quiros2019}. While 1~Hz is able to capture some conversation dynamics \cite{tanimwut2021}, it is insufficient to study fine-grained social phenomena such as back-channeling or mimicry that involve far lower latencies \cite[Sec.~3.3]{raman2020modular}. ConfLab provides data streams with higher sampling rates, synchronized at acquisition with our method shown to yield a $13$~ms latency at worst \cite{raman2020modular} (see Sec. \ref{sec:camwear}). Table~\ref{tab:dataset-comparison} summarizes the differences between ConfLab and other related datasets.

\section{Data Acquisition}
\label{sec:data-acquisition}
\begin{figure}[t]
\begin{minipage}{0.79\textwidth}
    \centering
    \includegraphics[height=3.3cm]{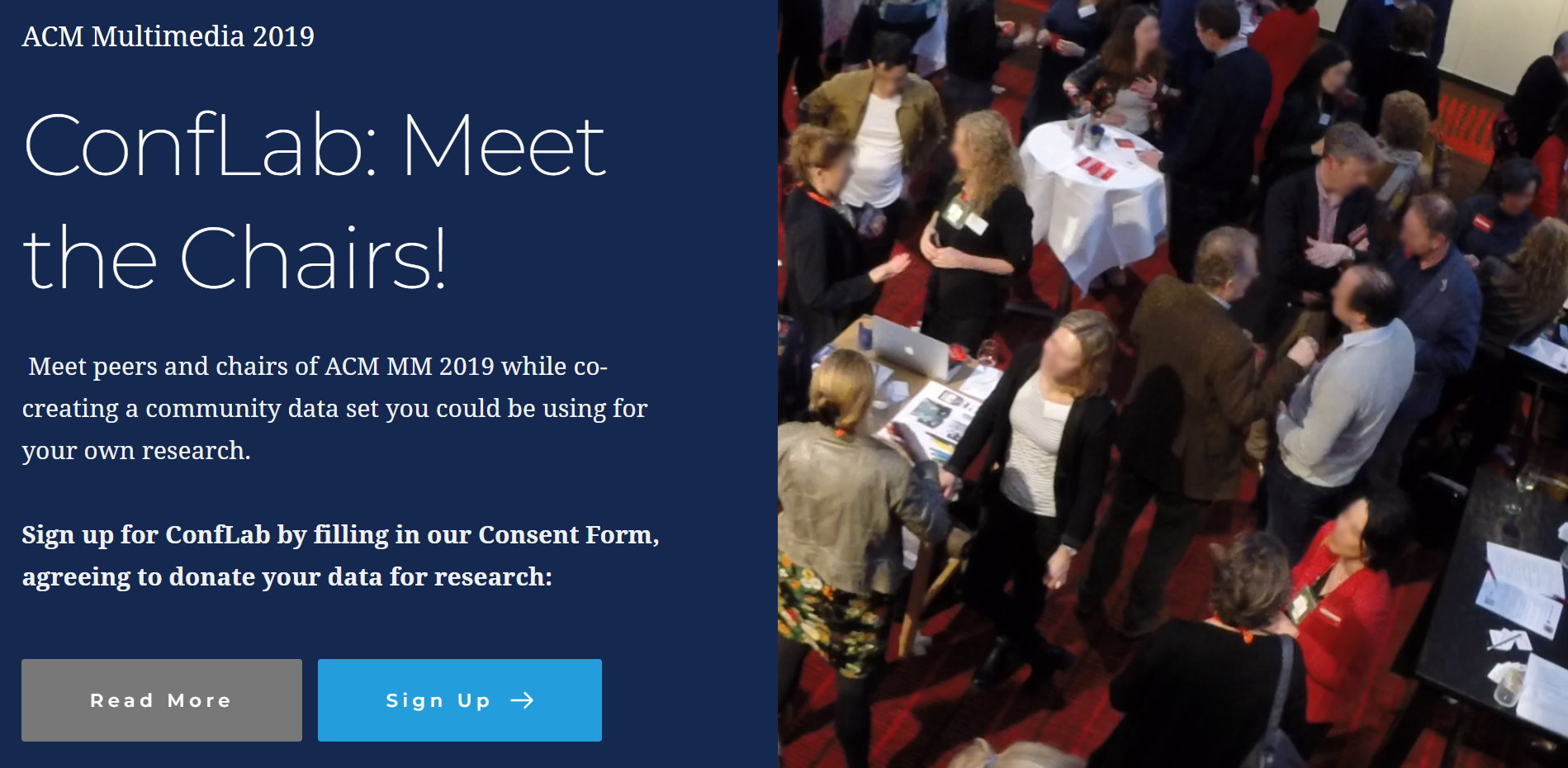}%
    \hspace{0.2cm}
    \includegraphics[height=3.3cm]{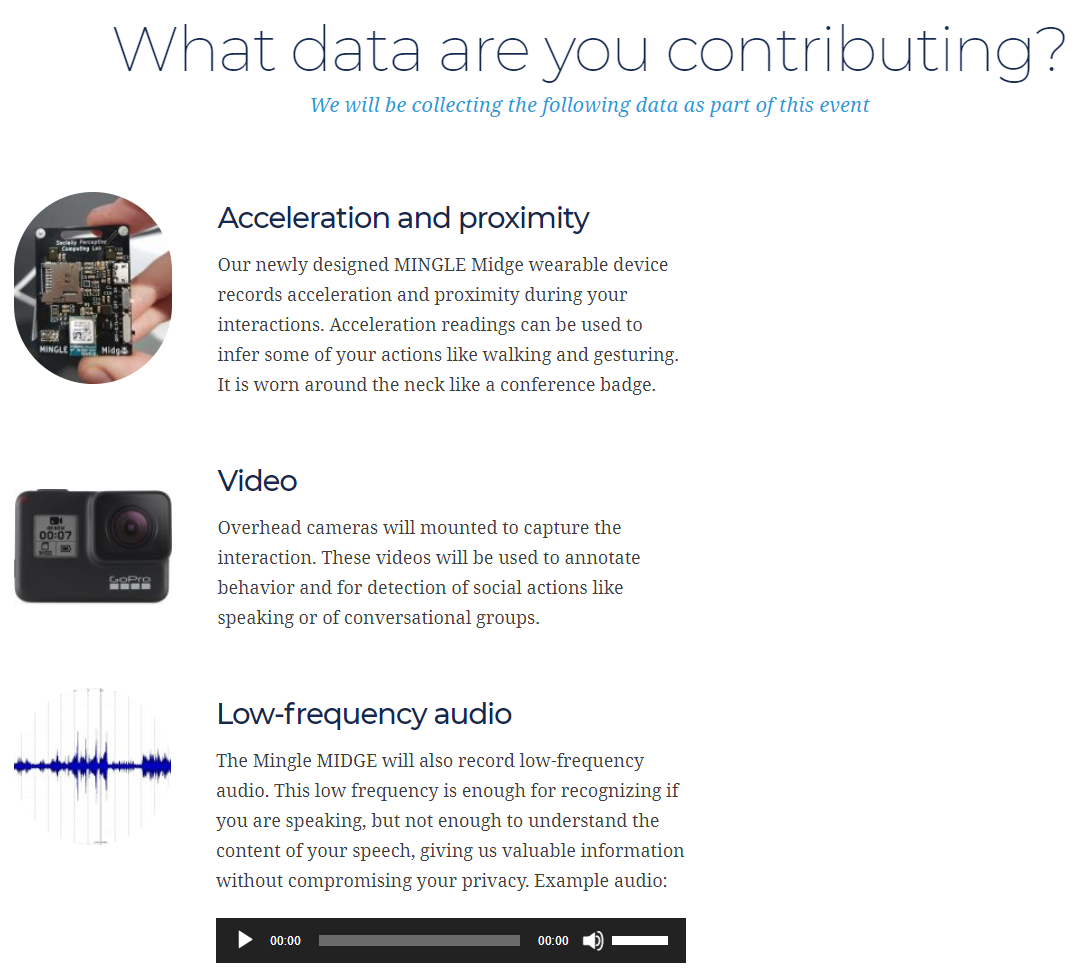}%
    \caption{Screenshots from the \textit{ConfLab: Meet the Chairs!} event website}
    \label{fig:event-website}
\end{minipage}\hfill
\begin{minipage}{0.21\textwidth}
    \centering 
    \includegraphics[height=3.3cm]{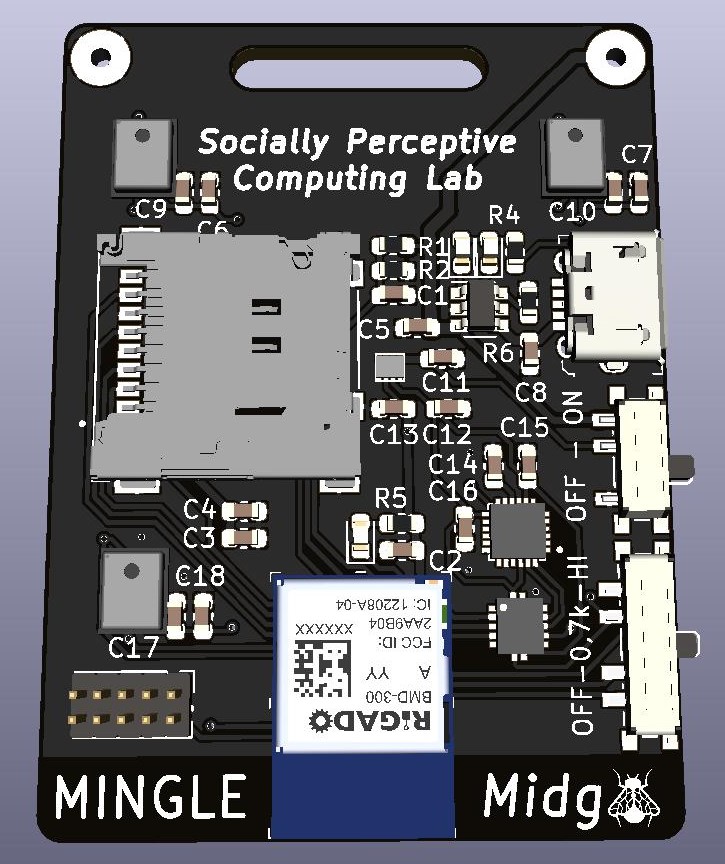}%
    \caption{The Midge}
    \label{fig:midge-design}
\end{minipage}
\vspace{-10pt}
\end{figure}
In this section we describe the considerations, design, and supporting community engagement activities for the first instantiation of ConfLab at ACM Multimedia 2019 (MM'19), to serve as a template and case study for other similar efforts.

\paragraph{Ecological Validity and Recruitment} An often-overlooked but crucial aspect of in-the-wild data collection is the design and ecological validity of the interaction setting \cite{andrade2018internal, labonte2018wild, hung2019complex}. To capture natural interactions in a professional setting and encourage mixed levels of status, acquaintance, and motivations to network, we co-designed a networking event with the MM'19 organizers called \textit{Meet the Chairs!} Our event website (\url{https://conflab.ewi.tudelft.nl/}) served to inform participants about the goals of a community created dataset, and transparently describe the data collection process (\figurename~\ref{fig:event-website}). 
During the conference, participants were recruited via word-of-mouth marketing, social media, conference announcements, and the event website. As an additional incentive beyond interacting with the Chairs and participating in a community-driven data endeavor, we provided attendees with post-hoc insights into their networking behavior from the collected wearable-sensors data. See Supplementary material for a sample participant report. 

\paragraph{Privacy and Ethics} The collection and sharing of ConfLab is GDPR compliant. The dataset design and process was approved by both, the Human Research Ethics Committee (HREC) at our institution (TUDelft) and the conference location's national authorities (France). All participants gave consent for the recording and sharing of their data at registration.(See the Datasheet in the Appendix for the consent form.) Given the involvement of private human data, ConfLab is only available for academic research purposes under an End User License Agreement. Such an \textit{as open as possible and as closed as necessary} ethos for open science acknowledges the limitation that personal data places on open sharing \cite{uyrdm, uurdm}.

\begin{figure}[!t]
    \centering
    \includegraphics[width=0.32\linewidth]{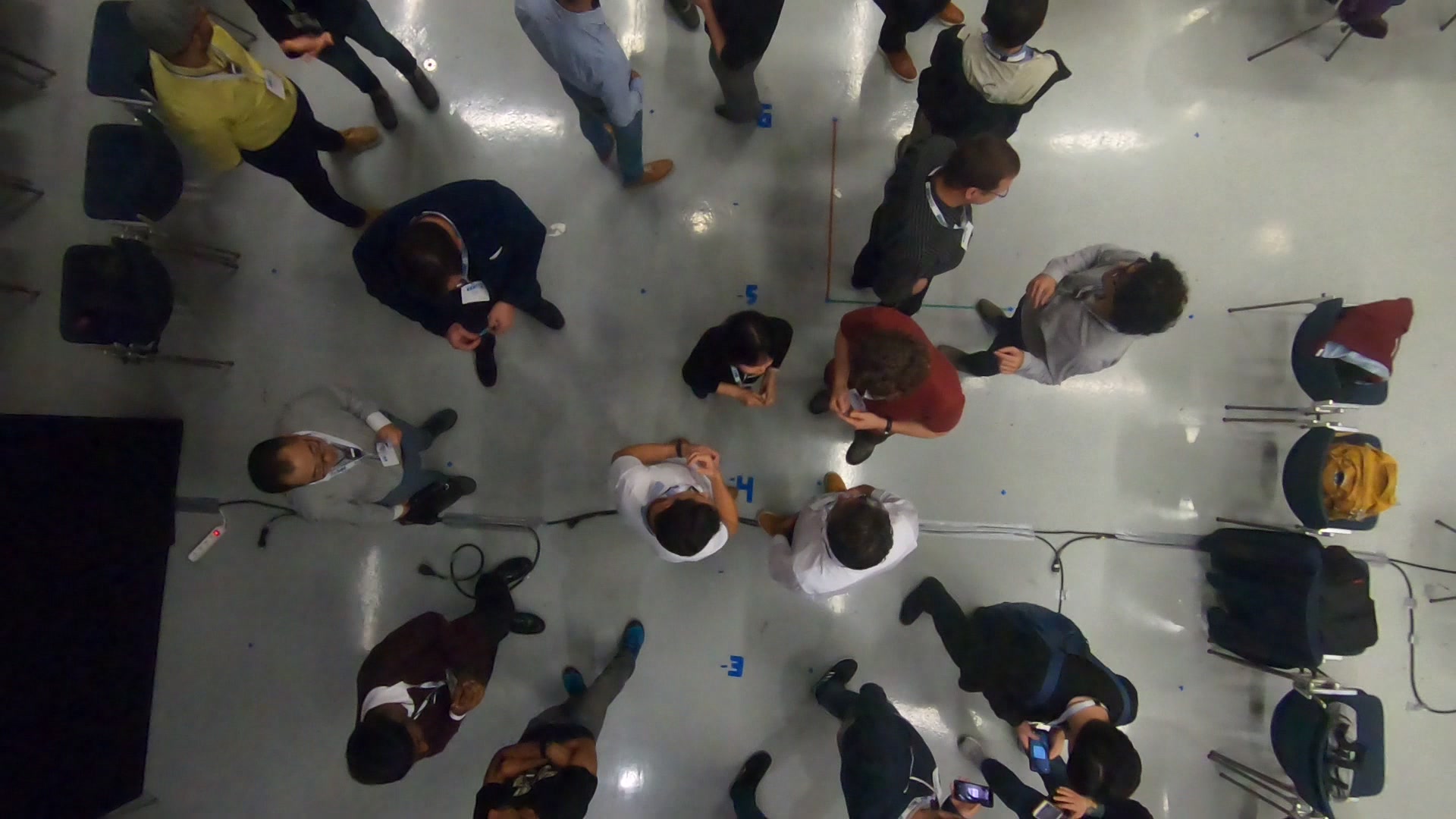}  \includegraphics[width=0.32\linewidth]{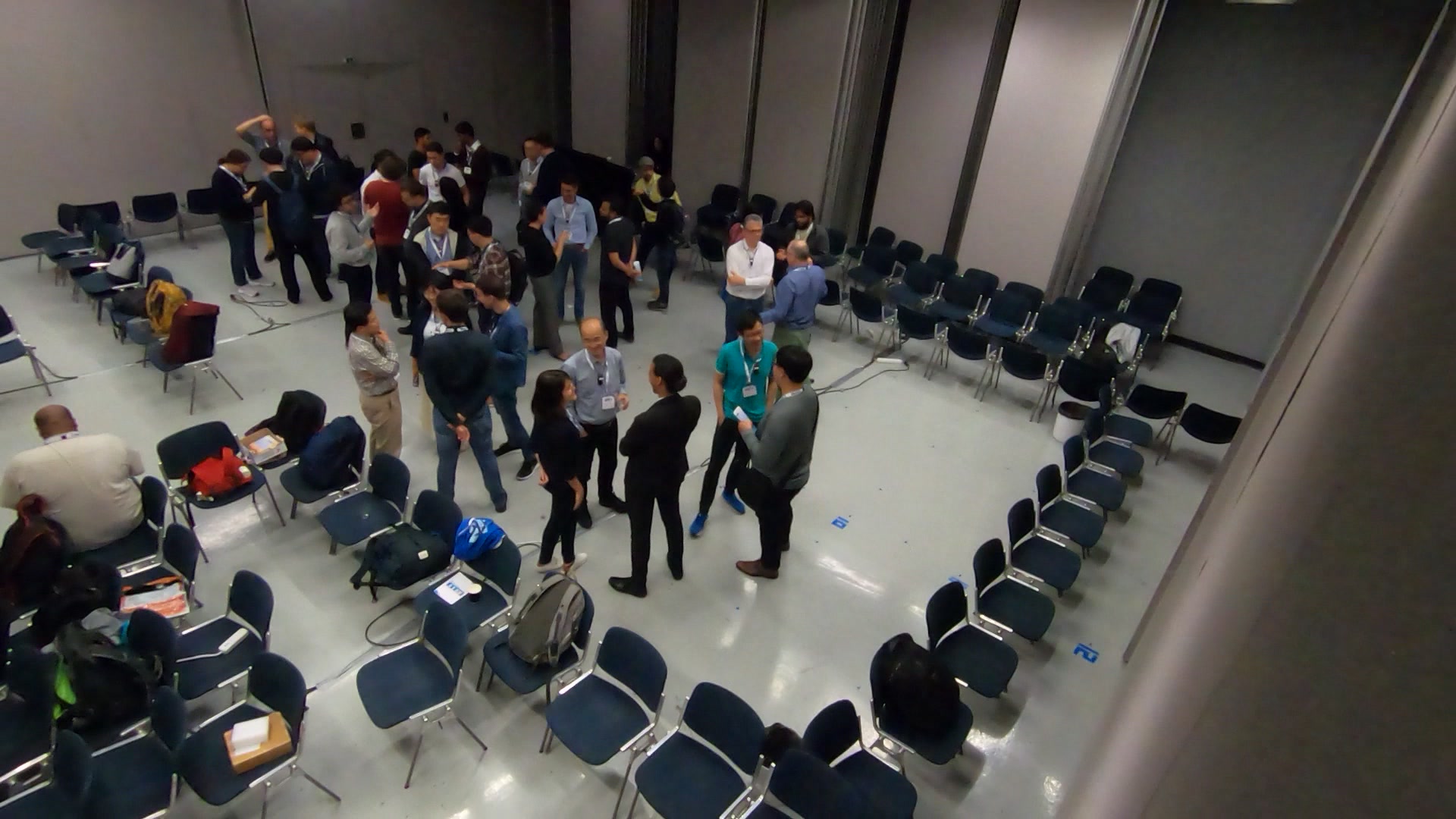}
    \includegraphics[width=0.32\linewidth]{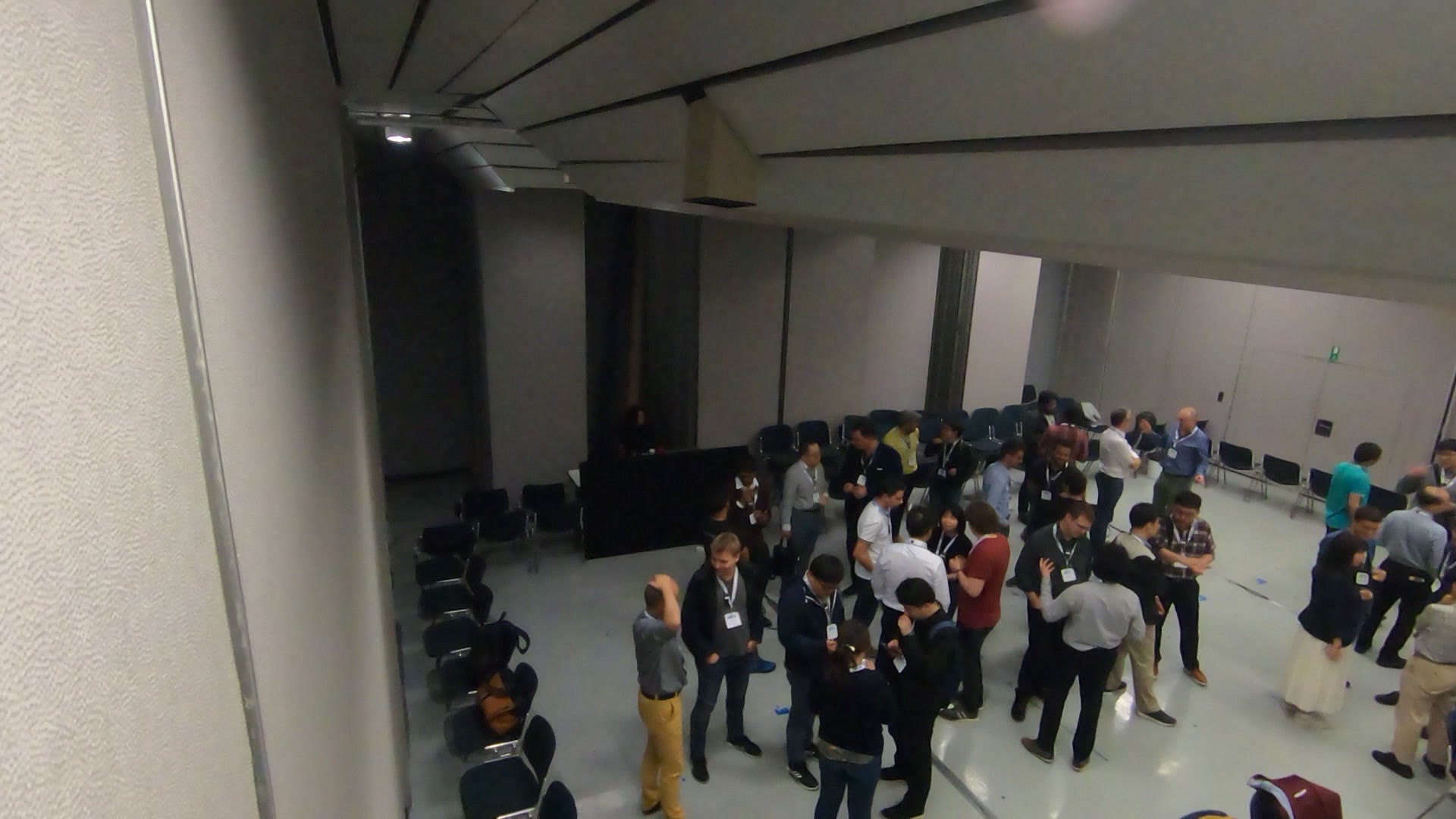}
    
    \includegraphics[width=0.32\linewidth]{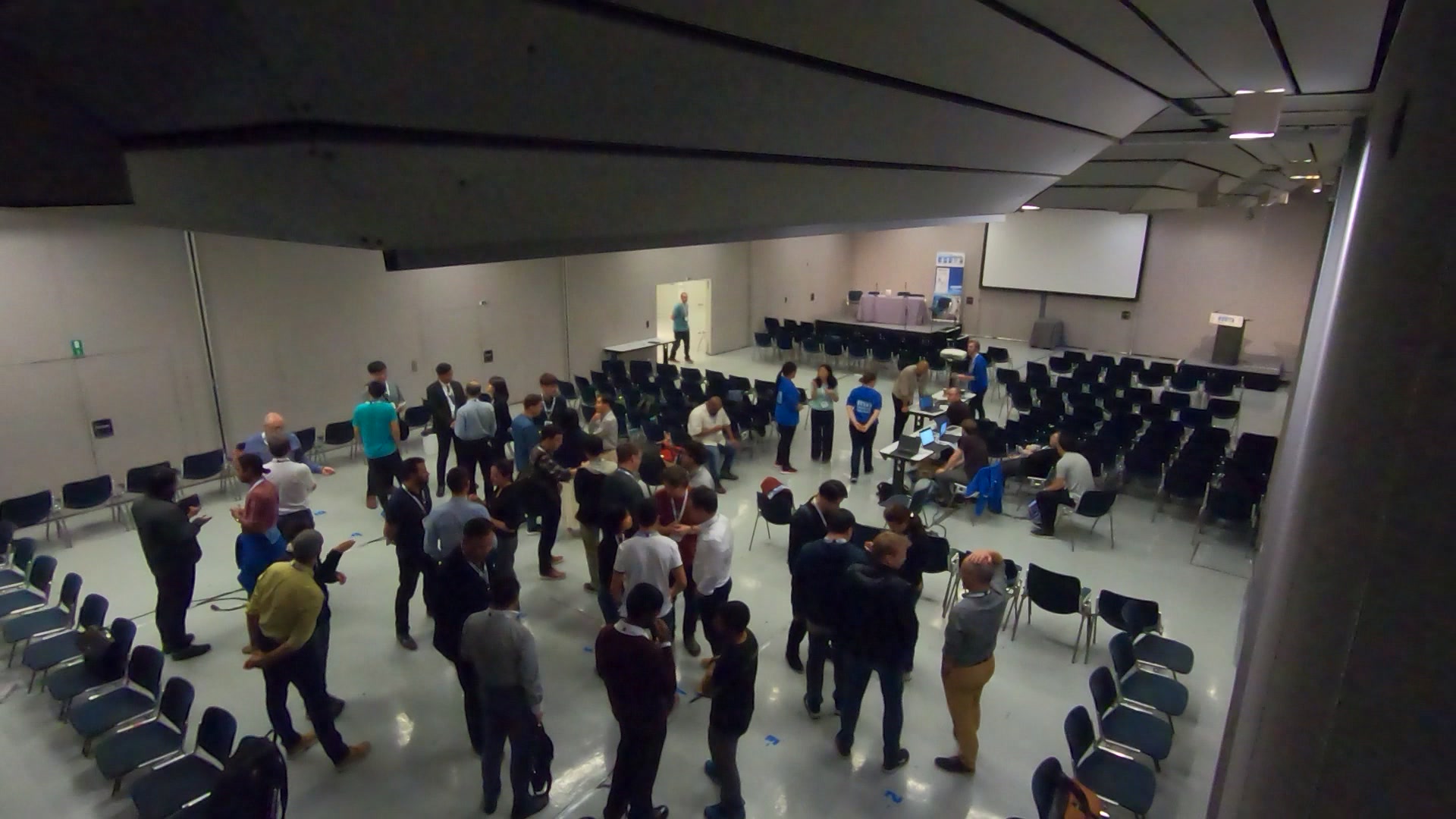}
    \includegraphics[width=0.32\linewidth]{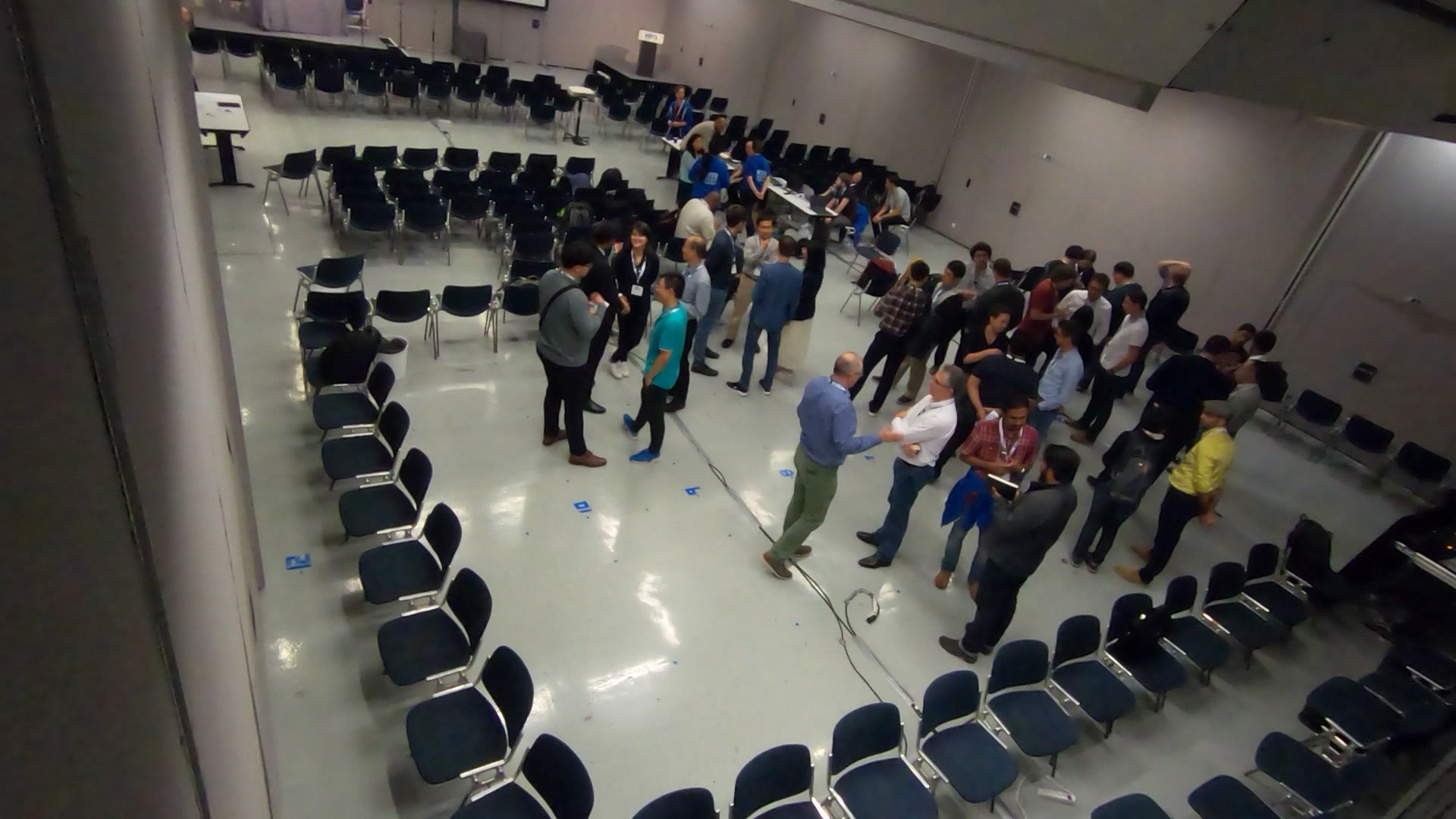}
    
    \caption{Comparing the top-down (top-left, camera 4) and elevated-side camera views (rest). Note how the top-down view is better at mitigating the capture of faces and suffers from fewer occlusions. This allows for a clearer capture of gestures and lower extremities for the most number of people while also preserving privacy.}
    \label{fig:cam_views}
\end{figure}

\paragraph{Data Capture Setup}
\label{sec:camwear}
Our goal while designing the capture setup was to find the best trade-off between maximizing data fidelity and interfering with the naturalness of the interaction (ecological validity) or violating participant privacy (ethical considerations). Through discussions with the HREC and General Chairs of MM'19 we decided to mitigate the capture of faces, which constitute one of the most sensitive personally-identifiable features. Avoiding the inclusion of faces serves two purposes. First, it safeguards against misuse in downstream tasks with potential negative societal impacts such as harmful surveillance. Such issues have led to the retraction of some person re-identification datasets \cite{ft2019whosusing}. Second, it protects the participants who are part of a real research community; since the dataset does not involve role-playing or scripted conversations, the dataset contains their actual behavior. Consequently, we chose an aerial perspective for the video modality (see \autoref{fig:cam_views}).
The $10$~m $\times$ $5$~m interaction area was recorded by $14$ GoPro Hero 7 Black cameras ($60$fps, $1080$p, Linear,
NTSC) \cite{gopro7}. $10$ of these were placed directly overhead at a height of $\sim3.5$~m at $1$~m intervals, with $4$ cameras at the corners providing an elevated-side-view perspective. (The HREC has suggested not sharing the elevated-side-view videos due to the presence of faces.) For capturing multimodal data streams, we designed a custom wearable multi-sensor pack called the Midge\footnote{Documentation and schematics: \url{https://github.com/TUDelft-SPC-Lab/spcl_midge_hardware}} (see \figurename~\ref{fig:midge-design} for a design render), based on the open-source Rhythm Badge designed for office environments \cite{lederman2018rhythm}. We improved upon the Rhythm Badge to achieve more fine-grained and flexible data capture (see Appendix~\ref{app:data-capture}). We designed the Midge in a conference badge form-factor for seamless integration. Unlike smartphones, wearable badges allow for a simple \textit{grab-and-go} setup and do not suffer from sensor/firmware differences across models. Popular human behavior datasets are synchronized by maximizing similarity scores around manually identified common events, such as infrared camera detections \cite{alameda2015salsa}, or speech plosives \cite{recola}. While recordings in lab settings can allow for fully wired recording setups, recording in-the-wild requires a distributed wireless solution. We developed a solution to synchronize the cameras and wearable sensors directly at acquisition while significantly lowering the cost of the recording setup \cite{raman2020modular}, making it easier for others to replicate our capture setup. See Appendix~\ref{app:data-capture} for synchronization and calibration details, and Appendix~\ref{app:datasheet} for images of the setup.

\paragraph{Data Association and Participant Protocol}
\label{sec:protocol}
One consideration for multimodal data recording is the data association problem\textemdash how can pixels corresponding to an individual be linked to their other data streams? 
To this end, we designed a participant registration protocol. Arriving participants were greeted and fitted with a Midge. The ID of the Midge acted as the participant's identifier. One team member took a picture of the participant while ensuring both the face of the participant and the ID on the Midge were visible. In practice, it is preferable to avoid this step by using a fully automated multimodal association approach. However this remains an open research challenge \cite{cabrera2016matching,cabrera2018hierarchical}.
During the event, participants mingled freely\textemdash they were allowed to carry bags or use mobile phones. Conference volunteers helped to fetch drinks for participants. Participants could leave before the end of the one hour session. 
\paragraph{Replicating Data Collection Setup and Community Engagement}
\label{sec:postevent}
After the event, we gave a tutorial at MM'19 \cite{hung2019multimodal} to demonstrate how our collection setup could be replicated, and to invite conference attendees and event participants to reflect on the broader considerations surrounding privacy-preserving data capture, sharing, and future directions such initiatives could take. 

\section{Data Annotation}
\label{sec:annotation}
\paragraph{Continuous Keypoints Annotation}

Existing datasets of in-the-wild social interactions have mainly focused on localizing subjects via bounding boxes \cite{alameda2015salsa,MnM2021}. However, richer information about the social dynamics such as gestures and changes in orientation cannot be retrieved from bounding boxes alone, and necessitates the labeling of multiple skeletal keypoints. The typical approach to keypoint annotation involves using tools such as Vatic \cite{Vondrick2013} or CVAT \cite{CvatGithub} to manually label every $N$ frames followed by interpolating over the rest of the frames. This one-frame-at-a-time annotation procedure makes obtaining keypoint annotations a labor- and cost-intensive process. Moreover, interpolation fails to capture the finer temporal dynamics of the underlying behavior, and reduces the benefits of higher-framerate video capture. Limited by existing tools, no related dataset of in-the-wild human behavior has included time-continuous pose or speaking status annotations. 




\begin{figure*}[t]

\sbox\twosubbox{%
  \resizebox{\dimexpr0.95\textwidth-1em}{!}{%
    \includegraphics[height=3cm]{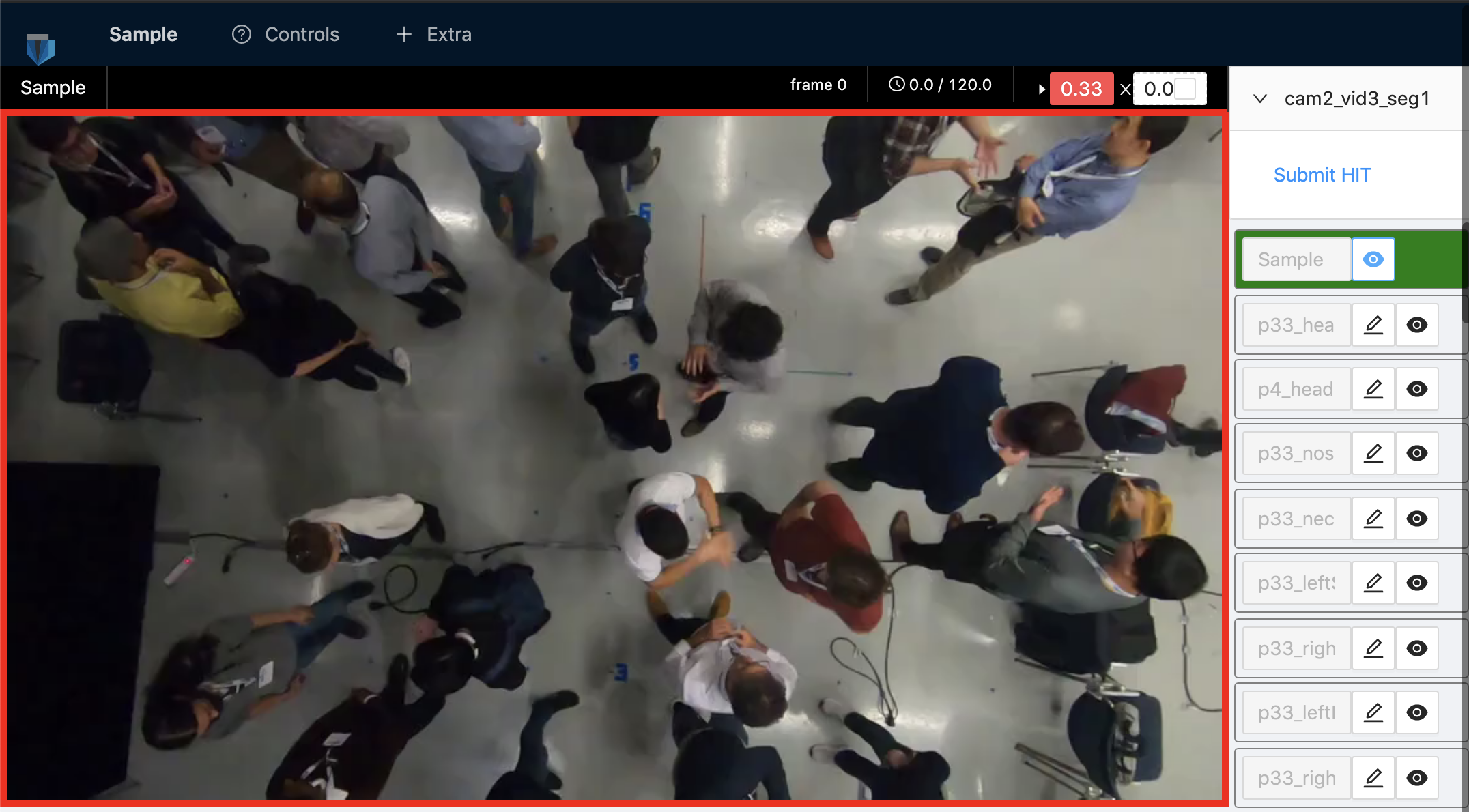}%
    \includegraphics[height=3cm]{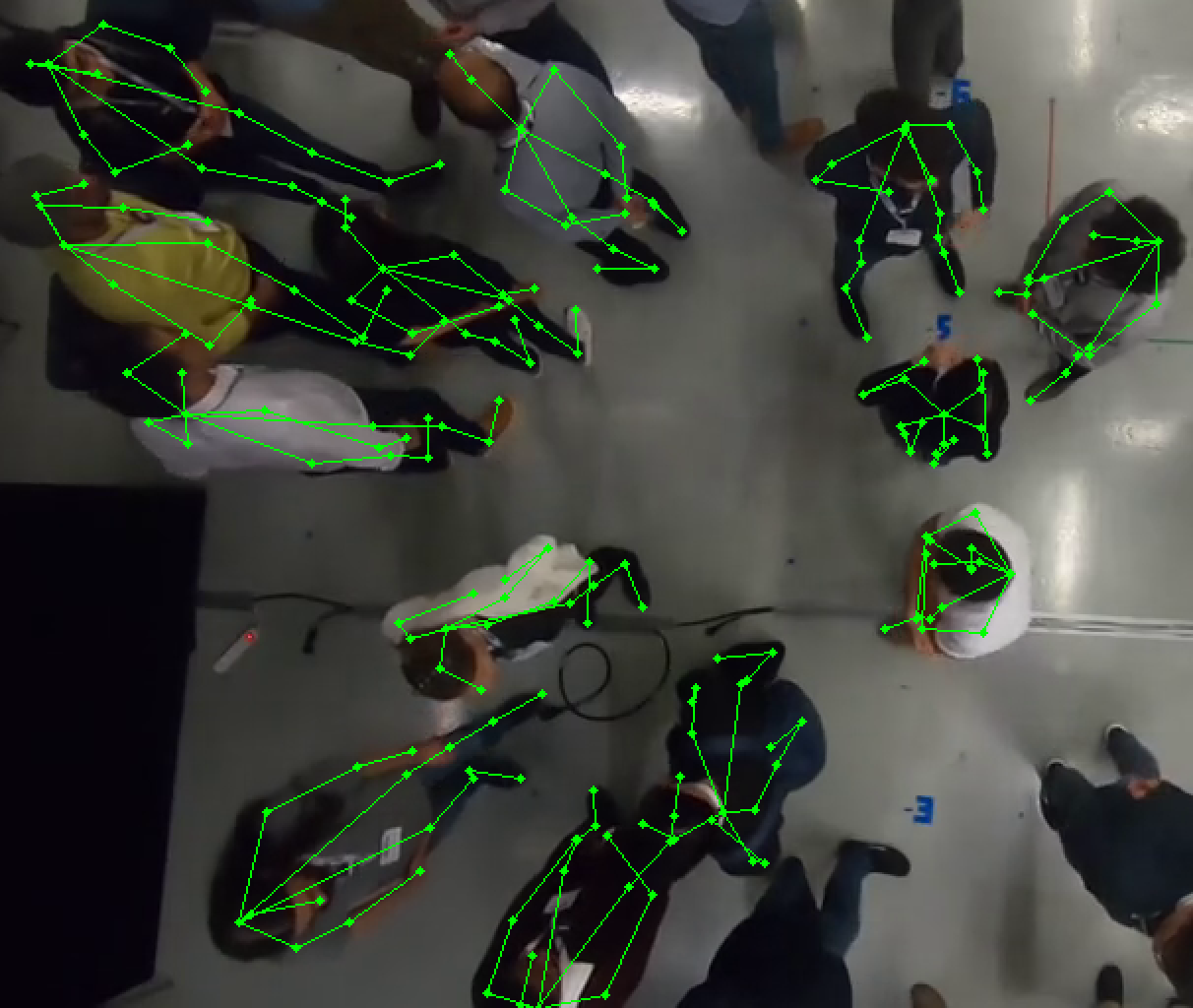}%
    \includegraphics[height=3cm]{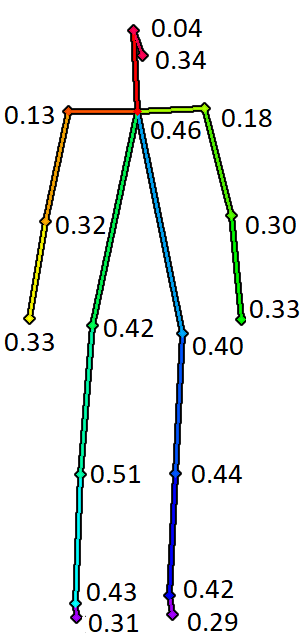}
  }%
}
\setlength{\twosubht}{\ht\twosubbox}

\centering

\subcaptionbox{Keypoint annotation interface in covfee \cite{CovfeeGithub} \label{fig:covfee}}{%
  \includegraphics[height=\twosubht]{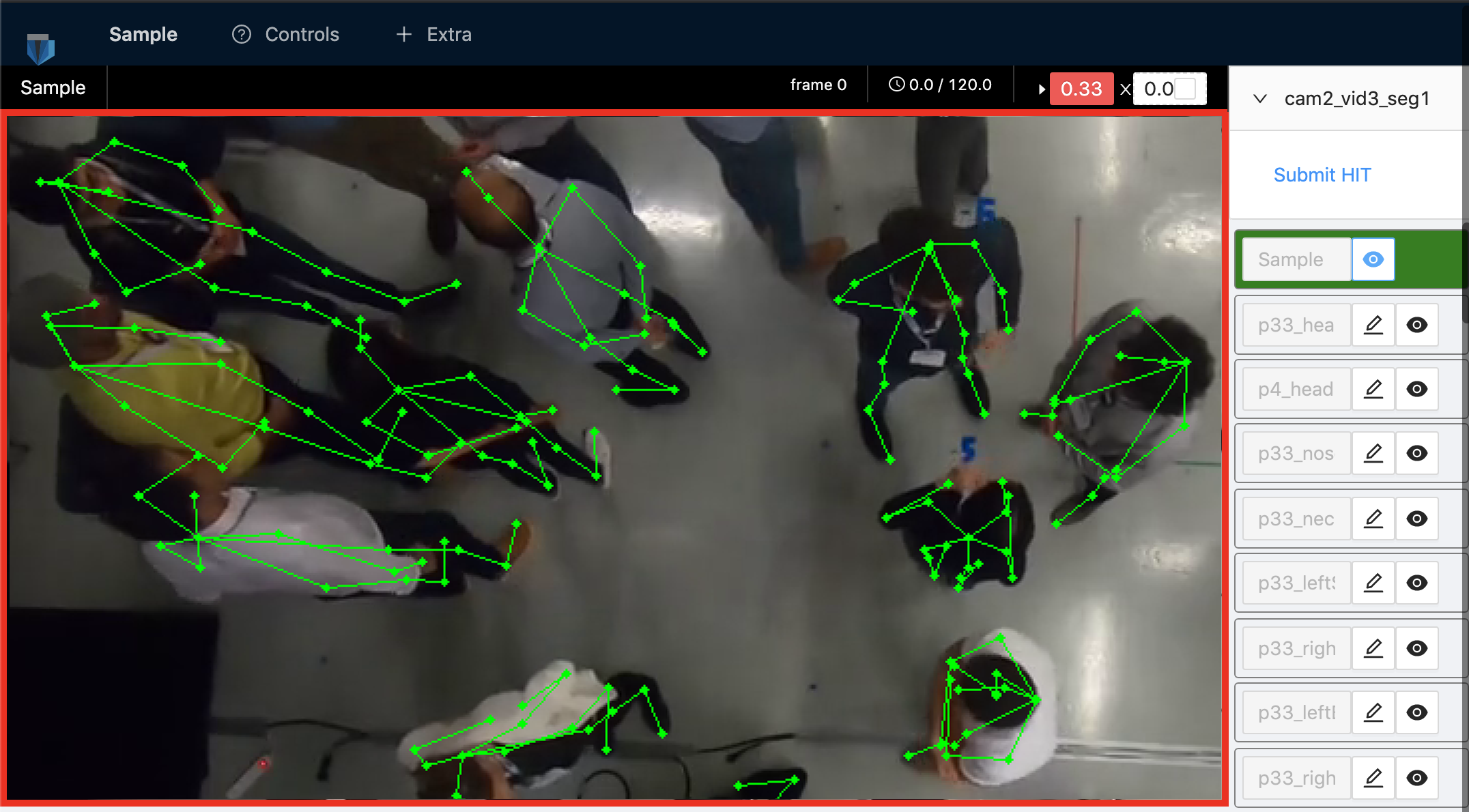}%
}\quad
\subcaptionbox{Gallery of identities\label{fig:annotated-kps}}{%
  \includegraphics[height=\twosubht]{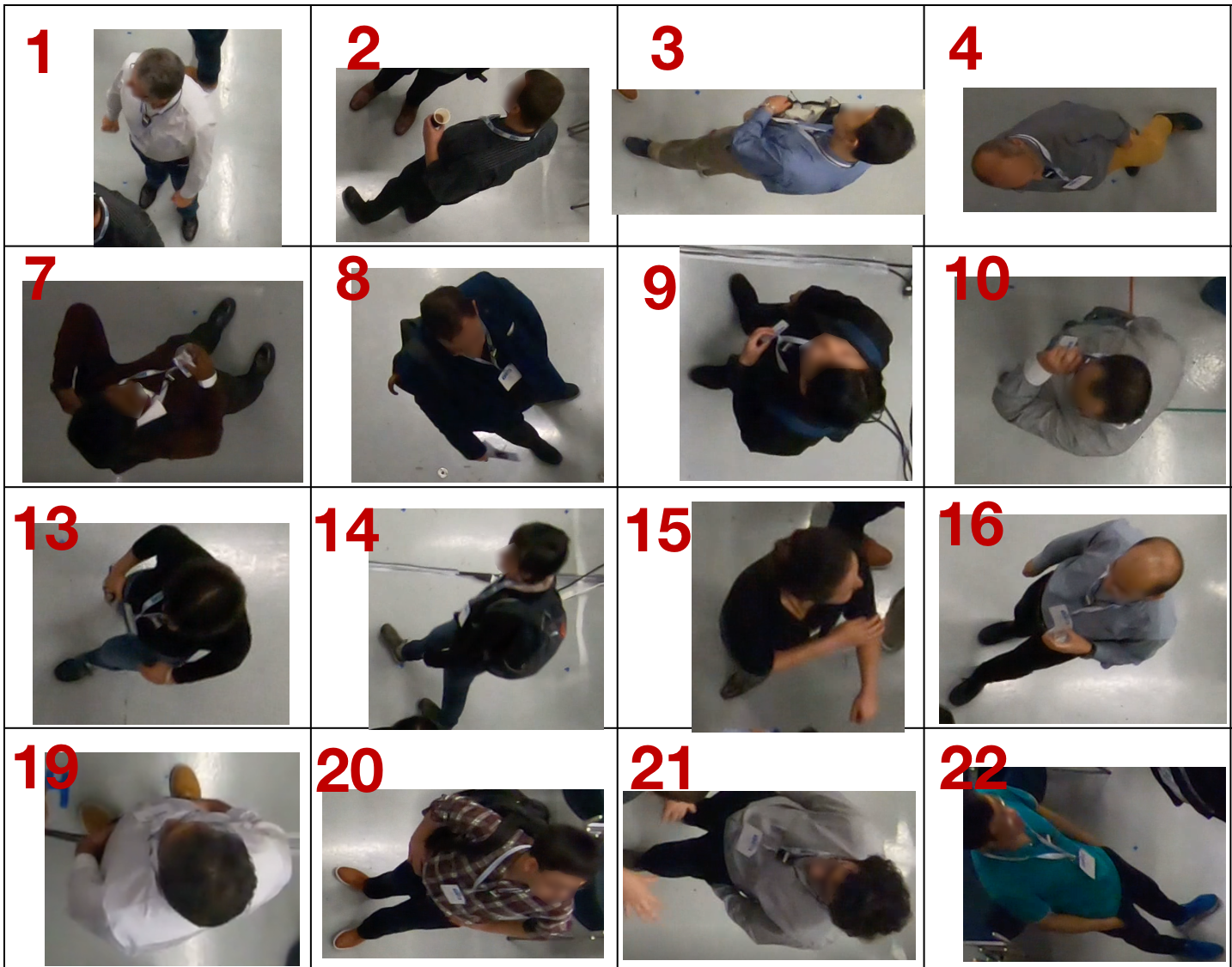}%
} \quad
\subcaptionbox{Occlusion\label{fig:skeleton}}{%
  \includegraphics[height=\twosubht]{imgs/skeleton.png}%
}
\caption{Illustration of the body keypoints annotation procedure: (a): our custom time continuous annotation interface; (b): the gallery of person identities used by annotators to identify people in the scene (faces blurred); and (c): the skeleton template with the fraction of occluded frames.}
\label{fig:keypoints}
\vspace{-10pt}
\end{figure*}

In contrast, to overcome these issues we collected fine-grained time-continuous annotations of keypoints via a web-based interface implemented as part of the Covfee framework \cite{covfee-paper}. Here, annotators follow individual joints using their mouse or trackpad while playing the video in their web browser. The playback speed of the video is automatically adjusted using an optical-flow-based technique to enable annotators to follow keypoints continuously without pausing the video. This design enables easy keypoint labeling in \textit{every} frame of the video ($60$~Hz). We also incorporated a binary \textit{occlusion} flag for every body keypoint. Annotators simultaneously controlled this flag to indicate when a body joint was not directly visible. Note that the flag is only an additional confidence indicator; we asked the annotators to label the occluded keypoint using their best estimate if it was deemed to be within the frame. Our pilot study on the efficacy of Covfee compared to non-continuous annotation via CVAT \cite{CvatGithub} is presented in \cite{covfee-paper}. For the pilot annotators, the continuous annotation methodology resulted in a 3$\times$ speedup with statistically indifferent error rates. 

We chose the top-down camera views for annotation since they suffer from fewer occlusions than the elevated-side views, enabling improved capture of gestures and lower extremities for more number of people (see \autoref{fig:cam_views}). Given the overlap in the camera views, we annotated keypoints in five of the ten overhead cameras (see \autoref{fig:scene}). Note that the same subject could be annotated in multiple cameras due to the overlap in even the five annotated cameras. Videos were split into two-minute segments to ease the annotation procedure. Each segment was annotated by one annotator by tracking the joints of all the people in the scene.

\paragraph{Continuous Speaking Status Annotations}
Speaking status is a key non-verbal cue for many social interaction analysis tasks \cite{4042075}.
We annotated the binary speaking status of every subject due to its importance as a key feature of social interaction \cite{Muller2018a,gedikhungIMWUT2018,ramanAutomaticEstimationConversation2019,hung2010estimating,5549893} and to contribute the existing community who are working on this task \cite{Beyan2020,Shahid2019,Gedik2017a}. 
Action annotations have traditionally been carried out using frame-wise techniques \cite{MnM2021}, where annotators find the start and end frame of the action of interest using a graphical interface. 
Given the speed enhancement of continuous annotation, we also annotated speaking status via a continuous technique. We implemented a binary annotation interface as part of Covfee \cite{covfee-paper}. We asked annotators to press a key when they perceived speaking starting or ending. In a pilot study with two annotators, we measured a frame-level agreement (Fleiss’ $\kappa$) of 0.552, comparable to previous work \cite{Cabrera-Quiros2018b}. 
Similar to \cite{MnM2021}, the annotations were made by watching the video. We provided the annotators with all overhead views to best capture visual behavior. 


\paragraph{F-formation Annotations}
Identifying who is likely to have social influence on whom is another important feature for analyzing social behavior. This is operationalised via the theory of F-formations, which are groups of people arranging themselves to converse or socially interact. Similar to prior datasets \cite{MnM2021,alameda2015salsa,ZenEtAl2010}, F-formations group membership were annotated using an approximation of Kendon's definition \cite{Kendon1990}. F-formation stands for Facing formation, which is a socio-spatial arrangement where people have direct, easy and equal access while excluding the space from others in the surroundings. The arrangement commonly maintains a convex space in the middle of all the participants (determined by the location and orientation of their lower body), although other spatial arrangements (e.g., side-by-side, L-shaped) are possible, especially for smaller-sized groups of people. Annotations were labeled by one annotator at 1 Hz, following this definition. Since this is a largely objective and common framework for defining F-formations, we deemed it sufficient to obtain one set of annotations. Further, since F-formations may span camera views, we always used the camera that captured each F-formation in its entirety for annotation.

\section{Dataset Statistics}
\label{sec:description}
\begin{figure}[t]

\sbox\twosubbox{%
  \resizebox{\dimexpr0.95\textwidth-1em}{!}{%
    \includegraphics[height=7cm]{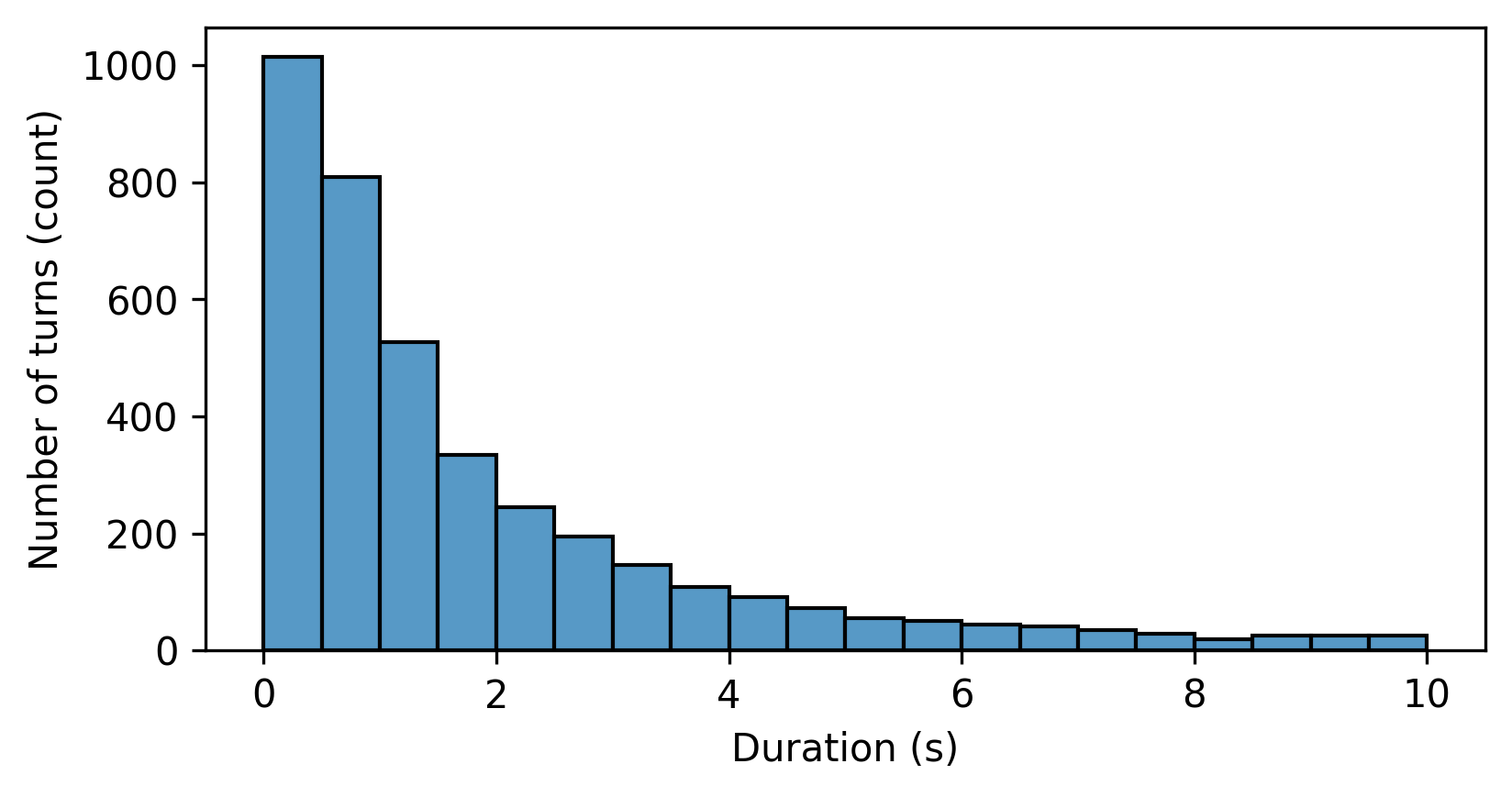}%
    \includegraphics[height=7cm]{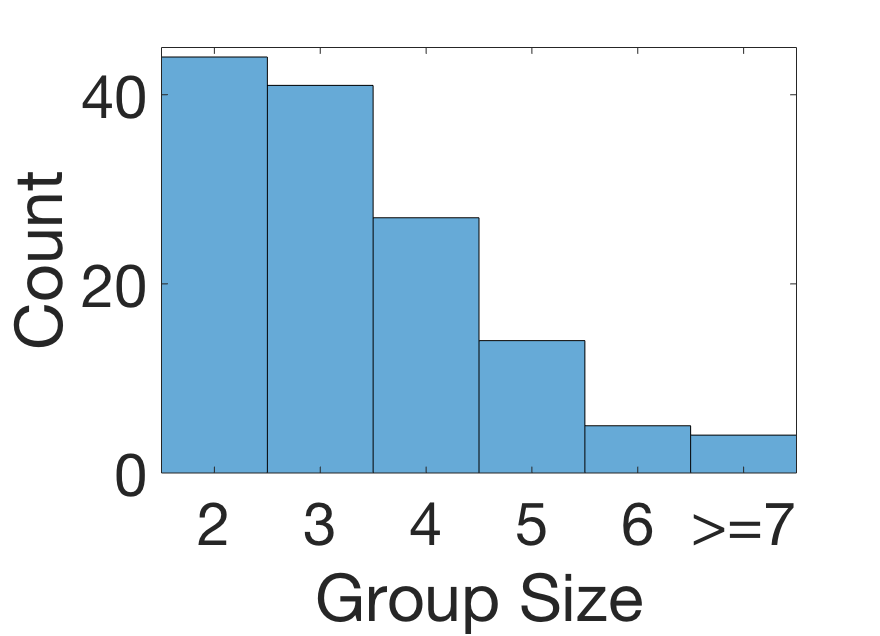}%
    \includegraphics[height=7cm]{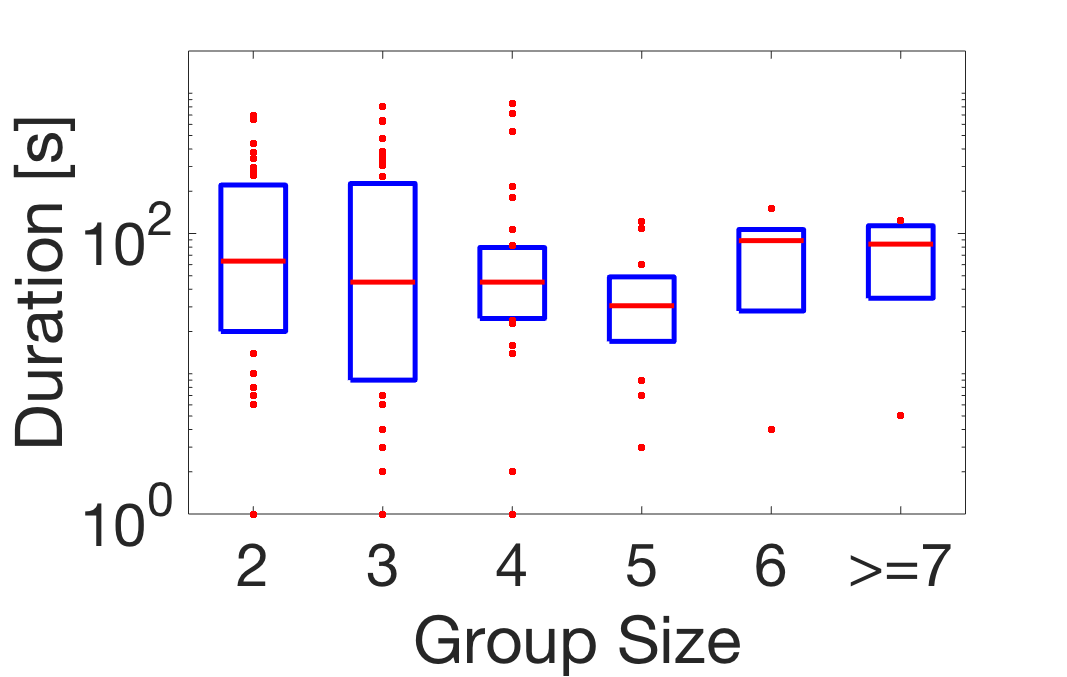}
    \includegraphics[height=7cm]{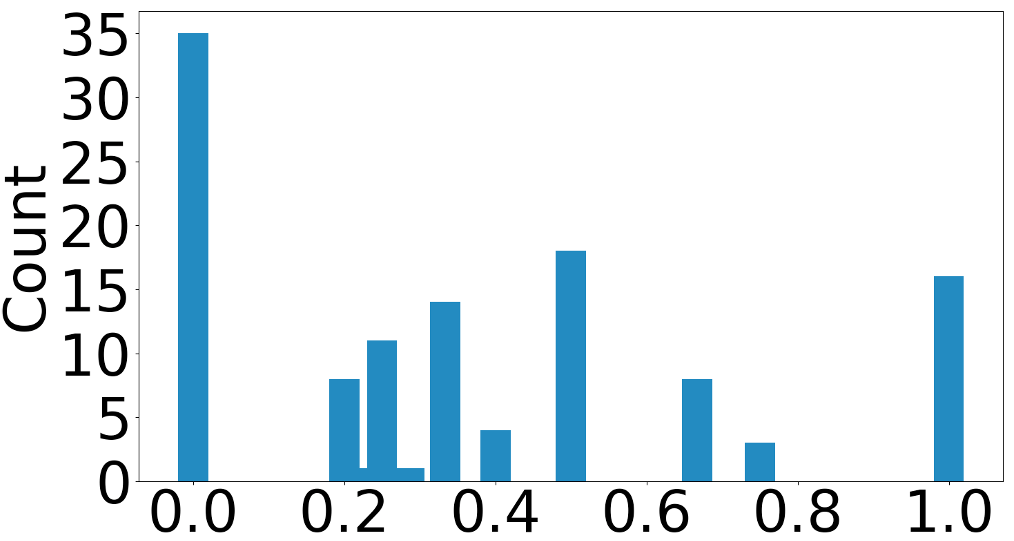}
  }%
}
\setlength{\twosubht}{\ht\twosubbox}

\centering

\subcaptionbox{\small speaking turn lengths\label{fig:ssstats}}{%
  \includegraphics[height=\twosubht]{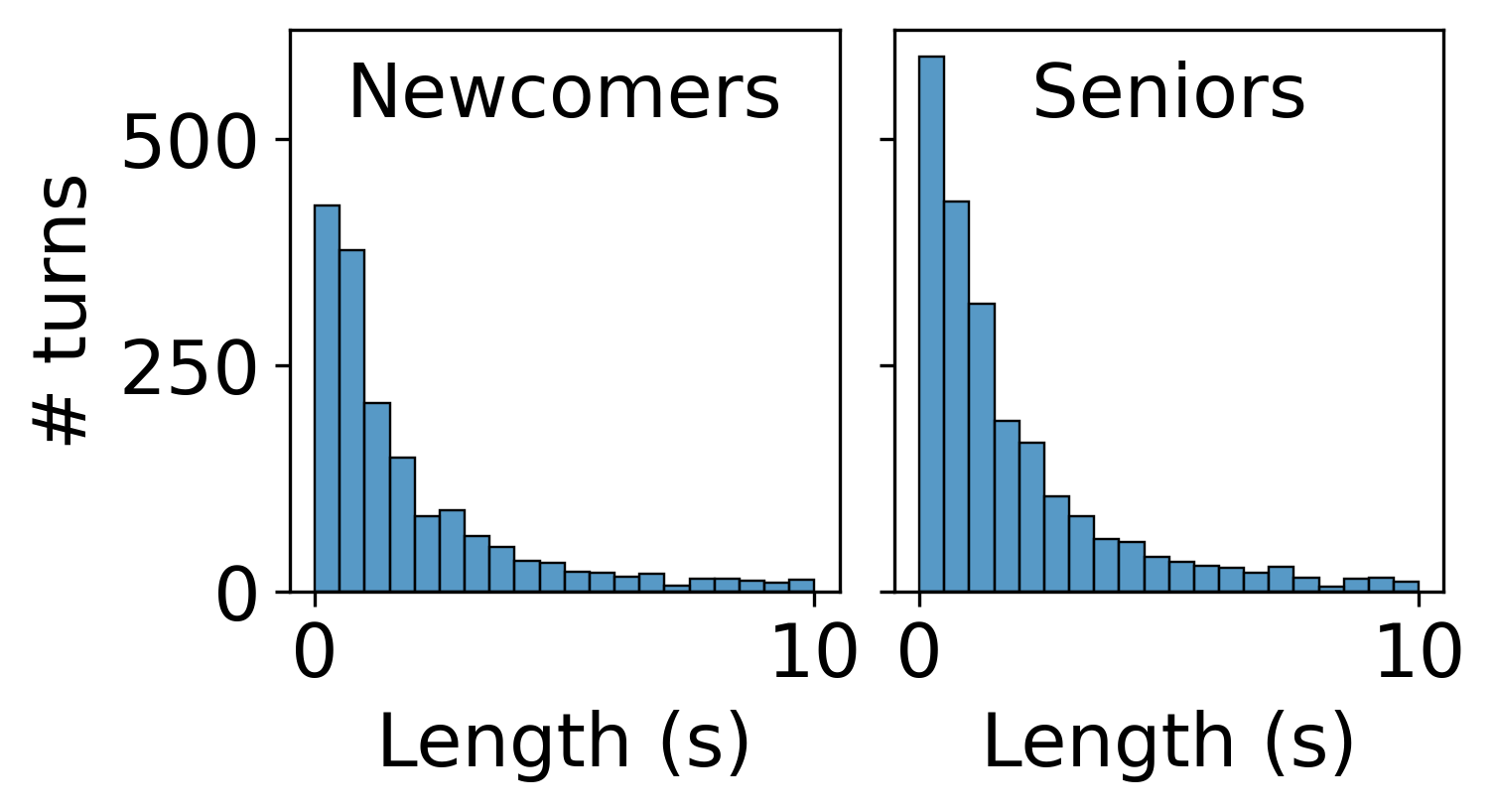}%
}\quad
\subcaptionbox{\small group size\label{fig:FFstats_size}}{%
  \includegraphics[height=\twosubht]{imgs/group_size_revised3.png}%
} \quad
\subcaptionbox{\small group duration\label{fig:FFstats_duration}}{%
  \includegraphics[height=\twosubht]{imgs/group_duration_revised3.png}%
}
\quad
\subcaptionbox{\small fraction of newcomers in groups \label{fig:newcomer_stats} }{%
  \includegraphics[scale=0.2]{imgs/newcomer_stats3.png}%
}
\caption{Data distributions for speaking status and conversation groups}
\label{fig:anno_stats}

\end{figure}
\paragraph{Individual-Level Statistics}
\autoref{fig:skeleton} shows the average occlusion values we obtained from annotators for each of the $17$ keypoints. In \autoref{fig:ssstats} we show the distribution of turn lengths in our speaking status annotations, for both newcomers and veterans, as per their self-reported newcomer status to the conference. 
We defined a turn to be a contiguous segment of positively-labeled speaking status, which resulted in a total of $4096$ turns annotated.

\paragraph{Group-Level Statistics}
We found $119$ distinct F-formations of size greater than or equal to two, and $38$ instances of singletons. Of these, there are $14$ F-formations and $2$ singletons that include member(s) using the mobile phone. The distributions for group size and duration per group size are shown in \autoref{fig:FFstats_size} and \autoref{fig:FFstats_duration}, respectively. 
Mean group duration doesn't seem to be influenced by group size although higher variations are seen at smaller group sizes. 
The fraction of community newcomers (first-time attending the conference) in groups is summarized in histogram in \autoref{fig:newcomer_stats}. The figure demonstrates two peaks on both sides of the spectrum (i.e., no newcomers vs. all newcomers in the same group). This spread over mixed and non-mixed seniority presents opportunities to study how acquaintance and seniority influence conversation dynamics.

\section{Research Tasks}
\label{sec:tasks}

We report experimental results on three baseline benchmark tasks: person and keypoints detection, speaking status detection, and F-formation detection. The first task is a fundamental building block for automatically analyzing human social behaviors. The other two demonstrate how learned body keypoints can be used in the behavior analysis pipeline. We chose these benchmarking tasks since they have been commonly studied on other in-the-wild behavior datasets. Code for all benchmark tasks is available at: \url{https://github.com/TUDelft-SPC-Lab/conflab}. See the \textit{Uses} section of the Datasheet in the Appendix for a discussion of the broader range of tasks ConfLab enables.

\subsection{Person and Keypoints Detection}
\label{subsec:kp-experiments}
This benchmark involves the tasks of person detection (identifying bounding boxes) and pose estimation (localizing skeletal keypoints). Since pre-trained SOTA methods struggle with a privacy-sensitive top-down perspective \cite{carissimi2018filling} (also see \figurename~\ref{fig:pretrained-rsn} and Appendix~\ref{app:kp_results} for ConfLab results), we finetune COCO-pretrained models on our dataset. We used Mask-RCNN \cite{he2017maskrcnn} (Detectron2 framework \cite{wu2019detectron2} implementation) with a ResNet-50 backbone for both tasks for benchmarking. Since keypoint annotations were made per camera, we used four of the overhead cameras for training (Cameras 2, 4, 8, 10) and one for testing (Camera 6). Implementation details are available in Appendix~\ref{app-subsec:kp-implementation}.

\paragraph{Evaluation Metrics}
We evaluated person-detection performance using the standard metrics in the MS-COCO dataset paper \cite{lin2014microsoftcoco}. We report average precision (AP) for intersection over union (IoU) thresholds of $0.50$ and $0.75$, and the mean AP from an IoU range from $0.50$ to $0.95$ in $0.05$ increments. For keypoint detection, we use object keypoint similarity (OKS) \cite{lin2014microsoftcoco}. AP$^\text{OKS}$ is a mean average precision for different OKS thresholds from $0.5$ to $0.95$. 

\begin{figure}[!t]
\begin{minipage}[t]{.43\linewidth}
\centering
\captionsetup{type=table}
\ra{1.3}
\caption{Mask-RCNN results for person bounding box detection and keypoint estimation.} 
\label{tab:mask_rcnn}
\begin{adjustbox}{max width=\linewidth}
\begin{tabular}{@{}lcccccc@{}} 
\toprule
\multirow{2}{*}{Model} &
 \multicolumn{3}{c}{Person Detection} & \multicolumn{3}{c}{Keypoint Estimation} \\ \cmidrule{2-7} & AP$_\text{50}$ & AP & AP$_\text{75}$
& AP$^\text{OKS}_\text{50}$ & AP$^\text{OKS}$ & AP$^\text{OKS}_\text{75}$ \\
\midrule
R50-FPN & 73.9 & 38.9 & 38.4 & 45.3 & 13.5 & 3.3  \\
\bottomrule
\end{tabular}
\end{adjustbox}
\end{minipage}\hfill
\begin{minipage}[t]{.55\linewidth}
    \centering
    \captionsetup{type=figure}
    \includegraphics[width=0.49\linewidth]{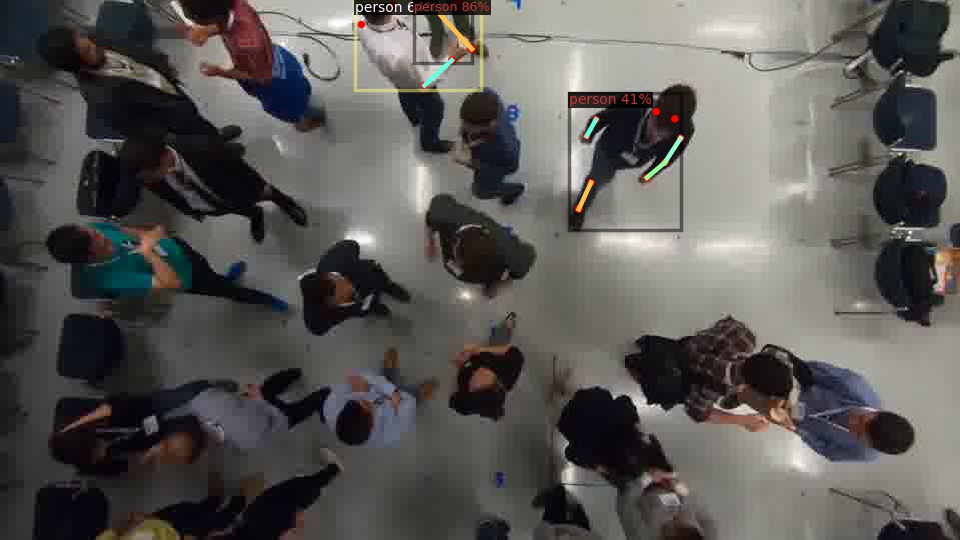}
    \includegraphics[width=0.49\linewidth]{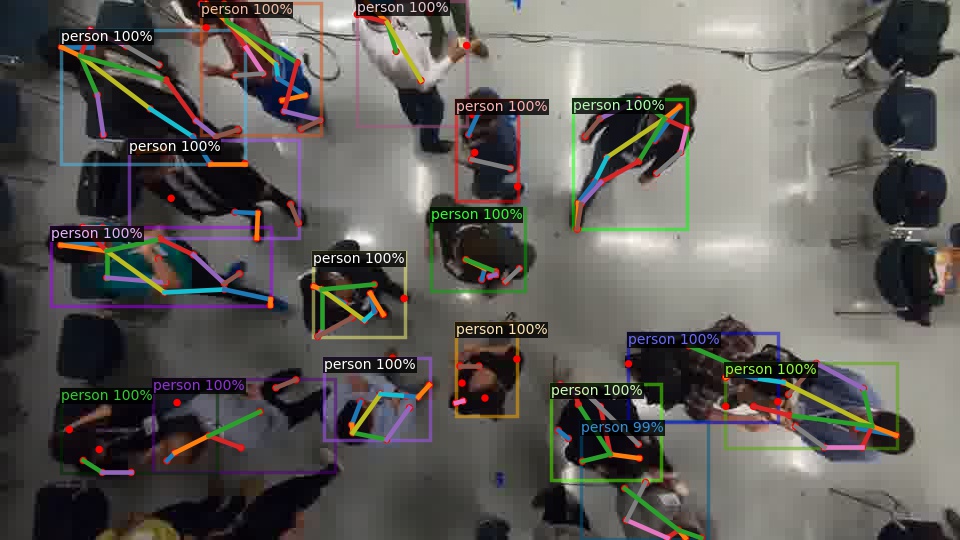}
    \caption{Predictions from the Mask-RCNN model; COCO pretrained (left), and ConfLab finetuned (right).}
    \label{fig:det2pred-corto}
\end{minipage}
\vspace{-1pt}
\end{figure}

\paragraph{Results and Analyses}
\autoref{tab:mask_rcnn} summarizes our person detection and joint estimation results. Our baseline achieves 73.9 AP$_\text{50}$ in detection and 45.3 AP$^\text{OKS}_\text{50}$ in keypoint estimation. Figure \ref{fig:det2pred-corto} shows qualitative results from our fine-tuned network.
For further insight we performed several analyses and ablations. In Appendix Table~\ref{tab:abl_n}, we depict the effect of varying the number of training samples on performance. For training, we use the same four cameras and only vary the number of frames for each camera. We evaluate on the same testing images from camera 6. We find that performance saturates at $16$\% training samples. We next investigated the effect of increasing training data size by adding specific cameras one at a time. We report results in Appendix Table~\ref{tab:abl_data_amount}. There is a $260$\% performance gain when first doubling the training samples to $69$~k with the addition of camera 4, and a $46$\% gain when adding another $43$~k samples from camera 8. 
Finally, since the lower body regions suffer from higher occlusion, we experiment with different sections of body for further insight and report results in Appendix Table~\ref{tab:num_kp}. 

\subsection{Speaking Status Detection}
\label{sec:speaking-status}

In data collected from real-life social settings, individual audio recordings can be hard to obtain due to privacy concerns \cite{Shen2018}. This has led to the exploration of other modalities to capture some of the motion characteristics of speaking-related gestures \cite{Cabrera-Quiros2018b,Quiros2019}. In this task we explore the use of body pose and wearable acceleration data for detecting the speaking status of a person in the scene.

\paragraph{Setup} We use the SOTA MS-G3D graph neural network for skeleton action recognition \cite{Gupta2020}, pre-trained on Kinetics Skeleton 400. For the acceleration modality, we evaluated three time series classifiers, each of which we trained from scratch: 1D Resnet \cite{resnet1d}, InceptionTime \cite{Wang2017a}, and Minirocket \cite{minirocket}. We performed late fusion by averaging the scores from both modalities.
Like prior work \cite{Quiros2019,Gedik2017a}, the task was set up as a binary classification problem. We divided our pose (skeleton) tracks into $3$-second windows with $1.5$~s overlap. A window was labeled positive if more than $50$\% of the continuous speaking status labels within it are positive. This resulted in an imbalanced dataset of $42882$ windows with $29.2$\% positive labels. 
Poses were pre-processed for training following \cite{Gupta2020}. Three of the keypoints (head, and feet tips) were discarded due to not being present in Kinetics. We adapted the network by freezing all layers except for the last fully connected layer and training for five extra epochs. Acceleration readings were not pre-processed, other than by interpolating the original variable-sampling-rate signals to a fixed $50$~Hz.

\paragraph{Evaluation} Evaluation was carried out via $10$-fold cross-validation at the subject level, ensuring that no examples from the test subjects were used in training. We used the area under the ROC curve (AUC) as main evaluation metric to account for the imbalance in the labels. 

\paragraph{Results} The results in Table \ref{tab:ss-baseline} indicate a better performance from the acceleration-based methods. One possible reason for the lower performance of the pose-based methods is the significant domain shift between Kinetics and Conflab, especially in camera viewpoint (frontal vs top-down). The acceleration performance is in line with previous work \cite{Gedik2017a}. Multimodal results were slightly higher than acceleration-only results, despite our naive fusion approach, a possible point to improve in future work \cite{MultimodalML2019}. Experiments with the rest of the IMU modalities are presented in Appendix \ref{app:ss_extra}.

\begin{table}[!tb]
\begin{minipage}{.48\linewidth}

\ra{1.2}
\caption{ROC AUC and accuracy of skeleton-based, acceleration-based and multimodal speaking status detection (10-fold cross-validation).}
\label{tab:ss-baseline}
\begin{adjustbox}{max width=\linewidth}
\begin{tabular}{@{}lccc@{}} 
 \toprule
Modality & Model & AUC & Acc. \\
 \midrule
 Pose & MS-G3D \cite{Liu2020} & 0.676 & 0.677 \\
 \multirow{3}{*}{Acceleration} & InceptionTime \cite{Wang2017a} & 0.798 & 0.768  \\
 & Resnet 1D \cite{resnet1d} & 0.801 & 0.767 \\
 & Minirocket \cite{minirocket} & 0.813 & 0.768 \\
 Multimodal & MS-G3D + Minirocket & 0.823 & 0.775 \\ 
 \bottomrule
\end{tabular}
\end{adjustbox}

\end{minipage}\hfill
\begin{minipage}{.48\linewidth}
\caption{Average F1 scores for F-formation detection comparing GTCG \cite{VasconEtAl2014} and GCFF \cite{cristani2013human} with the effect of different threshold and orientations (standard deviation in parenthesis).}
\label{tab:ff-baseline}

\setlength{\tabcolsep}{0.5\tabcolsep}
\begin{adjustbox}{max width=\linewidth}
\begin{tabular}{@{}lllll@{}}
\toprule
         & \multicolumn{2}{c}{GTCG}                                     & \multicolumn{2}{c}{GCFF}                            \\ \cmidrule{2-5} 
         & \multicolumn{1}{c}{T=2/3} & \multicolumn{1}{c}{T=1}         & \multicolumn{1}{c}{T=2/3} & \multicolumn{1}{c}{T=1} \\ \midrule
Head     & 0.51 (0.09)               & \multicolumn{1}{l}{0.40 (0.12)} & 0.47 (0.07)               & 0.31 (0.23)             \\
Shoulder & 0.46 (0.11)               & \multicolumn{1}{l}{0.38 (0.11)} & 0.56 (0.25)               & 0.36 (0.16)             \\
Hip      & 0.45 (0.10)               & \multicolumn{1}{l}{0.37 (0.12)} & 0.39 (0.06)               & 0.25 (0.11)             \\ \bottomrule
\end{tabular}

\end{adjustbox}
\end{minipage}
\end{table}

\subsection{F-formation Detection}
\label{sec: fformation}
\newcommand{\floor}[1]{\left\lfloor #1 \right\rfloor}
\newcommand{\ceil}[1]{\left\lceil #1 \right\rceil}
\paragraph{Setup} Like prior work \cite{hung2011detecting,SwoffordEtAl2020,SettiEtAl2015,VasconEtAl2014}, we operationalize interaction groups using the framework of F-formations \cite{Kendon1990}. We provide performance results for F-formation detection using GTCG \cite{VasconEtAl2014} and GCFF \cite{cristani2013human} as a baseline. Recent deep learning methods such as DANTE \cite{SwoffordEtAl2020} are not directly applicable since they depend on knowing the number of people in the scene, which is variable for ConfLab. We use pre-trained model parameters (reported in the original GTCG and GCFF papers on the Cocktail Party dataset \cite{ZenEtAl2010}) and tuned a subset of parameters more relevant to ConfLab attributes on camera 6. More details can be found in Appendix \ref{app-subsec:ff-implementation}. 
We derive three different sets of orientation features from (i) head, (ii) shoulder and (iii) hip keypoints. 

\paragraph{Evaluation Metrics} We use the standard F1 score as evaluation metric for group detection \cite{VasconEtAl2014,cristani2013human}. A group is correctly estimated (true positive) if at least $\ceil{\textit{T}*|\textit{G}|}$ of the members of group $\textit{G}$ are correctly identified, and no more than $1-\ceil{\textit{T}*|\textit{G}|} $ is incorrectly identified, where \textit{T} is the tolerance threshold. We report results for $T= \frac{2}{3}$ and $T= 1$ (more strict threshold) in Table \ref{tab:ff-baseline}.

\paragraph{Results} We show that different results are obtained using different sources of orientations. Different occlusion levels in keypoints due to camera viewpoint may have affected performance. Another factor influencing model performance is that F-formations (which are driven by lower-body orientations \cite{Kendon1990}) may have multiple conversations floors \cite{ramanAutomaticEstimationConversation2019}. Floors are indicated by coordinated speaker turn taking patterns and influence coordinated head orientations of the group.



\section{Conclusion and Discussion}
ConfLab contributes a new concept for real-life data collection in the wild and captures a high-fidelity dataset of mixed levels of acquaintance, seniority, and personal motivations.

\paragraph{ConfLab: the Dataset} We improved upon prior work by providing higher-resolution, fidelity, and synchronization across sensor networks. We also carefully designed our social interaction setup to enable a diverse mix of seniority, acquaintanceship, and motivations for mingling. The result is a rich set of $17$ body-keypoint annotations of $48$ people at $60$~Hz from overhead cameras for developing more robust estimation of keypoints, speaking status and F-formations for further analyses of more complex socio-relational phenomena. Our benchmark results for these tasks highlight how the improved fidelity of ConfLab can assist in the development of more robust methods for these key tasks. 
We hope that models trained on ConfLab for localizing keypoints would fill the gap in the cue extraction pipeline, enabling past datasets \cite{hung2011detecting,MnM2021} without articulated pose data to be reinvigorated; this would open the floodgates for more robust analysis of the social phenomena labeled in these other datasets. Finally, our baseline social tasks form the basis for further explorations into downstream prediction tasks of socially-related  constructs such as conversation quality \cite{raman2022perceived} 
, dominance \cite{5549893}, rapport \cite{Muller2018a}, influence \cite{dong2007using} etc. 

\paragraph{ConfLab: the Data-Collection Concept}
To relate an individual's behaviors to trends within their social network, further iterations of ConfLab are needed. These iterations would enable the study of behavioral patterns at different timescales, including multiple interactions in one day, multiple days at a conference, or across distinct conferences. This paper serves as a template for such future ventures. We hope that if the idea of a conference as a living lab gains traction, the effort and cost of data collection can be amortized across different research groups, even involving support from the conference organizers. This \textit{data by the community for the community} ethos can enable the generation of a corpus of related datasets enabling new research questions.

\paragraph{Societal Impact}
ConfLab's long-term vision is towards developing technology to assist individuals in navigating social interactions. In this work we have identified choices that maximize data fidelity while upholding ethical best practices: an overhead camera perspective that mitigates identifying faces, recording audio at a low-frequency, and using non-intrusive wearable sensors matching a conference badge form-factor. We argue this is an essential step towards a long-term goal of developing personalized and socially aware technologies that enhance social experiences. At the same time, such interventions could also affect a community in unintended ways: worsened social satisfaction, lack of agency, stereotyping; or benefit only those members of the community who make use of resulting applications at the expense of the rest. More nefarious uses involve exploiting the data for developing methods that harmfully surveil or profile people. Researchers must consider such inadvertent effects while developing downstream applications. Finally, since we recorded the dataset at a scientific conference and required voluntary participation, there is an implicit selection bias in the population represented in the data. Researchers should be aware that insights resulting from the data may not generalize to the general population.

\paragraph{Empowering Users Through an Agentist Rather Than Structurist Approach}
The analysis of human behavior in social settings has classically taken a more top-down perspective. For instance, the analysis of situated interactions (via only proximity networks) has provided insight into the process of making science in the field of Meta Science \cite{eberle2021initiating}. However, while social network science is a well-populated domain, it lacks a more individualized measurement of social behavior: see more discussion of the structure vs. agency debate \cite{pleasants2019free}. Relying on the network science approach jeopardizes an individual’s right to technologies that enable free will. We consider the agency in choosing such technologies to be a form of individual harm avoidance. ConfLab provides access to more than just proximity data about social interactions, enabling the study of context-specific social dynamics. These dynamics are a uniquely dependent not only on the individual, but also the group they are interacting with \cite{ramanSocialProcessesSelfSupervised2021}. We hope our highlighting of participatory design practices and these value-sensitive design principles promote social safety in developing socially assistive technologies.

\section*{Acknowledgements}
The authors would like to thank: the ACM Multimedia 2019 General Chairs Martha Larson,  Benoit Huet, and Laurent Amsaleg for their support in making the data collection at a major international conference a reality; Bernd Dudzik, Yeshwanth Napolean, and Ruud de Jong for their help in setting up the recording on-site; the participants and student volunteers for the \textit{Meet the Chairs!} event; the Amazon Mechanical Turk workers for their efforts in annotating the dataset; Rich Radke, Martin Atzmueller, Laura Cabrera Quiros, Alan Hanjalic, and Xucong Zhang for the insightful discussions; Santosh Ilamparuthi for the innumerable discussions and support towards strengthening the ethical soundness of recording and sharing ConfLab; Jan van der Heul for the incredibly responsive support in setting up the 4TU Data repository for ConfLab; and Bart Vastenhouw, Myrthe Tielman, and Catharine Oertel for help with the data sharing.

ConfLab was partially funded by Netherlands Organization for
Scientific Research (NWO) under project number 639.022.606 with associated Aspasia Grant, and also by the ACM Multimedia 2019 conference via student helpers, and crane hiring for camera mounting.

{\small
\bibliographystyle{unsrtnat}
\bibliography{references}

\begin{thebibliography}{87}
\providecommand{\natexlab}[1]{#1}
\providecommand{\url}[1]{\texttt{#1}}
\expandafter\ifx\csname urlstyle\endcsname\relax
  \providecommand{\doi}[1]{doi: #1}\else
  \providecommand{\doi}{doi: \begingroup \urlstyle{rm}\Url}\fi

\bibitem[Dudzik et~al.(2021)Dudzik, Columbus, Hrkalovic, Balliet, and
  Hung]{dudzik2021recognizing}
Bernd Dudzik, Simon Columbus, Tiffany~Matej Hrkalovic, Daniel Balliet, and
  Hayley Hung.
\newblock Recognizing perceived interdependence in face-to-face negotiations
  through multimodal analysis of nonverbal behavior.
\newblock In \emph{Proceedings of the 2021 International Conference on
  Multimodal Interaction}, pages 121--130, 2021.

\bibitem[Fleeson(2007)]{Fleeson2007}
William Fleeson.
\newblock Situation-based contingencies underlying trait-content manifestation
  in behavior.
\newblock \emph{Journal of personality}, 75:\penalty0 825--862, 8 2007.
\newblock ISSN 0022-3506.
\newblock \doi{10.1111/J.1467-6494.2007.00458.X}.
\newblock URL \url{https://pubmed.ncbi.nlm.nih.gov/17576360/}.

\bibitem[Guardia and Ryan(2007)]{GuardiaRyan2007}
Jennifer G.~La Guardia and Richard~M. Ryan.
\newblock Why identities fluctuate: Variability in traits as a function of
  situational variations in autonomy support.
\newblock \emph{Journal of Personality}, 75:\penalty0 1205--1228, 12 2007.
\newblock ISSN 00223506.
\newblock \doi{10.1111/j.1467-6494.2007.00473.x}.

\bibitem[Hall et~al.(2019)Hall, Horgan, and Murphy]{Hall2019}
Judith~A. Hall, Terrence~G. Horgan, and Nora~A. Murphy.
\newblock Nonverbal communication.
\newblock \emph{https://doi.org/10.1146/annurev-psych-010418-103145},
  70:\penalty0 271--294, 1 2019.
\newblock ISSN 15452085.
\newblock \doi{10.1146/ANNUREV-PSYCH-010418-103145}.
\newblock URL
  \url{https://www.annualreviews.org/doi/abs/10.1146/annurev-psych-010418-103145}.

\bibitem[Osborne-Crowley(2020)]{osborne2020social}
Katherine Osborne-Crowley.
\newblock Social cognition in the real world: reconnecting the study of social
  cognition with social reality.
\newblock \emph{Review of general psychology}, 24\penalty0 (2):\penalty0
  144--158, 2020.

\bibitem[Andrade(2018)]{andrade2018internal}
Chittaranjan Andrade.
\newblock Internal, external, and ecological validity in research design,
  conduct, and evaluation.
\newblock \emph{Indian journal of psychological medicine}, 40\penalty0
  (5):\penalty0 498--499, 2018.

\bibitem[Labonte-LeMoyne et~al.(2018)Labonte-LeMoyne, Courtemanche, Fredette,
  and L{\'e}ger]{labonte2018wild}
{\'E}lise Labonte-LeMoyne, Fran{\c{c}}ois Courtemanche, Marc Fredette, and
  Pierre-Majorique L{\'e}ger.
\newblock How wild is too wild: Lessons learned and recommendations for
  ecological validity in physiological computing research.
\newblock In \emph{PhyCS}, pages 123--130, 2018.

\bibitem[Hung et~al.(2019{\natexlab{a}})Hung, Gedik, and
  Quiros]{hung2019complex}
Hayley Hung, Ekin Gedik, and Laura~Cabrera Quiros.
\newblock Complex conversational scene analysis using wearable sensors.
\newblock In \emph{Multimodal Behavior Analysis in the Wild}, pages 225--245.
  Elsevier, 2019{\natexlab{a}}.

\bibitem[Cabrera-Quiros et~al.(2021)Cabrera-Quiros, Demetriou, Gedik, van~der
  Meij, and Hung]{MnM2021}
Laura Cabrera-Quiros, Andrew Demetriou, Ekin Gedik, Leander van~der Meij, and
  Hayley Hung.
\newblock The matchnmingle dataset: A novel multi-sensor resource for the
  analysis of social interactions and group dynamics in-the-wild during
  free-standing conversations and speed dates.
\newblock \emph{IEEE Transactions on Affective Computing}, 12\penalty0
  (1):\penalty0 113--130, 2021.

\bibitem[Hung and Kr{\"o}se(2011)]{hung2011detecting}
Hayley Hung and Ben Kr{\"o}se.
\newblock Detecting f-formations as dominant sets.
\newblock In \emph{Proceedings of the 13th international conference on
  multimodal interfaces}, pages 231--238, 2011.

\bibitem[Alameda-Pineda et~al.(2015)Alameda-Pineda, Staiano, Subramanian,
  Batrinca, Ricci, Lepri, Lanz, and Sebe]{alameda2015salsa}
Xavier Alameda-Pineda, Jacopo Staiano, Ramanathan Subramanian, Ligia Batrinca,
  Elisa Ricci, Bruno Lepri, Oswald Lanz, and Nicu Sebe.
\newblock Salsa: A novel dataset for multimodal group behavior analysis.
\newblock \emph{IEEE Transactions on Pattern Analysis and Machine
  Intelligence}, 38\penalty0 (8):\penalty0 1707--1720, 2015.

\bibitem[Cristani et~al.(2011)Cristani, Bazzani, Paggetti, Fossati, Tosato,
  Bue, Menegaz, and Murino]{BMVC.25.23}
Marco Cristani, Loris Bazzani, Giulia Paggetti, Andrea Fossati, Diego Tosato,
  Alessio~Del Bue, Gloria Menegaz, and Vittorio Murino.
\newblock Social interaction discovery by statistical analysis of f-formations.
\newblock In Jesse Hoey, Stephen~J. McKenna, and Emanuele Trucco, editors,
  \emph{British Machine Vision Conference, {BMVC} 2011, Dundee, UK, August 29 -
  September 2, 2011. Proceedings}, pages 1--12. {BMVA} Press, 2011.
\newblock \doi{10.5244/C.25.23}.
\newblock URL \url{https://doi.org/10.5244/C.25.23}.

\bibitem[Zen et~al.(2010)Zen, Lepri, Ricci, and Lanz]{ZenEtAl2010}
Gloria Zen, Bruno Lepri, Elisa Ricci, and Oswald Lanz.
\newblock Space speaks: towards socially and personality aware visual
  surveillance.
\newblock In \emph{Proceedings of the 1st ACM international workshop on
  Multimodal pervasive video analysis}, pages 37--42, 2010.

\bibitem[Murgia(2019)]{ft2019whosusing}
Madhumita Murgia.
\newblock Who’s using your face? the ugly truth about facial recognition.
\newblock \emph{Financial Times}, 2019.

\bibitem[Carissimi et~al.(2018)Carissimi, Rota, Beyan, and
  Murino]{carissimi2018filling}
Nicolo Carissimi, Paolo Rota, Cigdem Beyan, and Vittorio Murino.
\newblock Filling the gaps: Predicting missing joints of human poses using
  denoising autoencoders.
\newblock In \emph{Proceedings of the European Conference on Computer Vision
  (ECCV) Workshops}, pages 0--0, 2018.

\bibitem[Gedik and Hung(2018)]{gedikhungIMWUT2018}
Ekin Gedik and Hayley Hung.
\newblock Detecting conversing groups using social dynamics from wearable
  acceleration: Group size awareness.
\newblock \emph{Proc. ACM Interact. Mob. Wearable Ubiquitous Technol.},
  2\penalty0 (4), dec 2018.
\newblock \doi{10.1145/3287041}.
\newblock URL \url{https://doi.org/10.1145/3287041}.

\bibitem[Gedik and Hung(2017)]{Gedik2017a}
Ekin Gedik and Hayley Hung.
\newblock Personalised models for speech detection from body movements using
  transductive parameter transfer.
\newblock \emph{Personal and Ubiquitous Computing}, 21\penalty0 (4):\penalty0
  723--737, August 2017.
\newblock ISSN 1617-4909.
\newblock \doi{10.1007/s00779-017-1006-4}.

\bibitem[Raman et~al.(2020)Raman, Tan, and Hung]{raman2020modular}
Chirag Raman, Stephanie Tan, and Hayley Hung.
\newblock A modular approach for synchronized wireless multimodal multisensor
  data acquisition in highly dynamic social settings.
\newblock In \emph{Proceedings of the 28th ACM International Conference on
  Multimedia}, MM '20, page 3586–3594, New York, NY, USA, 2020. Association
  for Computing Machinery.
\newblock ISBN 9781450379885.
\newblock \doi{10.1145/3394171.3413697}.
\newblock URL \url{https://doi.org/10.1145/3394171.3413697}.

\bibitem[Cai et~al.(2020)Cai, Wang, Luo, Yin, Du, Wang, Zhang, Zhou, Zhou, and
  Sun]{cai2020learningRSN}
Yuanhao Cai, Zhicheng Wang, Zhengxiong Luo, Binyi Yin, Angang Du, Haoqian Wang,
  Xiangyu Zhang, Xinyu Zhou, Erjin Zhou, and Jian Sun.
\newblock Learning delicate local representations for multi-person pose
  estimation.
\newblock In \emph{European Conference on Computer Vision}, pages 455--472.
  Springer, 2020.

\bibitem[Birhane and Prabhu(2021)]{BirhaneP21}
Abeba Birhane and Vinay~Uday Prabhu.
\newblock Large image datasets: {A} pyrrhic win for computer vision?
\newblock In \emph{{IEEE} Winter Conference on Applications of Computer Vision,
  {WACV} 2021, Waikoloa, HI, USA, January 3-8, 2021}, pages 1536--1546. {IEEE},
  2021.
\newblock \doi{10.1109/WACV48630.2021.00158}.
\newblock URL \url{https://doi.org/10.1109/WACV48630.2021.00158}.

\bibitem[Setti et~al.(2015)Setti, Russell, Bassetti, and
  Cristani]{SettiEtAl2015}
Francesco Setti, Chris Russell, Chiara Bassetti, and Marco Cristani.
\newblock F-formation detection: Individuating free-standing conversational
  groups in images.
\newblock \emph{PloS one}, 10\penalty0 (5):\penalty0 e0123783, 2015.

\bibitem[Swofford et~al.(2020)Swofford, Peruzzi, Tsoi, Thompson,
  Mart{\'\i}n-Mart{\'\i}n, Savarese, and V{\'a}zquez]{SwoffordEtAl2020}
Mason Swofford, John Peruzzi, Nathan Tsoi, Sydney Thompson, Roberto
  Mart{\'\i}n-Mart{\'\i}n, Silvio Savarese, and Marynel V{\'a}zquez.
\newblock Improving social awareness through dante: Deep affinity network for
  clustering conversational interactants.
\newblock \emph{Proceedings of the ACM on Human-Computer Interaction},
  4\penalty0 (CSCW1):\penalty0 1--23, 2020.

\bibitem[Vascon et~al.(2014)Vascon, Mequanint, Cristani, Hung, Pelillo, and
  Murino]{VasconEtAl2014}
Sebastiano Vascon, Eyasu~Zemene Mequanint, Marco Cristani, Hayley Hung,
  Marcello Pelillo, and Vittorio Murino.
\newblock A game-theoretic probabilistic approach for detecting conversational
  groups.
\newblock In \emph{Asian conference on computer vision}, pages 658--675.
  Springer, 2014.

\bibitem[Joo et~al.(2017)Joo, Simon, Li, Liu, Tan, Gui, Banerjee, Godisart,
  Nabbe, Matthews, Kanade, Nobuhara, and Sheikh]{Joo_2017_TPAMI}
Hanbyul Joo, Tomas Simon, Xulong Li, Hao Liu, Lei Tan, Lin Gui, Sean Banerjee,
  Timothy~Scott Godisart, Bart Nabbe, Iain Matthews, Takeo Kanade, Shohei
  Nobuhara, and Yaser Sheikh.
\newblock Panoptic studio: A massively multiview system for social interaction
  capture.
\newblock \emph{IEEE Transactions on Pattern Analysis and Machine
  Intelligence}, 2017.

\bibitem[Vondrick et~al.(2013)Vondrick, Patterson, and Ramanan]{Vondrick2013}
Carl Vondrick, Donald Patterson, and Deva Ramanan.
\newblock Efficiently {{Scaling}} up {{Crowdsourced Video Annotation}}.
\newblock \emph{International Journal of Computer Vision}, 101\penalty0
  (1):\penalty0 184--204, 2013.
\newblock ISSN 0920-5691.
\newblock \doi{10.1007/s11263-012-0564-1}.

\bibitem[Ricci et~al.(2015)Ricci, Varadarajan, Subramanian, Bulò, Ahuja, and
  Lanz]{7410886}
Elisa Ricci, Jagannadan Varadarajan, Ramanathan Subramanian, Samuel~Rota Bulò,
  Narendra Ahuja, and Oswald Lanz.
\newblock Uncovering interactions and interactors: Joint estimation of head,
  body orientation and f-formations from surveillance videos.
\newblock In \emph{2015 IEEE International Conference on Computer Vision
  (ICCV)}, pages 4660--4668, 2015.

\bibitem[Bazzani et~al.(2013)Bazzani, Cristani, Tosato, Farenzena, Paggetti,
  Menegaz, and Murino]{bazzani2013social}
Loris Bazzani, Marco Cristani, Diego Tosato, Michela Farenzena, Giulia
  Paggetti, Gloria Menegaz, and Vittorio Murino.
\newblock Social interactions by visual focus of attention in a
  three-dimensional environment.
\newblock \emph{Expert Systems}, 30\penalty0 (2):\penalty0 115--127, 2013.

\bibitem[Carletta et~al.(2006)Carletta, Ashby, Bourban, Flynn, Guillemot, Hain,
  Kadlec, Karaiskos, Kraaij, Kronenthal, Lathoud, Lincoln, Lisowska, McCowan,
  Post, Reidsma, and Wellner]{amicorpus}
Jean Carletta, Simone Ashby, Sebastien Bourban, Mike Flynn, Mael Guillemot,
  Thomas Hain, Jaroslav Kadlec, Vasilis Karaiskos, Wessel Kraaij, Melissa
  Kronenthal, Guillaume Lathoud, Mike Lincoln, Agnes Lisowska, Iain McCowan,
  Wilfried Post, Dennis Reidsma, and Pierre Wellner.
\newblock The ami meeting corpus: A pre-announcement.
\newblock In Steve Renals and Samy Bengio, editors, \emph{Machine Learning for
  Multimodal Interaction}, pages 28--39, Berlin, Heidelberg, 2006. Springer
  Berlin Heidelberg.

\bibitem[Cattuto et~al.(2010)Cattuto, Broeck, Barrat, Colizza, Pinton, and
  Vespignani]{Cattuto2010DynamicsOP}
C.~Cattuto, W.~V.~D. Broeck, A.~Barrat, V.~Colizza, J.~Pinton, and Alessandro
  Vespignani.
\newblock Dynamics of person-to-person interactions from distributed rfid
  sensor networks.
\newblock \emph{PLoS ONE}, 5, 2010.

\bibitem[Hoffman et~al.(2020)Hoffman, Block, Elmer, and
  Stadtfeld]{hoffman_block_elmer_stadtfeld_2020}
Marion Hoffman, Per Block, Timon Elmer, and Christoph Stadtfeld.
\newblock A model for the dynamics of face-to-face interactions in social
  groups.
\newblock \emph{Network Science}, 8\penalty0 (S1):\penalty0 S4–S25, 2020.
\newblock \doi{10.1017/nws.2020.3}.

\bibitem[Atzmueller and Lemmerich(2018)]{Atzmuellerwww18}
Martin Atzmueller and Florian Lemmerich.
\newblock Homophily at academic conferences.
\newblock In \emph{Companion Proceedings of the The Web Conference 2018}, WWW
  '18, page 109–110, Republic and Canton of Geneva, CHE, 2018. International
  World Wide Web Conferences Steering Committee.

\bibitem[Olgu{\'\i}n et~al.(2008)Olgu{\'\i}n, Waber, Kim, Mohan, Ara, and
  Pentland]{olguin2008sensible}
Daniel~Olgu{\'\i}n Olgu{\'\i}n, Benjamin~N Waber, Taemie Kim, Akshay Mohan,
  Koji Ara, and Alex Pentland.
\newblock Sensible organizations: Technology and methodology for automatically
  measuring organizational behavior.
\newblock \emph{IEEE Transactions on Systems, Man, and Cybernetics, Part B
  (Cybernetics)}, 39\penalty0 (1):\penalty0 43--55, 2008.

\bibitem[Chaffin et~al.(2017)Chaffin, Heidl, Hollenbeck, Howe, Yu, Voorhees,
  and Calantone]{chaffin2017promise}
Daniel Chaffin, Ralph Heidl, John~R Hollenbeck, Michael Howe, Andrew Yu, Clay
  Voorhees, and Roger Calantone.
\newblock The promise and perils of wearable sensors in organizational
  research.
\newblock \emph{Organizational Research Methods}, 20\penalty0 (1):\penalty0
  3--31, 2017.

\bibitem[Rosatelli et~al.(2019)Rosatelli, Gedik, and Hung]{8925179}
Alessio Rosatelli, Ekin Gedik, and Hayley Hung.
\newblock Detecting f-formations roles in crowded social scenes with wearables:
  Combining proxemics dynamics using lstms.
\newblock In \emph{2019 8th International Conference on Affective Computing and
  Intelligent Interaction Workshops and Demos (ACIIW)}, pages 147--153, 2019.
\newblock \doi{10.1109/ACIIW.2019.8925179}.

\bibitem[{Cabrera-Quiros} et~al.(2018){Cabrera-Quiros}, M.J.~Tax, and
  Hung]{Cabrera-Quiros2018b}
Laura {Cabrera-Quiros}, David M.J.~Tax, and Hayley Hung.
\newblock Gestures in-the-wild : Detecting conversational hand gestures in
  crowded scenes using a multimodal fusion of bags of video trajectories and
  body worn acceleration.
\newblock pages 1--10, 2018.

\bibitem[Quiros and Hung(2019)]{Quiros2019}
J.~V. Quiros and H.~Hung.
\newblock {{CNNs}} and {{Fisher Vectors}} for {{No}}-{{Audio Multimodal Speech
  Detection}}.
\newblock In \emph{{{MediaEval}}}, 2019.

\bibitem[Tan et~al.(2021{\natexlab{a}})Tan, Tax, and Hung]{tanimwut2021}
Stephanie Tan, David M.~J. Tax, and Hayley Hung.
\newblock Multimodal joint head orientation estimation in interacting groups
  via proxemics and interaction dynamics.
\newblock \emph{Proc. ACM Interactive, Mobile, Wearable, and Ubiquitous
  Technology}, 5\penalty0 (1), March 2021{\natexlab{a}}.

\bibitem[uyr()]{uyrdm}
University of york research data management.
\newblock
  \url{https://www.york.ac.uk/library/info-for/researchers/data/sharing/access/}.

\bibitem[uur()]{uurdm}
Utrecht university research data management.
\newblock
  \url{https://www.uu.nl/en/research/research-data-management/guides/handling-personal-data}.

\bibitem[gop()]{gopro7}
Go pro hero 7 black.
\newblock
  \url{https://gopro.com/en/nl/shop/cameras/hero7-black/CHDHX-701-master.html}.

\bibitem[Lederman et~al.(2018)Lederman, Mohan, Calacci, and
  Pentland]{lederman2018rhythm}
Oren Lederman, Akshay Mohan, Dan Calacci, and Alex~Sandy Pentland.
\newblock Rhythm: A unified measurement platform for human organizations.
\newblock \emph{IEEE MultiMedia}, 25\penalty0 (1):\penalty0 26--38, 2018.

\bibitem[{Ringeval} et~al.(2013){Ringeval}, {Sonderegger}, {Sauer}, and
  {Lalanne}]{recola}
F.~{Ringeval}, A.~{Sonderegger}, J.~{Sauer}, and D.~{Lalanne}.
\newblock Introducing the recola multimodal corpus of remote collaborative and
  affective interactions.
\newblock In \emph{2013 10th IEEE International Conference and Workshops on
  Automatic Face and Gesture Recognition (FG)}, pages 1--8, 2013.

\bibitem[Cabrera-Quiros and Hung(2016)]{cabrera2016matching}
Laura Cabrera-Quiros and Hayley Hung.
\newblock Who is where? matching people in video to wearable acceleration
  during crowded mingling events.
\newblock In \emph{Proceedings of the 24th ACM international conference on
  Multimedia}, pages 267--271, 2016.

\bibitem[Cabrera-Quiros and Hung(2018)]{cabrera2018hierarchical}
Laura Cabrera-Quiros and Hayley Hung.
\newblock A hierarchical approach for associating body-worn sensors to video
  regions in crowded mingling scenarios.
\newblock \emph{IEEE Transactions on Multimedia}, 21\penalty0 (7):\penalty0
  1867--1879, 2018.

\bibitem[Hung et~al.(2019{\natexlab{b}})Hung, Raman, Gedik, Tan, and
  Vargas~Quiros]{hung2019multimodal}
Hayley Hung, Chirag Raman, Ekin Gedik, Stephanie Tan, and Jose Vargas~Quiros.
\newblock Multimodal data collection for social interaction analysis
  in-the-wild.
\newblock In \emph{Proceedings of the 27th ACM International Conference on
  Multimedia}, pages 2714--2715, 2019{\natexlab{b}}.

\bibitem[Cva()]{CvatGithub}
Computer {{Vision Annotation Tool}} ({{CVAT}}).

\bibitem[Cov()]{CovfeeGithub}
Covfee: {{Continuous Video Feedback Tool}}.
\newblock Jose Vargas.

\bibitem[Vargas~Quiros et~al.(2022)Vargas~Quiros, Tan, Raman, Cabrera-Quiros,
  and Hung]{covfee-paper}
Jose Vargas~Quiros, Stephanie Tan, Chirag Raman, Laura Cabrera-Quiros, and
  Hayley Hung.
\newblock Covfee: an extensible web framework for continuous-time annotation of
  human behavior.
\newblock In Cristina Palmero, Julio C.~S. Jacques~Junior, Albert Clapés,
  Isabelle Guyon, Wei-Wei Tu, Thomas~B. Moeslund, and Sergio Escalera, editors,
  \emph{Understanding Social Behavior in Dyadic and Small Group Interactions},
  volume 173 of \emph{Proceedings of Machine Learning Research}, pages
  265--293. PMLR, 16 Oct 2022.
\newblock URL \url{https://proceedings.mlr.press/v173/vargas-quiros22a.html}.

\bibitem[Gatica-Perez(2006)]{4042075}
Daniel Gatica-Perez.
\newblock Analyzing group interactions in conversations: a review.
\newblock In \emph{2006 IEEE International Conference on Multisensor Fusion and
  Integration for Intelligent Systems}, pages 41--46, 2006.
\newblock \doi{10.1109/MFI.2006.265658}.

\bibitem[M{\"u}ller et~al.(2018)M{\"u}ller, Huang, and Bulling]{Muller2018a}
Philipp M{\"u}ller, Michael~Xuelin Huang, and Andreas Bulling.
\newblock Detecting {{Low Rapport During Natural Interactions}} in {{Small
  Groups}} from {{Non}}-{{Verbal Behaviour}}.
\newblock In \emph{23rd {{International Conference}} on {{Intelligent User
  Interfaces}}}. {ACM}, 2018.
\newblock ISBN 978-1-4503-4945-1.
\newblock \doi{10.1145/3172944.3172969}.

\bibitem[Raman and Hung(2019)]{ramanAutomaticEstimationConversation2019}
Chirag Raman and Hayley Hung.
\newblock Towards automatic estimation of conversation floors within
  {{F}}-formations.
\newblock \emph{arXiv:1907.10384 [cs]}, July 2019.

\bibitem[Hung and Gatica-Perez(2010)]{hung2010estimating}
Hayley Hung and Daniel Gatica-Perez.
\newblock Estimating cohesion in small groups using audio-visual nonverbal
  behavior.
\newblock \emph{IEEE Transactions on Multimedia}, 12\penalty0 (6):\penalty0
  563--575, 2010.

\bibitem[Hung et~al.(2011)Hung, Huang, Friedland, and Gatica-Perez]{5549893}
Hayley Hung, Yan Huang, Gerald Friedland, and Daniel Gatica-Perez.
\newblock {Estimating Dominance in Multi-Party Meetings Using Speaker
  Diarization}.
\newblock \emph{IEEE Transactions on Audio, Speech, and Language Processing},
  19\penalty0 (4):\penalty0 847--860, may 2011.

\bibitem[Beyan et~al.(2020)Beyan, Shahid, and Murino]{Beyan2020}
Cigdem Beyan, Muhammad Shahid, and Vittorio Murino.
\newblock {{RealVAD}}: {{A Real}}-world {{Dataset}} and {{A Method}} for
  {{Voice Activity Detection}} by {{Body Motion Analysis}}.
\newblock \emph{x}, 9210\penalty0 (c):\penalty0 1--16, 2020.
\newblock \doi{10.1109/tmm.2020.3007350}.

\bibitem[Shahid et~al.(2019)Shahid, Beyan, and Murino]{Shahid2019}
Muhammad Shahid, Cigdem Beyan, and Vittorio Murino.
\newblock Voice activity detection by upper body motion analysis and
  unsupervised domain adaptation.
\newblock \emph{Proceedings - 2019 International Conference on Computer Vision
  Workshop, ICCVW 2019}, pages 1260--1269, 2019.
\newblock \doi{10.1109/ICCVW.2019.00159}.

\bibitem[Kendon(1990)]{Kendon1990}
Adam Kendon.
\newblock \emph{Conducting interaction: Patterns of behavior in focused
  encounters}, volume~7.
\newblock CUP Archive, 1990.

\bibitem[He et~al.(2017)He, Gkioxari, Doll{\'a}r, and Girshick]{he2017maskrcnn}
Kaiming He, Georgia Gkioxari, Piotr Doll{\'a}r, and Ross Girshick.
\newblock Mask r-cnn.
\newblock In \emph{Proceedings of the IEEE international conference on computer
  vision}, pages 2961--2969, 2017.

\bibitem[Wu et~al.(2019)Wu, Kirillov, Massa, Lo, and
  Girshick]{wu2019detectron2}
Yuxin Wu, Alexander Kirillov, Francisco Massa, Wan-Yen Lo, and Ross Girshick.
\newblock Detectron2.
\newblock \url{https://github.com/facebookresearch/detectron2}, 2019.

\bibitem[Lin et~al.(2014)Lin, Maire, Belongie, Hays, Perona, Ramanan,
  Doll{\'a}r, and Zitnick]{lin2014microsoftcoco}
Tsung-Yi Lin, Michael Maire, Serge Belongie, James Hays, Pietro Perona, Deva
  Ramanan, Piotr Doll{\'a}r, and C~Lawrence Zitnick.
\newblock Microsoft coco: Common objects in context.
\newblock In \emph{European conference on computer vision}, pages 740--755.
  Springer, 2014.

\bibitem[Shen et~al.(2018)Shen, Lederman, Cao, Berg, Tang, and
  Pentland]{Shen2018}
Jiaxing Shen, Oren Lederman, Jiannong Cao, Florian Berg, Shaojie Tang, and
  Alex~Sandy Pentland.
\newblock {{GINA}}: Group {{Gender Identification Using Privacy}}-{{Sensitive
  Audio Data}}.
\newblock \emph{Proceedings - IEEE International Conference on Data Mining,
  ICDM}, 2018-Novem:\penalty0 457--466, 2018.
\newblock ISSN 15504786.
\newblock \doi{10.1109/ICDM.2018.00061}.

\bibitem[Gupta et~al.(2020)Gupta, Thatipelli, Aggarwal, Maheshwari, Trivedi,
  Das, and Sarvadevabhatla]{Gupta2020}
Pranay Gupta, Anirudh Thatipelli, Aditya Aggarwal, Shubh Maheshwari, Neel
  Trivedi, Sourav Das, and Ravi~Kiran Sarvadevabhatla.
\newblock Quo {{Vadis}}, {{Skeleton Action Recognition}} ?
\newblock \emph{arXiv:2007.02072 [cs]}, July 2020.

\bibitem[Wang et~al.(2016)Wang, Yan, and Oates]{resnet1d}
Zhiguang Wang, Weizhong Yan, and Tim Oates.
\newblock Time series classification from scratch with deep neural networks: A
  strong baseline, 2016.
\newblock URL \url{https://arxiv.org/abs/1611.06455}.

\bibitem[Wang et~al.(2017)Wang, Yan, and Oates]{Wang2017a}
Zhiguang Wang, Weizhong Yan, and Tim Oates.
\newblock Time series classification from scratch with deep neural networks:
  {{A}} strong baseline.
\newblock \emph{Proceedings of the International Joint Conference on Neural
  Networks}, 2017-May:\penalty0 1578--1585, 2017.
\newblock \doi{10.1109/IJCNN.2017.7966039}.

\bibitem[Tan et~al.(2021{\natexlab{b}})Tan, Dempster, Bergmeir, and
  Webb]{minirocket}
Chang~Wei Tan, Angus Dempster, Christoph Bergmeir, and Geoffrey~I. Webb.
\newblock Multirocket: Multiple pooling operators and transformations for fast
  and effective time series classification, 2021{\natexlab{b}}.
\newblock URL \url{https://arxiv.org/abs/2102.00457}.

\bibitem[Baltrusaitis et~al.(2019)Baltrusaitis, Ahuja, and
  Morency]{MultimodalML2019}
Tadas Baltrusaitis, Chaitanya Ahuja, and Louis-Philippe Morency.
\newblock Multimodal machine learning: A survey and taxonomy.
\newblock \emph{IEEE Trans. Pattern Anal. Mach. Intell.}, 41\penalty0
  (2):\penalty0 423–443, February 2019.

\bibitem[Liu et~al.(2020)Liu, Shahroudy, Perez, Wang, Duan, and Kot]{Liu2020}
Jun Liu, Amir Shahroudy, Mauricio Perez, Gang Wang, Ling-Yu Duan, and Alex~C.
  Kot.
\newblock {{NTU RGB}}+{{D}} 120: {{A Large}}-{{Scale Benchmark}} for {{3D Human
  Activity Understanding}}.
\newblock \emph{IEEE Transactions on Pattern Analysis and Machine
  Intelligence}, 42\penalty0 (10):\penalty0 2684--2701, October 2020.
\newblock ISSN 0162-8828, 2160-9292, 1939-3539.
\newblock \doi{10.1109/TPAMI.2019.2916873}.

\bibitem[Cristani et~al.(2013)Cristani, Raghavendra, Del~Bue, and
  Murino]{cristani2013human}
Marco Cristani, Ramachandra Raghavendra, Alessio Del~Bue, and Vittorio Murino.
\newblock Human behavior analysis in video surveillance: A social signal
  processing perspective.
\newblock \emph{Neurocomputing}, 100:\penalty0 86--97, 2013.

\bibitem[Raman et~al.(2022{\natexlab{a}})Raman, Prabhu, and
  Hung]{raman2022perceived}
Chirag Raman, Navin~Raj Prabhu, and Hayley Hung.
\newblock Perceived conversation quality in spontaneous interactions,
  2022{\natexlab{a}}.
\newblock URL \url{https://arxiv.org/abs/2207.05791}.

\bibitem[Dong et~al.(2007)Dong, Lepri, Cappelletti, Pentland, Pianesi, and
  Zancanaro]{dong2007using}
Wen Dong, Bruno Lepri, Alessandro Cappelletti, Alex~Sandy Pentland, Fabio
  Pianesi, and Massimo Zancanaro.
\newblock Using the influence model to recognize functional roles in meetings.
\newblock In \emph{Proceedings of the 9th international conference on
  Multimodal interfaces}, pages 271--278, 2007.

\bibitem[Eberle et~al.(2021)Eberle, Stegmann, Barrat, Fischer, and
  Lund]{eberle2021initiating}
Julia Eberle, Karsten Stegmann, Alain Barrat, Frank Fischer, and Kristine Lund.
\newblock Initiating scientific collaborations across career levels and
  disciplines--a network analysis on behavioral data.
\newblock \emph{International Journal of Computer-Supported Collaborative
  Learning}, 16\penalty0 (2):\penalty0 151--184, 2021.

\bibitem[Pleasants(2019)]{pleasants2019free}
Nigel Pleasants.
\newblock Free will, determinism and the “problem” of structure and agency
  in the social sciences.
\newblock \emph{Philosophy of the Social Sciences}, 49\penalty0 (1):\penalty0
  3--30, 2019.

\bibitem[Raman et~al.(2021)Raman, Hung, and
  Loog]{ramanSocialProcessesSelfSupervised2021}
Chirag Raman, Hayley Hung, and Marco Loog.
\newblock Social {{Processes}}: {{Self-Supervised Meta-Learning}} over
  {{Conversational Groups}} for {{Forecasting Nonverbal Social Cues}}.
\newblock \emph{arXiv:2107.13576 [cs]}, July 2021.

\bibitem[Gebru et~al.(2021)Gebru, Morgenstern, Vecchione, Vaughan, Wallach,
  Iii, and Crawford]{gebru2021datasheets}
Timnit Gebru, Jamie Morgenstern, Briana Vecchione, Jennifer~Wortman Vaughan,
  Hanna Wallach, Hal~Daum{\'e} Iii, and Kate Crawford.
\newblock Datasheets for datasets.
\newblock \emph{Communications of the ACM}, 64\penalty0 (12):\penalty0 86--92,
  2021.

\bibitem[Pineau et~al.(2021)Pineau, Vincent-Lamarre, Sinha, Larivi{\`e}re,
  Beygelzimer, d’Alch{\'e} Buc, Fox, and Larochelle]{pineau2021improving}
Joelle Pineau, Philippe Vincent-Lamarre, Koustuv Sinha, Vincent Larivi{\`e}re,
  Alina Beygelzimer, Florence d’Alch{\'e} Buc, Emily Fox, and Hugo
  Larochelle.
\newblock Improving reproducibility in machine learning research: a report from
  the neurips 2019 reproducibility program.
\newblock \emph{Journal of Machine Learning Research}, 22, 2021.

\bibitem[Barquero et~al.(2022)Barquero, N{\'u}nez, Escalera, Xu, Tu, Guyon, and
  Palmero]{barquero2022didn}
German Barquero, Johnny N{\'u}nez, Sergio Escalera, Zhen Xu, Wei-Wei Tu,
  Isabelle Guyon, and Cristina Palmero.
\newblock Didn’t see that coming: a survey on non-verbal social human
  behavior forecasting.
\newblock In \emph{Understanding Social Behavior in Dyadic and Small Group
  Interactions}, pages 139--178. PMLR, 2022.

\bibitem[Raman et~al.(2022{\natexlab{b}})Raman, Hung, and Loog]{raman2022did}
Chirag Raman, Hayley Hung, and Marco Loog.
\newblock Why did this model forecast this future? closed-form temporal
  saliency towards causal explanations of probabilistic forecasts.
\newblock \emph{arXiv preprint arXiv:2206.00679}, 2022{\natexlab{b}}.

\bibitem[Hung et~al.(2013)Hung, Englebienne, and Kools]{hung2013classifying}
Hayley Hung, Gwenn Englebienne, and Jeroen Kools.
\newblock Classifying social actions with a single accelerometer.
\newblock In \emph{Proceedings of the 2013 ACM international joint conference
  on Pervasive and ubiquitous computing}, pages 207--210, 2013.

\bibitem[Raj~Prabhu et~al.(2020)Raj~Prabhu, Raman, and Hung]{raj2020defining}
Navin Raj~Prabhu, Chirag Raman, and Hayley Hung.
\newblock Defining and quantifying conversation quality in spontaneous
  interactions.
\newblock In \emph{Companion Publication of the 2020 International Conference
  on Multimodal Interaction}, pages 196--205, 2020.

\bibitem[Quiros et~al.(2021)Quiros, Kapcak, Hung, and
  Cabrera-Quiros]{quiros2021individual}
Jose David~Vargas Quiros, Oyku Kapcak, Hayley Hung, and Laura Cabrera-Quiros.
\newblock Individual and joint body movement assessed by wearable sensing as a
  predictor of attraction in speed dates.
\newblock \emph{IEEE Transactions on Affective Computing}, 2021.

\bibitem[OpenCV(2015)]{2015opencv}
OpenCV.
\newblock Open source computer vision library.
\newblock \url{https://github.com/opencv/opencv}, 2015.

\bibitem[idi()]{idiap-multicam}
Idiap multi camera calibration suite.
\newblock \url{https://github.com/idiap/multicamera-calibration}.

\bibitem[imu()]{imu}
Tdkicm20948.
\newblock
  \url{https://invensense.tdk.com/products/motion-tracking/9-axis/icm-20948/}.
\newblock Accessed: 2021-10-15.

\bibitem[Ba and Odobez(2010)]{BaAndOdobez}
Sileye~O Ba and Jean-Marc Odobez.
\newblock Multiperson visual focus of attention from head pose and meeting
  contextual cues.
\newblock \emph{IEEE Transactions on Pattern Analysis and Machine
  Intelligence}, 33\penalty0 (1):\penalty0 101--116, 2010.

\bibitem[Li et~al.(2019)Li, Wang, Yin, Peng, Du, Xiao, Yu, Lu, Wei, and
  Sun]{li2019rethinkingmspn}
Wenbo Li, Zhicheng Wang, Binyi Yin, Qixiang Peng, Yuming Du, Tianzi Xiao, Gang
  Yu, Hongtao Lu, Yichen Wei, and Jian Sun.
\newblock Rethinking on multi-stage networks for human pose estimation.
\newblock \emph{arXiv preprint arXiv:1901.00148}, 2019.

\bibitem[Cheng et~al.(2020)Cheng, Xiao, Wang, Shi, Huang, and
  Zhang]{Cheng_2020_CVPRhigherhrnet}
Bowen Cheng, Bin Xiao, Jingdong Wang, Honghui Shi, Thomas~S. Huang, and Lei
  Zhang.
\newblock Higherhrnet: Scale-aware representation learning for bottom-up human
  pose estimation.
\newblock In \emph{Proceedings of the IEEE/CVF Conference on Computer Vision
  and Pattern Recognition (CVPR)}, June 2020.

\bibitem[Newell et~al.(2017)Newell, Huang, and
  Deng]{newell2017associative_posehglass}
Alejandro Newell, Zhiao Huang, and Jia Deng.
\newblock Associative embedding: End-to-end learning for joint detection and
  grouping.
\newblock \emph{Advances in neural information processing systems}, 30, 2017.

\bibitem[Oguiza(2022)]{tsai}
Ignacio Oguiza.
\newblock tsai - a state-of-the-art deep learning library for time series and
  sequential data.
\newblock Github, 2022.
\newblock URL \url{https://github.com/timeseriesAI/tsai}.

\end{thebibliography}
}

\clearpage
\appendix

\setcounter{page}{1}

\onecolumn
\begin{center}
\Large
\textbf{ConfLab: A Data Collection Concept, Dataset, and\\Benchmark for Machine Analysis of Free-Standing\\Social Interactions in the Wild} \\
\vspace{0.2em}Appendices \\
\vspace{0.3em}
\end{center}

The Appendices are organized as follows:
\begin{itemize}
    \item \hyperref[app:hosting]{Section A}: Hosting, licensing, and organization information for ConfLab
    \item \hyperref[app:datasheet]{Section B}: Documentation for ConfLab, following Datasheets for Datasets \cite{gebru2021datasheets}
    \item \hyperref[app:participant_reports]{Section C}: Sample post-hoc behavioral analysis report sent to each ConfLab participant
    \item \hyperref[app:data-capture]{Section D}: Details about out data-capture setup
    \item \hyperref[app:implementation-details]{Section E}: Implementation details for models used in our benchmark research tasks
    \item \hyperref[app:extra_results]{Section F}: Additional experimental results and ablations
    \item \hyperref[app:reproducibility]{Section G}: Details for reproducibility following the ML Reproducibility Checklist \cite{pineau2021improving} 
\end{itemize}

\section{Hosting, Licensing, and Organization}
\label{app:hosting}

The dataset is hosted by 4TU.ResearchData, available at \url{https://doi.org/10.4121/c.6034313}. 

The dataset itself is available under restricted access defined by an End-User License Agreement (EULA). The EULA itself is available under a CC0 license. The code (\url{https://github.com/TUDelft-SPC-Lab/conflab}) for the benchmark baseline tasks, and the schematics and data associated with the design of our custom wearable sensor called the Midge (\url{https://github.com/TUDelft-SPC-Lab/spcl_midge_hardware}) are available under the MIT License.

\begin{figure}
    \centering
    \includegraphics[scale=0.8]{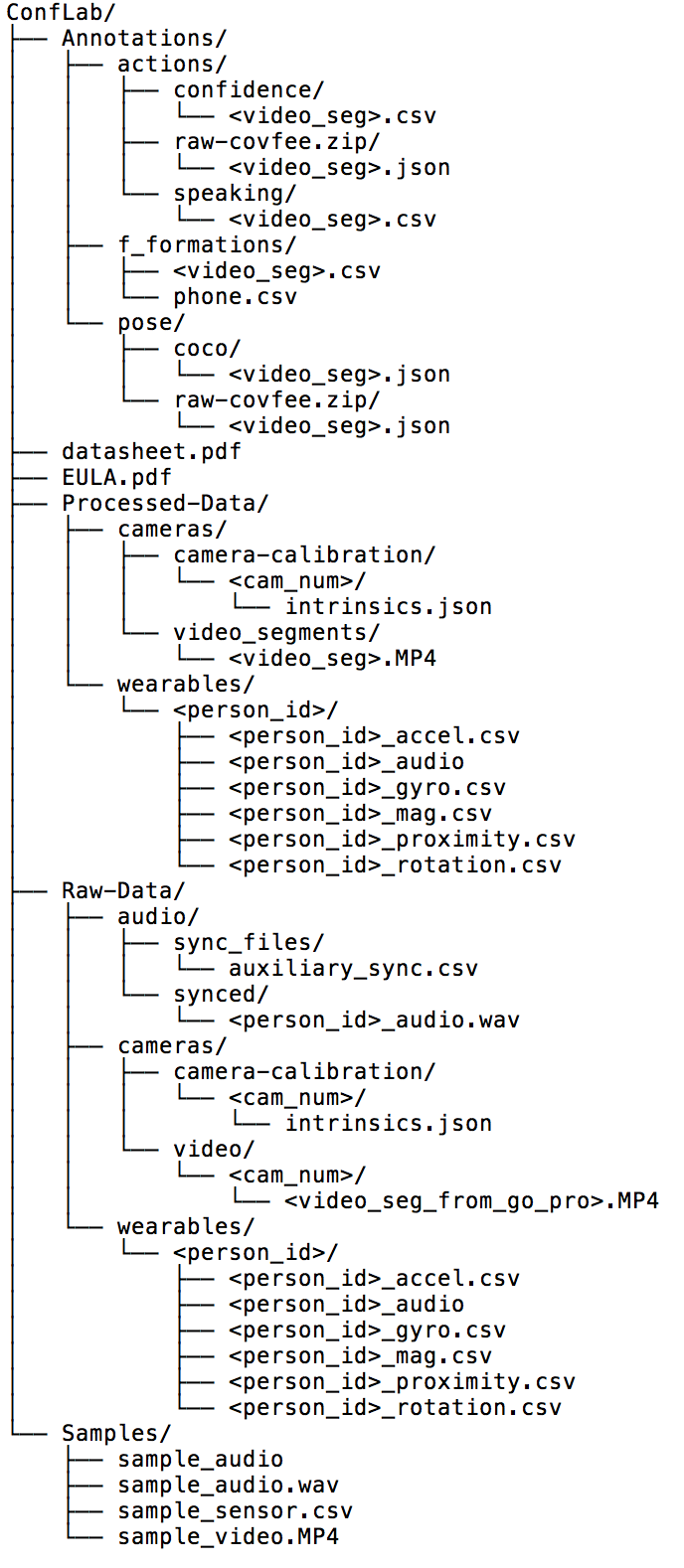}
    \caption{File structure of the ConfLab dataset}
    \label{fig:tree_structure}
\end{figure}

\figurename~\ref{fig:tree_structure} on the next page illustrates the organization of the ConfLab dataset on 4TU.ResearchData. The components are as follows:

\begin{itemize}
    \item Annotations (restricted, \url{https://doi.org/10.4121/20017664}):\\annotations of pose, speaking status, and F-formations 
    \item Datasheet for ConfLab (public, \url{https://doi.org/10.4121/20017559}):\\documentation of the dataset following Datasheets for Datasets \cite{gebru2021datasheets} (see Appendix~\ref{app:datasheet})
    \item EULA (public, \url{https://doi.org/10.4121/20016194}):\\End User License Agreement to be signed for requesting access to the restricted components
    \item Processed-Data (restricted, \url{https://doi.org/10.4121/20017805}):\\processed video and wearable sensor used for annotations
    \item Raw-Data (restricted, \url{https://doi.org/10.4121/20017748}):\\raw video and wearable sensor data
    \item Data Samples (restricted, \url{https://doi.org/10.4121/20017682}):\\samples of the sensor, audio, and video data
\end{itemize}

\clearpage
\definecolor{mypink}{HTML}{f93459} 
\newtcolorbox{dsheetsec}{
    colback=black,
    colframe=black,
    sharp corners,
    fontupper=\color{white}\sffamily\bfseries\centering
}
\newcommand{\questioncolor}{black}

\section{Datasheet For ConfLab} \label{app:datasheet}
\setlength{\parindent}{0pt}

\textbf{This document is based on \textit{Datasheets for Datasets} by Gebru \textit{et
al.} \cite{gebru2021datasheets}. Please see the most updated version
\underline{\textcolor{blue}{\href{http://arxiv.org/abs/1803.09010}{here}}}.}


\begin{dsheetsec} 
MOTIVATION 
\end{dsheetsec}
    
\textcolor{\questioncolor}{\textbf{Q. For what purpose was
the dataset created?} Was there a specific task in mind? Was there
a specific gap that needed to be filled? Please provide a description.
} 

There are two broad motivations for creating this dataset: first, to enable the privacy-preserving, multimodal study of \textit{real-life} social conversation dynamics; second, to bring the higher fidelity of wired in-the-lab recording setups to in-the-wild scenarios, enabling the study of \textit{fine time-scale} social dynamics in-the-wild.  

We propose the Conference Living Lab (ConfLab) with the following goals: (i) a data collection effort that follows a \textit{by the community for the community} ethos: the more volunteers, the more data, (ii) volunteers who potentially use the data can experience first-hand potential privacy and ethical considerations related to sharing their own data, (iii) in light of recent data sourcing issues \cite{BirhaneP21}, we incorporated privacy and invasiveness considerations directly into the decision-making process regarding sensor type, positioning, and sample-rates. 

From a technical perspective,
closest related datasets (see Table~\ref{tab:dataset-comparison} in the main paper) suffer from several technical limitations precluding the analysis and modeling of fine-grained social behavior: (i) lack of articulated pose annotations; (ii) a limited number of people in the scene, preventing complex interactions such as group splitting/merging behaviors, and (iii) an inadequate data sampling-rate and synchronization-latency to study time-sensitive social phenomena \cite[Sec.~3.3]{raman2020modular}. This often requires modeling simplifications such as the summarizing of features over rolling windows \cite{Gedik2017a, Cabrera-Quiros2018b, Quiros2019}. On the other hand, past high-fidelity datasets have largely involved role-played or scripted interactions in lab settings, with often a single-group in the scene. 

This dataset wasn't created with a specific task in mind, but intends to support a wide variety of multimodal modeling and analysis tasks across research domains (see the \textit{Uses} section).

    
\textcolor{\questioncolor}{\textbf{Q. Who created this dataset (e.g., which team, research group) and on behalf
of which entity (e.g., company, institution, organization)?
}} 

ConfLab was initiated by the Socially Perceptive Computing Lab, Delft University of Technology in cooperation and support from the general chairs of ACM Multimedia 2019 (Martha Larson, Benoit Huet, and Laurent Amsaleg), Nice, France. Since this dataset was by the community, for the community, members of the Multimedia community contributed as subjects in the dataset.
    
\textcolor{\questioncolor}{\textbf{Q. What support was needed to make this dataset?
}
(e.g.who funded the creation of the dataset? If there is an associated
grant, provide the name of the grantor and the grant name and number, or if
it was supported by a company or government agency, give those details.)
}

ConfLab was partially funded by Netherlands Organization for
Scientific Research (NWO) under project number 639.022.606 with associated Aspasia Grant, and also by the ACM Multimedia 2019 conference via student helpers, and crane hiring for camera mounting.

\textcolor{\questioncolor}{\textbf{Q. Any other comments?}} \\
None.

\begin{dsheetsec} 
COMPOSITION 
\end{dsheetsec}

\textcolor{\questioncolor}{\textbf{Q. What do the instances that comprise the dataset represent (e.g., documents, photos, people, countries)?} Are there multiple types of instances (e.g., movies, users, and ratings; people and interactions between them; nodes and edges)? Please provide a description.}

The dataset contains multimodal recordings of people interacting during a networking event embedded in an international multimodal machine learning conference.

\begin{figure}[h!]
\centering
\includegraphics[width=0.85\columnwidth,keepaspectratio=true, scale = 0.75]{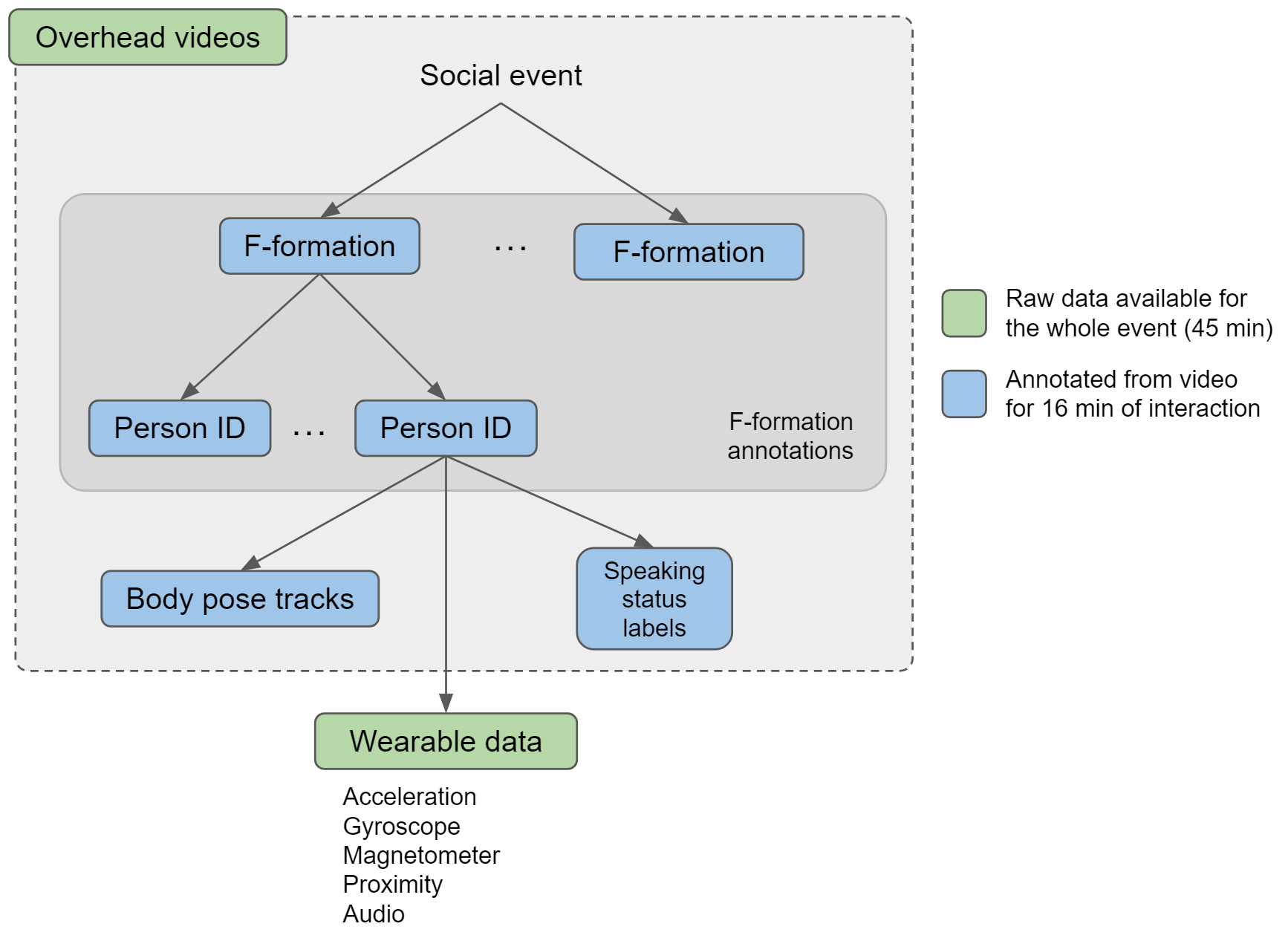}
\caption{Structure of some of the instances in the dataset and their relationships. The interaction space was captured via overhead videos, in which f-formations (conversation groups) were annotated. An F-formation consists of set of people interacting for a variable period of time, and identified via a subject ID. Each person in the F-formation can be associated to their pose (annotated in the videos), their wearable sensor (IMU) data, and their action (speaking status) labels.}
\label{fig:instances}
\end{figure}

Overall, the interaction scene contained conversation groups (operationalized as f-formations), composed of individual subjects, each of which had individual data associated to their wearable sensors. The complete interaction scene was additionally captured by overhead cameras. Figure \ref{fig:instances} shows the structure of these instances and their relationships.

Note however that the precise notion of what constitutes an instance in the dataset is very much task-specific. In our baseline tasks we considered the following instances:

\begin{description}
     \item[Person and Keypoints Detection] Frames, containing pose annotations ($17$ body keypoints per person per frame @$60$~Hz) from $5$ overhead videos ($1920\times1080, 60$~fps) for $16$ minutes of interaction.
    \item[Speaking Status Detection] Windows ($3$~seconds) of wearable sensor data and speaking status annotations ($60$~Hz) extracted from each subject's data.
    \item[F-formations] Operationalized conversation groups, annotated at $1$~Hz from the $16$ minutes of annotated data, and the pose data associated to the people in the F-formation.
\end{description}




\textcolor{\questioncolor}{\textbf{Q. How many instances are there in total (of each type, if appropriate)?}
} \\
The notion of instance is very much dependent on how a user intends to use the data. Regarding the instances in Figure \ref{fig:instances}, our full dataset consist of 45 minutes of:

\begin{description}
    \item[Video recordings] from $10$ overhead cameras placed over the interaction area. Five of these videos, enough to cover the complete interaction area, were used in annotation.
    \item[Individual wearable sensor data] For the $48$ subjects in the interaction area, a chest-worn conference-type badge recorded: audio ($1250$~Hz), and Inertial Measurement Unit (IMU) readings (accelerometer @ $56$~Hz, gyroscope @$56$~Hz, magnetometer @$56$~Hz and Bluetooth RSSI-based proximity @$5$~Hz)
    \item[Conference experience label] For each of the $48$ subjects, an associated self-report label indicating whether it was their first time in the conference.
\end{description}

The instances in the annotated $16$ minutes segment out of the $45$ minutes of interaction contain:

\begin{description}
    \item[2D body poses] For each of the $48$ subjects, full body pose tracks annotated at $60$Hz ($17$ keypoints per person). These were annotated using $5$ of the $10$ overhead cameras due to the significant overlap in views (cameras $2$, $4$, $6$, $8$, and $10$). Annotations were done separately for each camera by annotating all of the people visible in each video, for each of the $5$ cameras, and tagged with a participant ID. We made use of a novel continuous technique for annotation of keypoints. We chose this approach via a pilot study with 3 annotators, comparing our technique to annotations done using the non-continuous CVAT tool. We found no statistically significant differences in errors per-frame (as measured using Mean Squared Error across annotators), despite a 3x speed-up in annotation time in the continuous condition. The details of the technique and this pilot study can be found in \cite{covfee-paper}. 
    
    \item[Speaking status annotations] For each of the $48$ subjects, these include a) a binary signal ($60$~Hz) indicating whether the person is perceived to be speaking or not; b) continuous confidence value ($60$~Hz) indicating the degree of confidence of the annotator in their speaking status assessment. These annotations were done without access to audio due to issues with the synchronization of the audio recordings at the time of annotation. The confidence assessment is therefore largely based on the visibility of the target person and their speaking-associated gestures (eg. occlusion, orientation w.r.t. camera, visibility of the face)? We measured inter-annotator agreement for speaking status in a pilot where two annotators labeled three data subjects for 2 minutes each. We measured a frame-level agreement (Fleiss’ $\kappa$) of 0.552, comparable to previous work 
    \cite{Cabrera-Quiros2018b}.
    
    \item[F-formation annotations] These annotations label the conversing groups in the scene following previous work. Each individual belongs to one F-formation at a time or is a singleton in the interaction scene. The membership is binary. The annotations were done by one of the authors at $1$~Hz by watching the video. The time-stamped usage of mobile phones are available as auxiliary annotations, which are useful for the study of the role of mobile phone users as associates of F-formations. Since Kendon's theories date back to before the widespread use of mobile phones, their influence on F-formation membership remains an open question. 
\end{description}

In our baseline tasks, which made use of the complete annotated section of the dataset, the instance numbers were the following:

\begin{description}
     \item[Person and Keypoints Detection] $119$k frames ($60$fps) containing $1967$k person instances (poses) in total, from $48$ subjects recorded in $5$ cameras ($16$ minutes of annotated segment).
    \item[Speaking Status Detection] $42884$ $3$-second windows, extracted from the $48$ participants' wearable data and speaking status annotations.
    \item[F-formations] $119$ conversation groups. Details are in Section~\ref{sec:description}.
\end{description}

\textcolor{\questioncolor}{\textbf{Q. Does the dataset contain all possible instances or is it a sample (not necessarily random) of instances from a larger set?}
If the dataset is a sample, then what is the larger set? Is the sample
representative of the larger set (e.g., geographic coverage)? If so, please
describe how this representativeness was validated/verified. If it is not
representative of the larger set, please describe why not (e.g., to cover a
more diverse range of instances, because instances were withheld or
unavailable).
} 

The participants in our data collection are a sample of the conference attendees. Participants were recruited via the conference website, social media posting, and approaching them in person during the conference. Because participation in such a data collection can only be voluntary, the sample was not pre-designed and may not be representative of the larger set.

Additionally, $16$ minutes of sensor data has been annotated for keypoints, speaking status and F-formations out of the total of $45$ minutes recorded. The remaining part (across all modalities) is provided with no labels. For privacy reasons, the elevated cameras (distinct from the previously mentioned $8$ overhead cameras) and also individual frontal headshots that were used for manually associating the video data to the wearable sensor data is not being shared. 








\textcolor{\questioncolor}{\textbf{Q. Is any information missing from individual instances?
}
If so, please provide a description, explaining why this information is
missing (e.g., because it was unavailable). This does not include
intentionally removed information, but might include, e.g., redacted text.
} 

Camera $5$ failed early during the recording, but the space underneath it was captured by the adjacent cameras due to the high overlap in the camera field-of-views. Nevertheless we share what was recorded before the failure from camera $5$, bringing the total number of cameras to $9$. 

\textcolor{\questioncolor}{\textbf{Q. Are relationships between individual instances made explicit (e.g., users’
movie ratings, social network links)?
}
If so, please describe how these relationships are made explicit.
} 

The F-formations, subjects, and their associated data relate as shown in Figure \ref{fig:instances}. These associations are made explicit in the dataset via anonymous subject IDs, associated to pose tracks, speaking status annotations, and wearable sensor data. These same IDs were used to annotate the F-formations.

Pre-existing personal relationships between the subjects were not requested for privacy reasons.



\textcolor{\questioncolor}{\textbf{Q. Are there recommended data splits (e.g., training, development/validation,
testing)?
}
} 

Since the dataset can be used to study a variety of tasks, the answer to this question is task dependent. Please refer to our reproducibility details (Appendix~\ref{app:reproducibility} of our associated paper) for information about the splits that we used in out baselines. 

\textcolor{\questioncolor}{\textbf{Q. Are there any errors, sources of noise, or redundancies in the dataset?
}
If so, please provide a description.
} 

\begin{description}
    \item[Individual audio] Because audio was recorded by a front-facing wearable device worn on the chest, it contains a significant amount of cocktail party noise and cross-contamination from other people in the scene. In our experience this means that automatic speaking status detection is challenging with existing algorithms but manual annotation is possible.
    
    \item[Videos and 2D body poses] It is important to consider that the same person may appear in multiple videos at the same time if the person was in view of multiple cameras. Because 2D poses were annotated per video, the same is true of pose annotations. Each skeleton was tagged with a person ID, which should serve to identify such cases when necessary.
\end{description}

\textcolor{\questioncolor}{\textbf{Q. Is the dataset self-contained, or does it link to or otherwise rely on
external resources (e.g., websites, tweets, other datasets)?}} 

The dataset is self-contained.

\textcolor{\questioncolor}{\textbf{Q. Does the dataset contain data that might be considered confidential (e.g.,
data that is protected by legal privilege or by doctor-patient
confidentiality, data that includes the content of individuals’ non-public
communications)?
}
} 

The data contains personal data under GDPR in the form of video and audio recordings of subjects. The dataset is shared under an End User License Agreement for research purposes, to ensure that the data is not made public, and to protect the privacy of data subjects.

\textcolor{\questioncolor}{\textbf{Q. Does the dataset contain data that, if viewed directly, might be offensive,
insulting, threatening, or might otherwise cause anxiety?}} 

No.

\textcolor{\questioncolor}{\textbf{Q. Does the dataset relate to people?}} 

Yes, the dataset contains recordings of human subjects.

\textcolor{\questioncolor}{\textbf{Q. Does the dataset identify any subpopulations (e.g., by age, gender)?
}
If so, please describe how these subpopulations are identified and
provide a description of their respective distributions within the dataset.
} 

Data subjects answered the following questions before the start of the data collection event, after filling in their consent form:

\begin{itemize}
    \item Is this your first time attending ACM MM?
    \item Select the area(s) that describes best your research interest(s) in recent years. Descriptions of each theme are listed here: \url{https://acmmm.org/call-for-papers/}
\end{itemize}

\begin{figure}[t!]
\centering
\includegraphics[width=0.8\columnwidth,keepaspectratio=true, scale = 0.75]{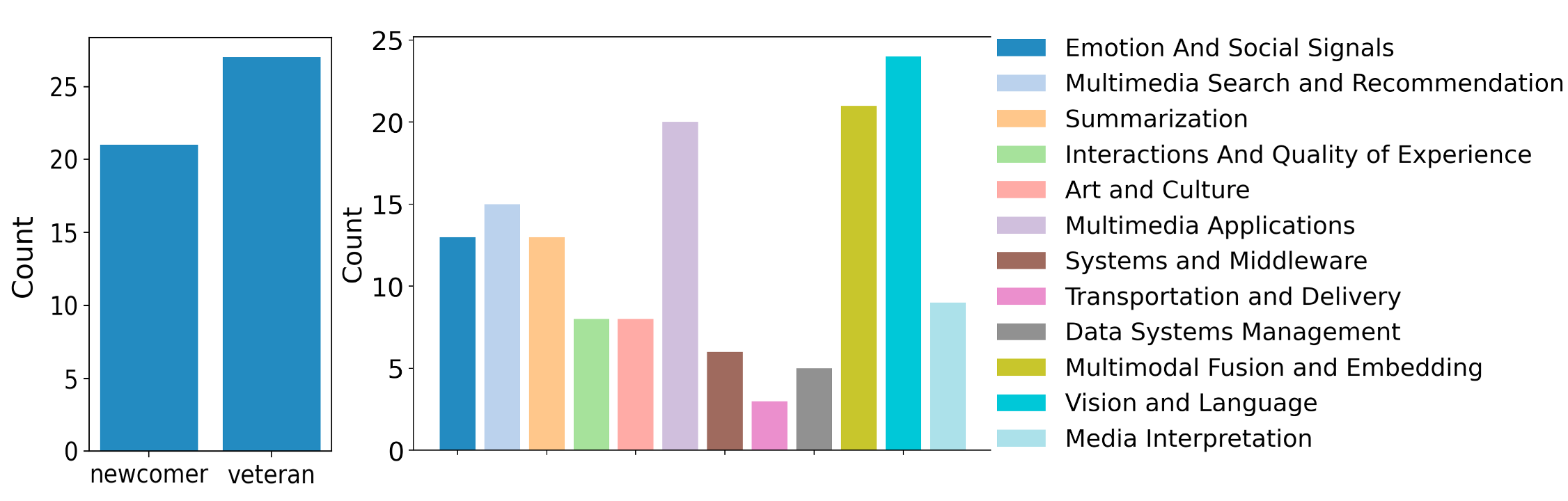}
\caption{Distribution of participant seniority (left) and research interests (right) in percentage.}
\label{fig:participant_stats}
\end{figure}

Figure \ref{fig:participant_stats} shows the distribution of the responses / populations.

\textcolor{\questioncolor}{\textbf{Q. Is it possible to identify individuals (i.e., one or more natural persons),
either directly or indirectly (i.e., in combination with other data) from
the dataset?
}} 

We do not share any directly identifiable information as part of the dataset.
However, individuals may be identified in the video recordings if the observer knows the participants in the recordings personally. Otherwise, individuals in the dataset may potentially be identified in combination with publicly available pictures or videos (from conference attendees or conference official photographer) from other media from the conference the dataset was recorded at. In any case, re-identifying the subjects is strictly against the End User License Agreement under which we share the dataset. 

\textcolor{\questioncolor}{\textbf{Q. Does the dataset contain data that might be considered sensitive in any way
(e.g., data that reveals racial or ethnic origins, sexual orientations,
religious beliefs, political opinions or union memberships, or locations;
financial or health data; biometric or genetic data; forms of government
identification, such as social security numbers; criminal history)?
}
} 

We did not request any such information from data participants. Here, the ACM Multimedia '19 General Chair Martha Larson also helped advocate on behalf of the attendees during the survey-design stage. As a result of these discussions, information such as participant gender, ethnicity, or country of origin was not asked. 

\textcolor{\questioncolor}{\textbf{Q. Any other comments?
}} \\
None. 

\begin{dsheetsec} 
COLLECTION 
\end{dsheetsec}

\textcolor{\questioncolor}{\textbf{Q. How was the data associated with each instance acquired?
}
Was the data directly observable (e.g., raw text, movie ratings),
reported by subjects (e.g., survey responses), or indirectly
inferred/derived from other data (e.g., part-of-speech tags, model-based
guesses for age or language)? If data was reported by subjects or
indirectly inferred/derived from other data, was the data
validated/verified? If so, please describe how.
} 

The collected data is directly observable, containing video recordings, low-frequency audio recordings and wearable sensing signals (inertial motion unit (IMU) and Bluetooth proximity sensors) of individuals in the interaction scenes. Accompanying data includes self-reported binary categorization of experience level which is available upon request from the authors. The self-reported interests categories are not shared because of privacy concerns.  

Video recordings capture the whole interaction floor where the association from multi-modal data to individual is done manually by annotators by referring to frontal (not-shared) and overhead views. The rest of the data was acquired from the wearable sensing badges, which is person-specific (i.e., no participant shared the device). Video and audio data were verified in playback. Wearable sensing data was verified through plots after parsing.

\textcolor{\questioncolor}{\textbf{Q. Over what timeframe was the data collected?
}
Does this timeframe match the creation timeframe of the data associated
with the instances (e.g., recent crawl of old news articles)? If not,
please describe the timeframe in which the data associated with the
s was created. Finally, list when the dataset was first published.
} 

All data was collected on October 24, 2019, except the self-reported experience level and research interest topics which are either obtained on the same day or not more than one week before the data collection day. This time frame matches the creation time frame of the data association for wearable sensing data. Video data was associated with individual during annotation stage (2020-2021), but all information used for association was obtained on the data collection day.

\textcolor{\questioncolor}{\textbf{Q. What mechanisms or procedures were used to collect the data (e.g., hardware
apparatus or sensor, manual human curation, software program, software
API)?
}
} 

To record videos, we used 14 GoPro Hero 7 Black cameras. 
The wearable sensor hardware has been documented and open-sourced at \url{https://github.com/TUDelft-SPC-Lab/spcl_midge_hardware}. The validation of the sensors was completed through an external contractor engineer.  
The data collection software was documented and published in \cite{covfee-paper}, which includes validation of the system. These hardwares and mechanisms have been open-sourced along with their respective publication. 

The synchronization setup for data collection (intramodal and intermodal) was documented and published in \cite{raman2020modular}, which includes validation of the system.

\begin{figure}[!t]
\centering
\begin{subfigure}[b]{0.32\textwidth}
    \centering
    \includegraphics[angle=270,totalheight=3cm]{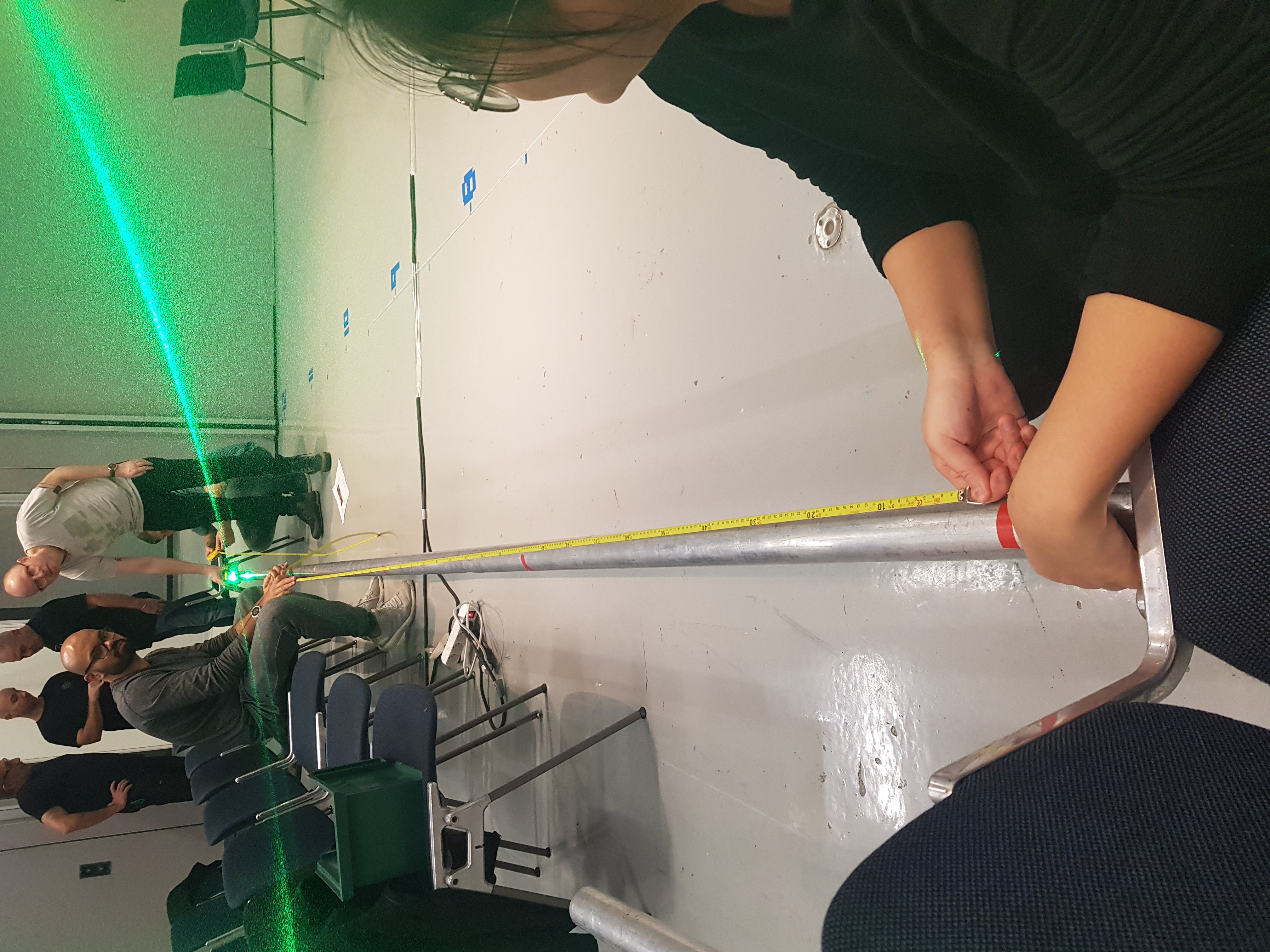}
    \caption{Aligning cameras}
    \label{fig:cam-align}
\end{subfigure}\hfill
\begin{subfigure}[b]{0.32\textwidth}
    \centering
    \includegraphics[height=3cm]{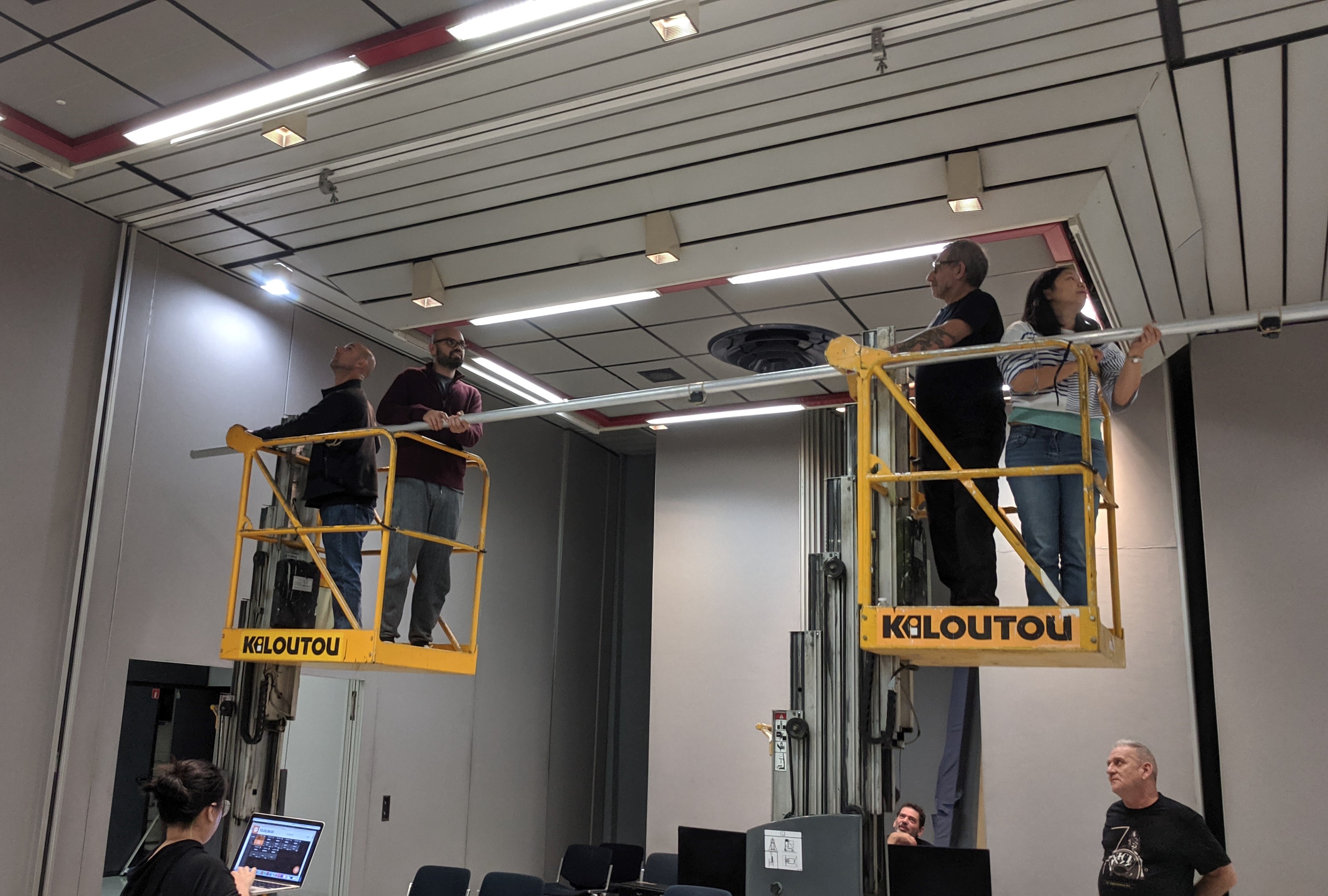}
    \caption{Affixing the mounting beam}
    \label{fig:cam-mount}
\end{subfigure}\hfill
\begin{subfigure}[b]{0.32\textwidth}
    \centering
    \includegraphics[angle=270,totalheight=3cm]{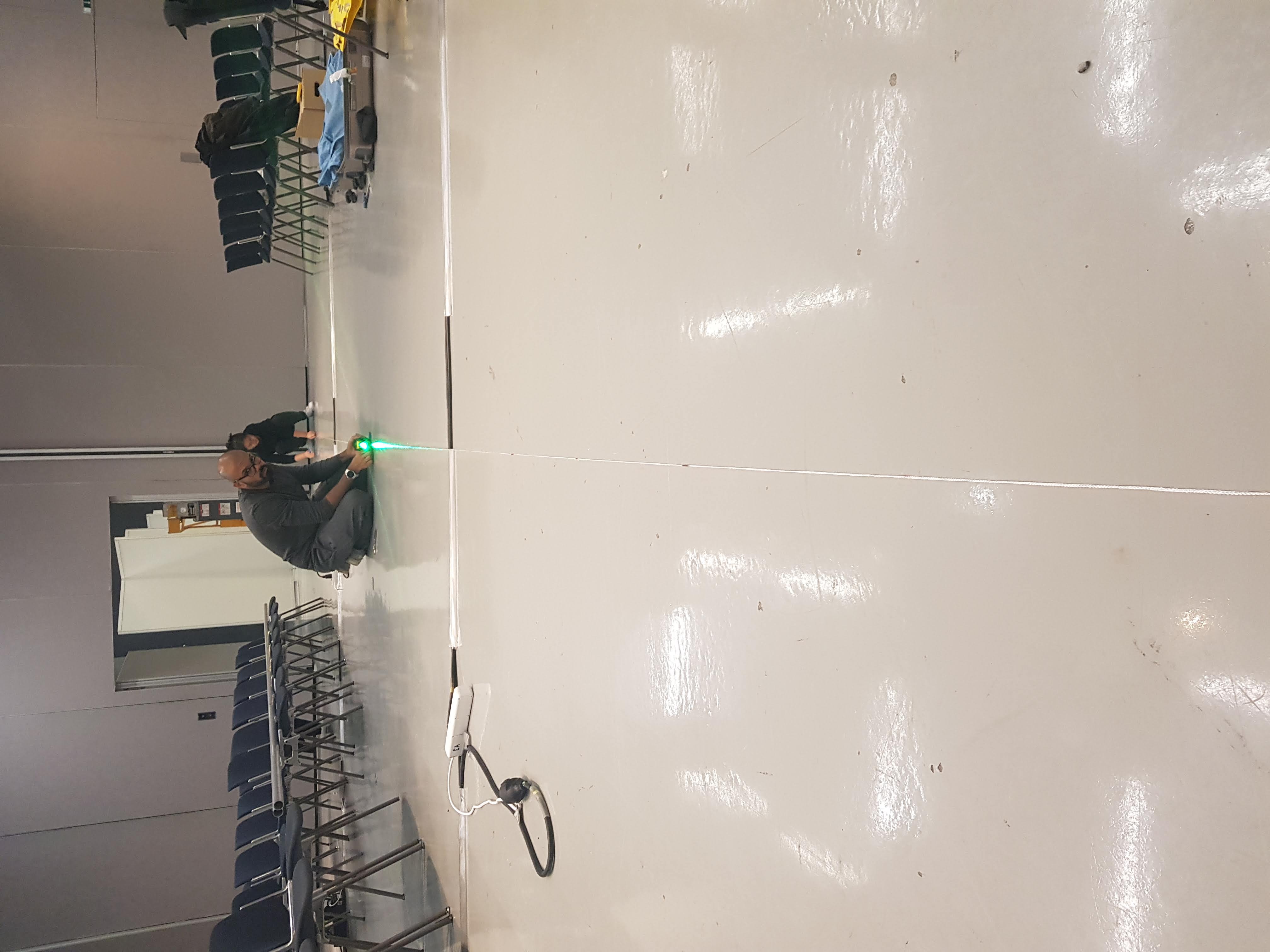}
    \caption{Aligning floor markers}
    \label{fig:align-grid}
\end{subfigure}\vfill
\begin{subfigure}[b]{0.32\textwidth}
    \centering
    \includegraphics[height=3cm]{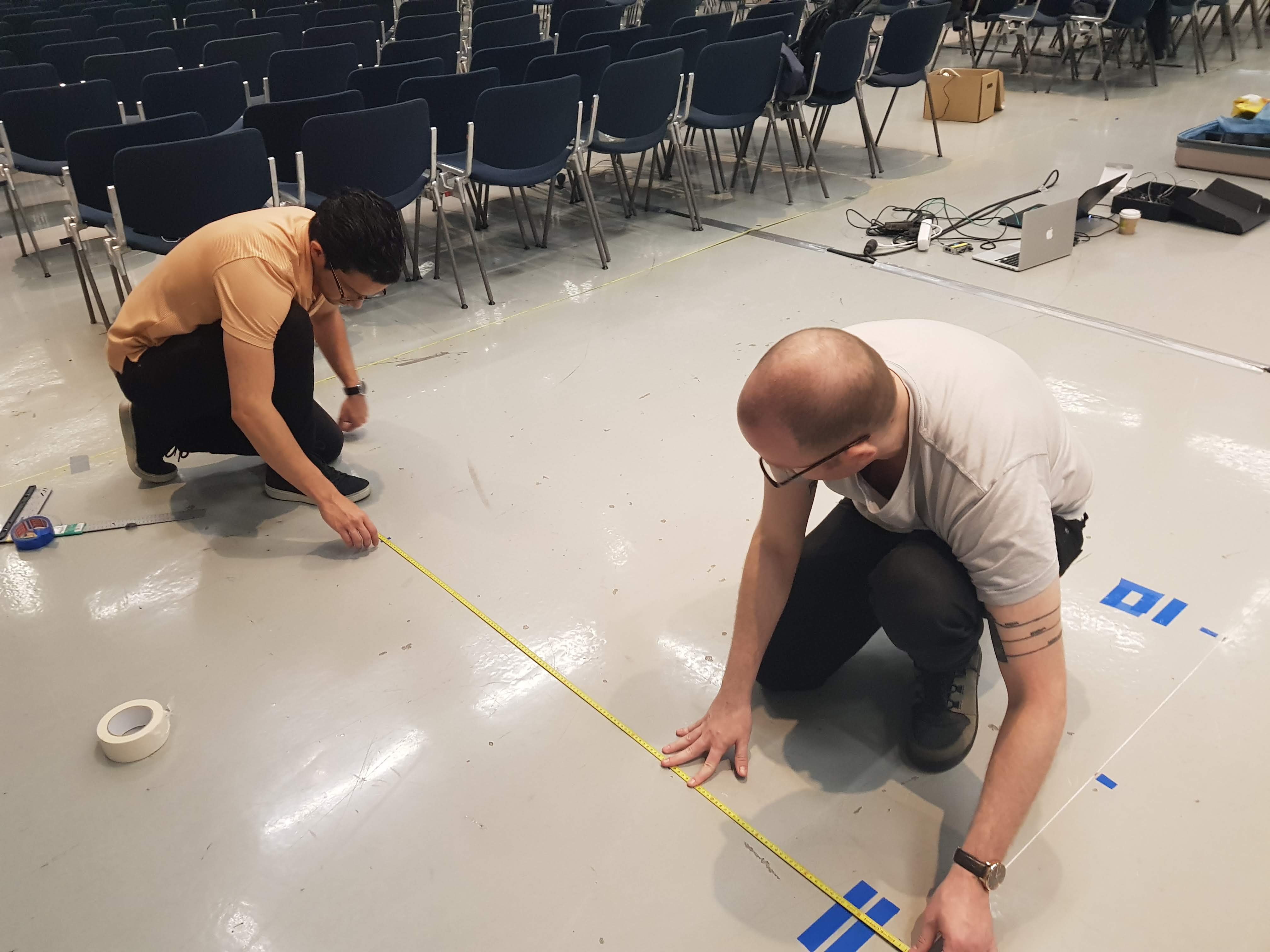}
    \caption{Marking the floor grid}
    \label{fig:mark-grid}
\end{subfigure}\hfill
\begin{subfigure}[b]{0.32\textwidth}
    \centering
    \includegraphics[height=3cm]{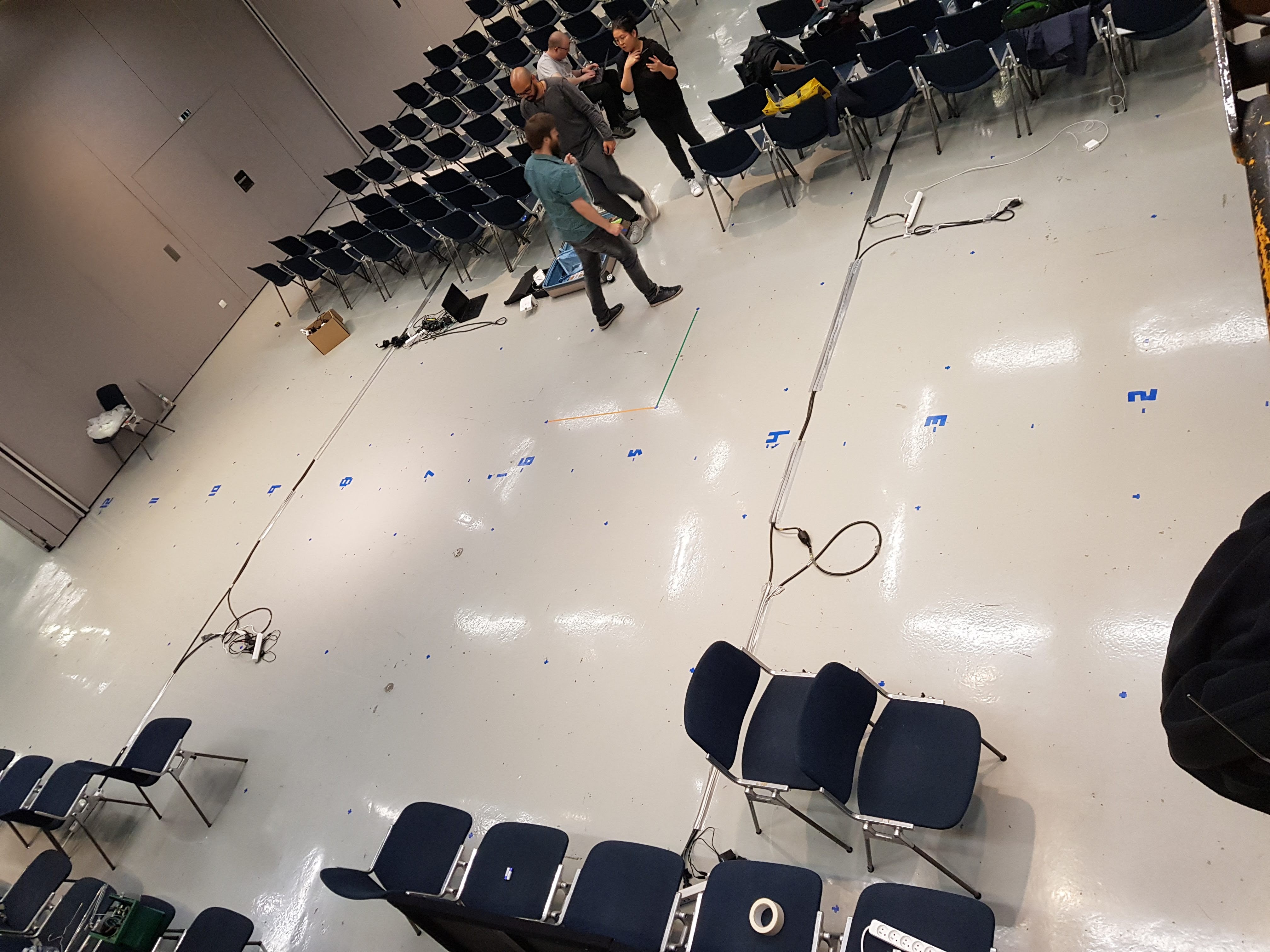}
    \caption{Interaction area}
    \label{fig:recording-area}
\end{subfigure}\hfill
\begin{subfigure}[b]{0.32\textwidth}
    \centering
    \includegraphics[height=3cm]{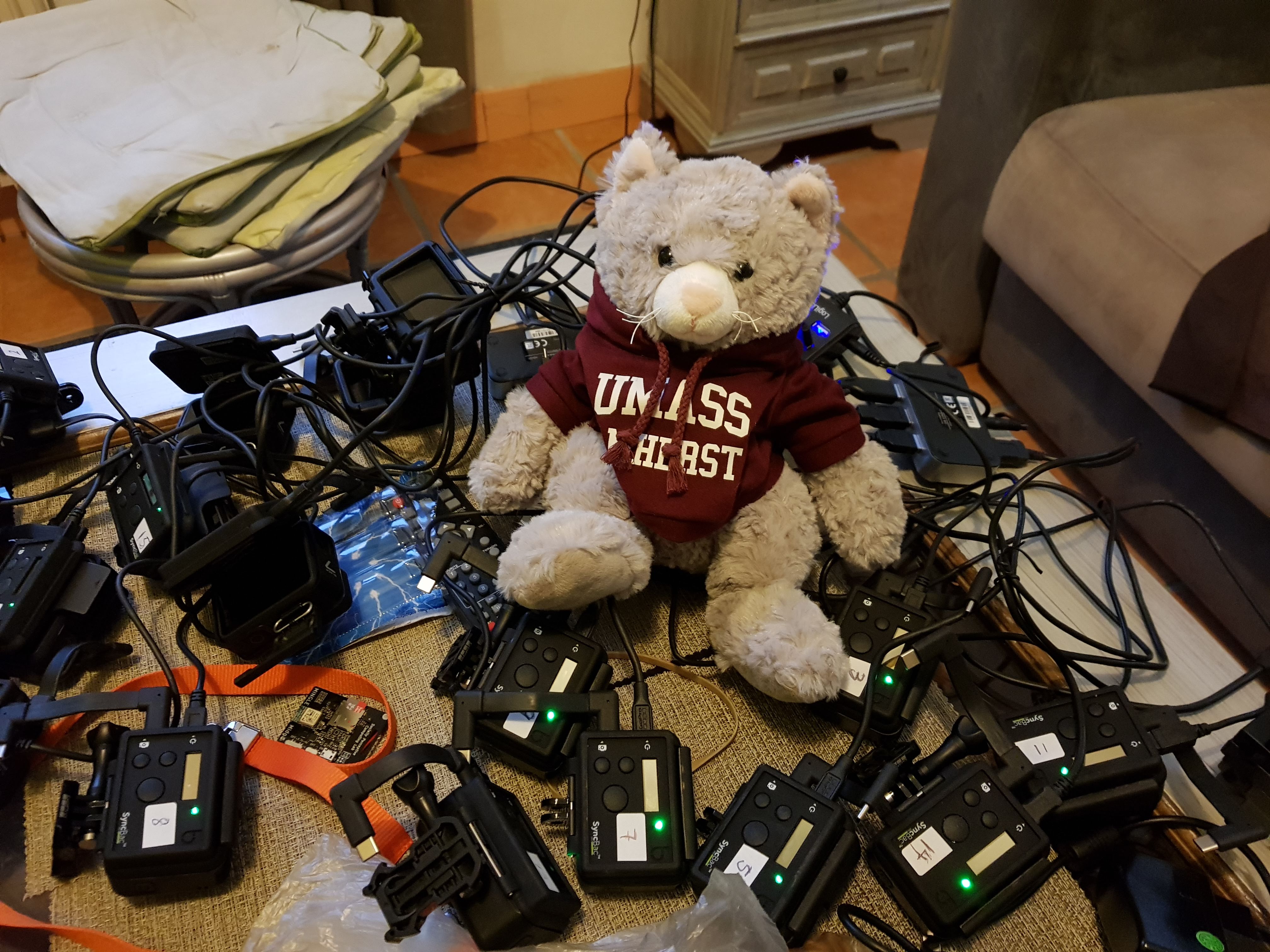}
    \caption{Verifying camera sync.}
    \label{fig:cam-sync}
\end{subfigure}\vfill
\begin{subfigure}[b]{0.32\textwidth}
    \centering
    \includegraphics[height=3cm]{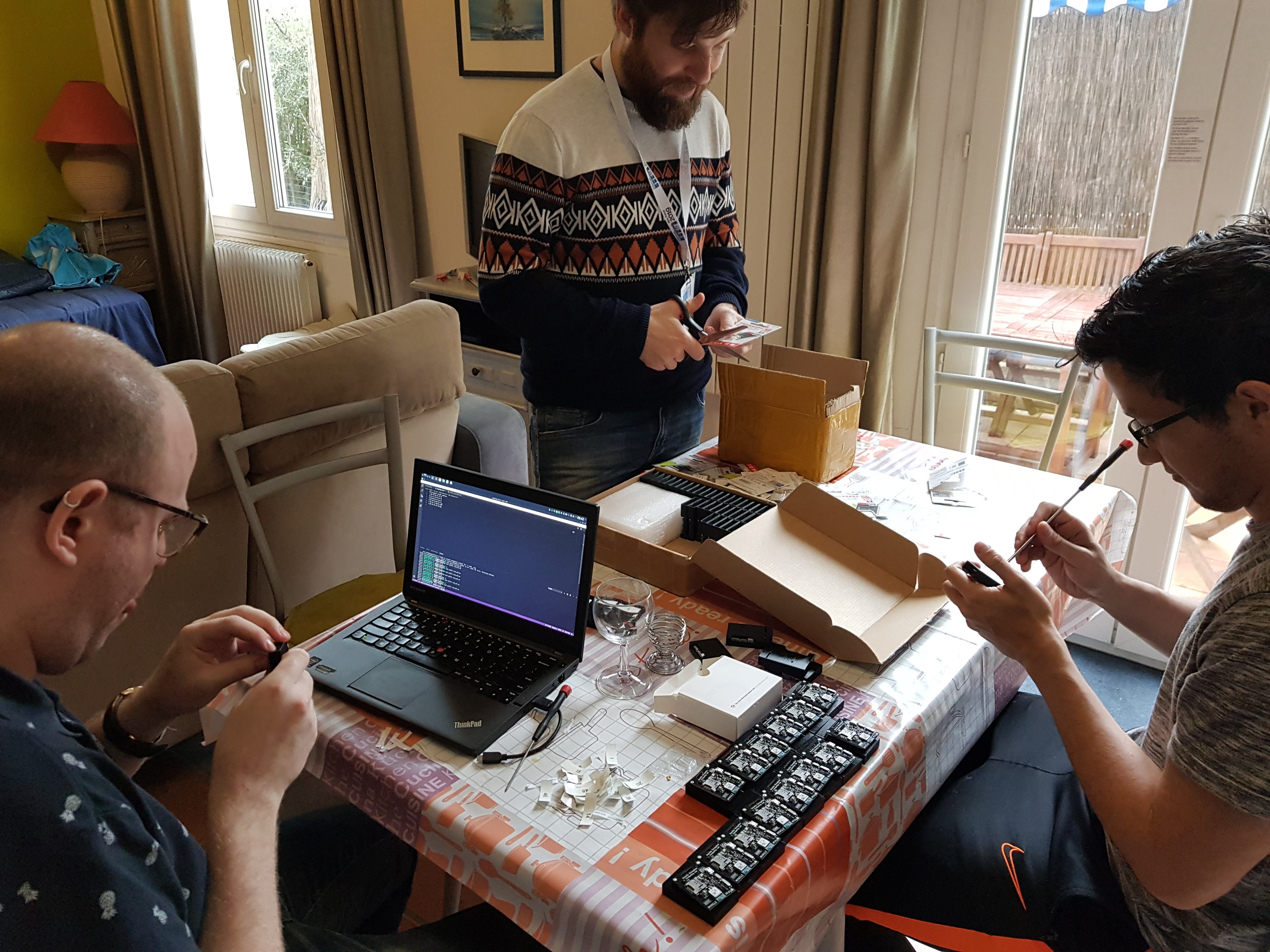}
    \caption{Assembling Midges}
    \label{fig:assemling-midges}
\end{subfigure}\hfill
\begin{subfigure}[b]{0.32\textwidth}
    \centering
    \includegraphics[height=3cm]{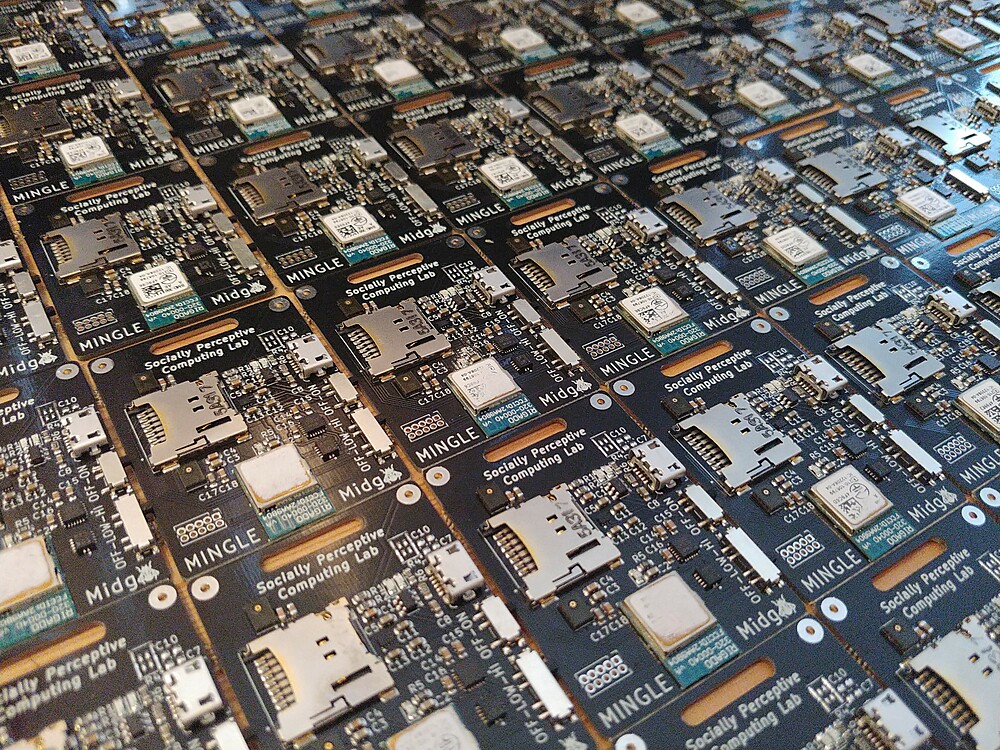}
    \caption{Midges}
    \label{fig:midges}
\end{subfigure}\hfill
\begin{subfigure}[b]{0.32\textwidth}
    \centering
    \includegraphics[height=3cm]{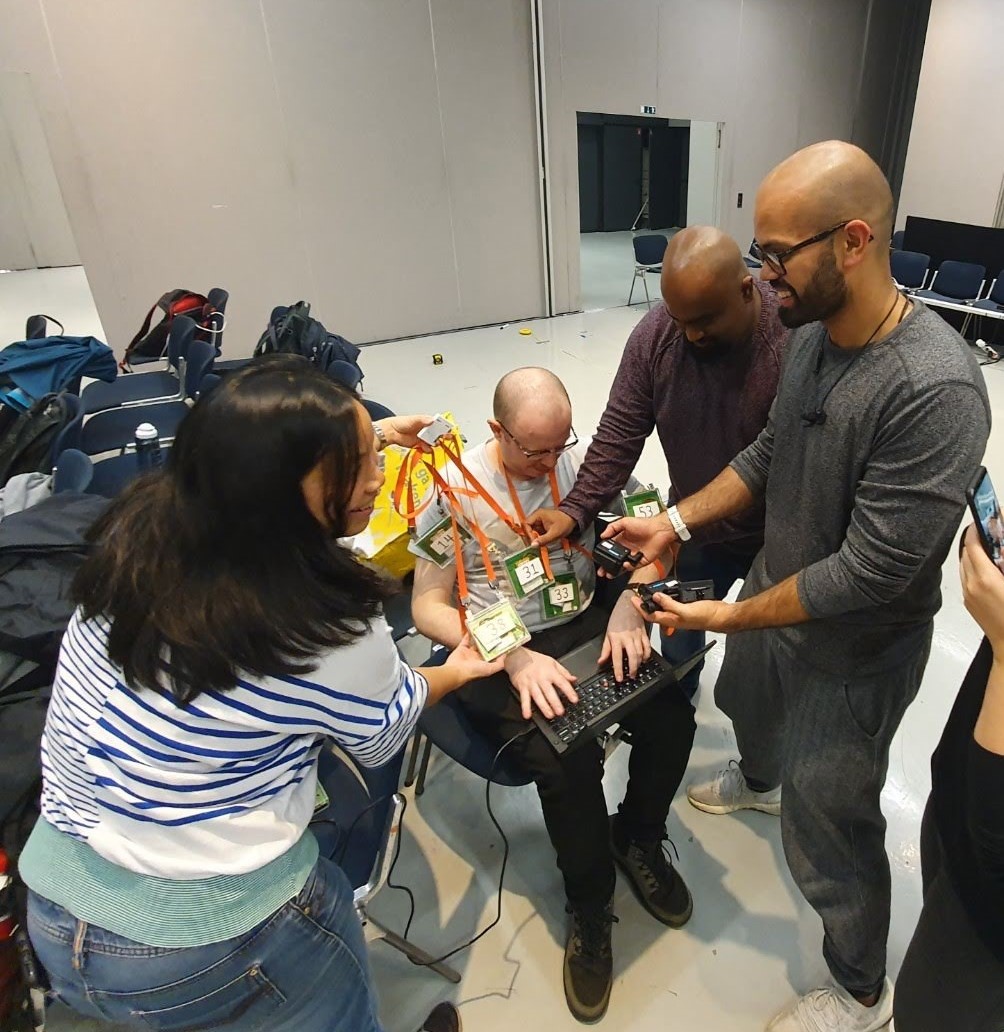}
    \caption{Verifying crossmodal sync.}
    \label{fig:cross-modal}
\end{subfigure}\hfill
\caption{Illustrating the process of setting up the data recording.}
\label{fig:recording-setup}
\end{figure}

To lend the reader further insight into the process of setting up the recording of such datasets in-the-wild, we share images of our process in Figure~\ref{fig:recording-setup}.

\textcolor{\questioncolor}{\textbf{Q. What was the resource cost of collecting the data?
}
} 
\begin{table}[]
\centering
\caption{Itemized costs associated with recording ConfLab}
\begin{tabular}{@{}llr@{}}
\toprule
\textbf{Item} & & \textbf{Cost (USD)} \\ 
\midrule
\textbf{Travel (total for 6 people)} & &  \\
\hspace{4mm}Flights & & 1800 \\
\hspace{4mm}Accommodation & & 1500 \\ 
\textbf{Equipment (one time)} & &      \\ 
\hspace{4mm}Mounting scaffold & & 2000  \\
\hspace{4mm}14 $\times$ GoPro Hero 7 Black & & 4900 \\
\hspace{4mm}Designing the Midge (custom wearable, now made open source) & & 26000 \\
\hspace{4mm}110 $\times$ Midges (boards, batteries, 4~GB sd cards, cases)  & & 3660 \\
\hspace{4mm}Multimodal synchronization setup & & 730  \\ 
\textbf{Annotations} & & 8000 \\ 
\textbf{Computational cost for experiments} & & 500  \\ 
\bottomrule
\end{tabular}
\label{tab:resource_cost}
\end{table}

The resources required to run this first edition of ConfLab include equipment, logistics, and travel costs. Table \ref{tab:resource_cost} shows the full breakdown of the costs. The equipment expenses are fixed one-time costs since the same equipment can be used for future iterations of ConfLab. The on-site costs at the conference venue were toward renting a crane for a day to mount the cameras on a scaffold on the ceiling. We have open-sourced the Midge (our custom wearable) schematics so that others don’t need to spend on the design and development.

No additional energy consumption was incurred for collecting the data. However, the ancillary activities (e.g., flights, accommodation) resulted in energy consumption. Flights from the Netherlands to France round-trip for six passengers results in 1020 kg carbon emissions. Accommodation for six members resulted in 22 kWh energy consumption. 

\textcolor{\questioncolor}{\textbf{Q. If the dataset is a sample from a larger set, what was the sampling
strategy (e.g., deterministic, probabilistic with specific sampling
probabilities)?
}
} 

ConfLab contains both annotated and unannotated segments of multi-modal data. The segment where the articulated pose and speaking status were annotated is selected to maximize crowd density in the scenes. The annotated segment is $16$ minutes; the whole set is roughly $1$ hour of recordings.

\textcolor{\questioncolor}{\textbf{Q. Who was involved in the data collection process (e.g., students,
crowdworkers, contractors) and how were they compensated (e.g., how much
were crowdworkers paid)?
}
} 

The Conflab dataset was captured during a special social event called \textit{Meet the Chairs!} at an international conference on signal processing and machine learning. Newcomers and old-timers to the conference freely donated their social behaviour data as part of a \textit{by the community, for the community} data collection effort. Aside from the chance to meet the chairs and create a community dataset, the attendees also received a personalised report of their social behaviour from the wearable sensors (see Appendix \ref{app:participant_reports}) Conference student volunteers were involved in assisting the set-up of the event. Conference organizers (mentioned in the \textit{Motivation} section) assisted in connecting us with conference venue contacts to mount our technical set-ups in the room. Volunteers and conference organizers were not paid by us. Conference venue contacts were paid by the conference organizers.

Data annotations were completed by crowdsourced workers. The crowdsourced workers were paid \$0.20 for qualification assignment (note that typically requesters do not pay for qualification tasks). Depending on the submitted results, workers earn qualification to access of the actual tasks. The annotation tasks were categorized into low-effort (\$150), medium-effort (\$300), and high-effort (\$450), corresponding to the amount of estimated time each would take. The duration of the tasks was determined by the crowd density and through timing of the pilot studies. The average hourly payment to workers is around \$8. 

\textcolor{\questioncolor}{\textbf{Q. Were any ethical review processes conducted (e.g., by an institutional
review board)?
}
If so, please provide a description of these review processes, including
the outcomes, as well as a link or other access point to any supporting
documentation.
}  

The data collection was approved by the Human Research Ethics Committee (HREC) of our university (Delft University of Technology), which reviews all research involving human subjects. The data collection protocol is also compliant to the conference location's national authorities (France). The review process included addressing privacy concerns to ensure compliance with GDPR and university guidelines, review of our informed consent form, data management plan, and end user license agreement for the dataset and a safety check of our custom wearable devices.

\textcolor{\questioncolor}{\textbf{Q. Does the dataset relate to people?
}
} 

Yes. 
\begin{figure*}[t!]
\centering
\begin{tabular}{ c c }
     \includegraphics[width=0.41\textwidth]{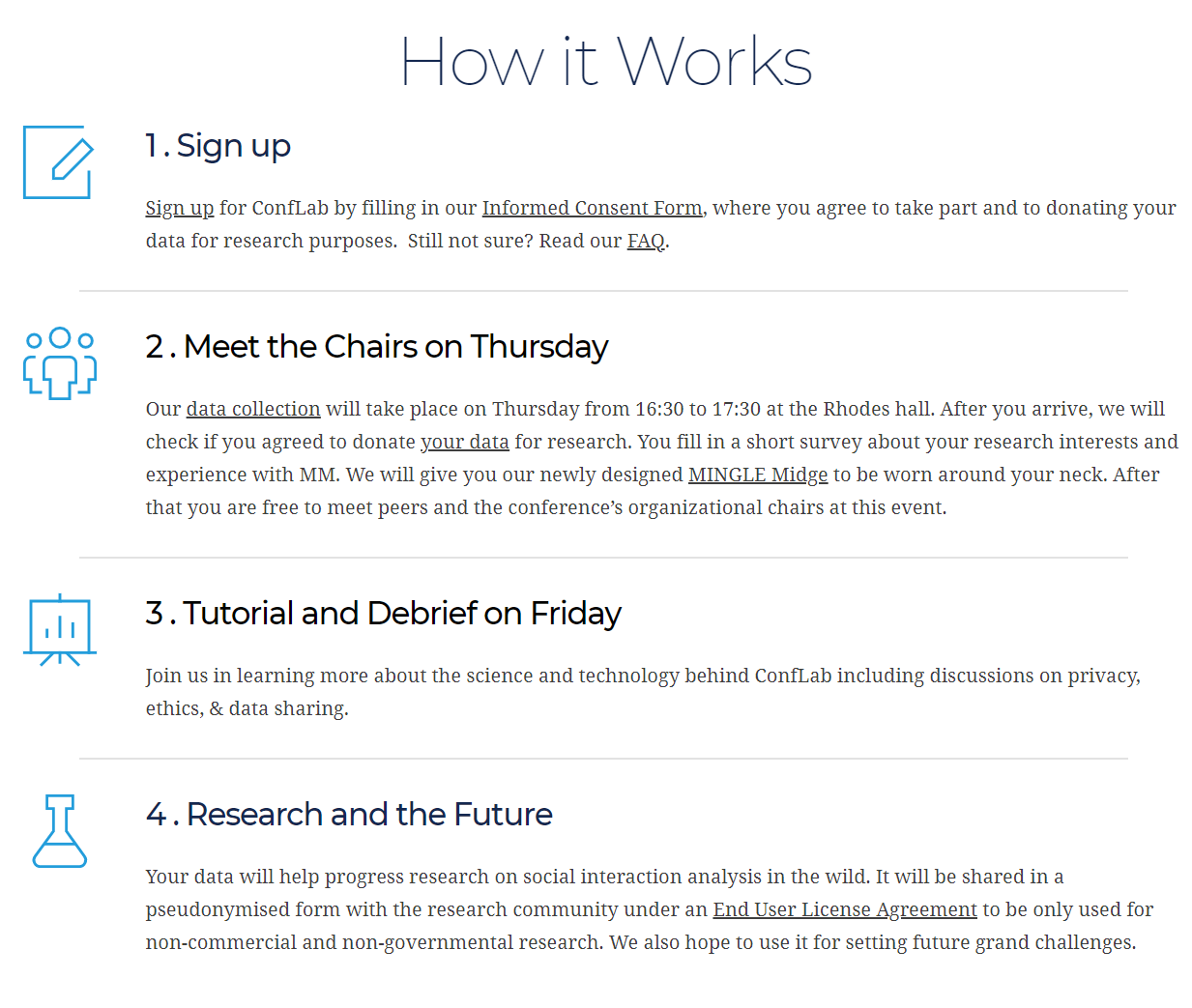} & \includegraphics[width=0.41\textwidth]{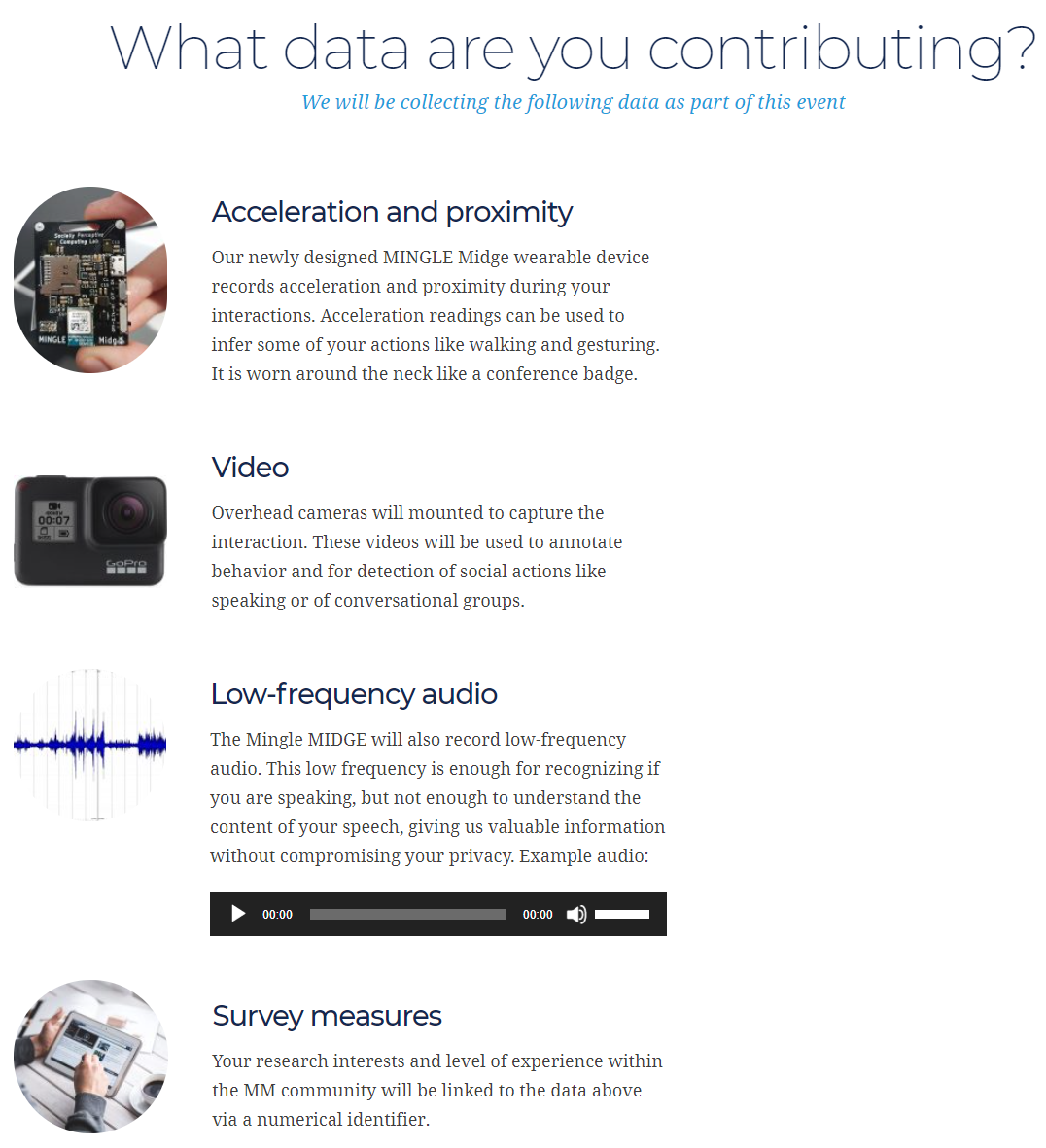} \\ 
\end{tabular}
\caption{Screenshots of the ConfLab web-page used for participant recruitment and registration.}
\label{fig:conflabwebpage}
\end{figure*}

\textcolor{\questioncolor}{\textbf{Q. Did you collect the data from the individuals in question directly, or
obtain it via third parties or other sources (e.g., websites)?
}
} 

We collected the data from individuals directly. 

\textcolor{\questioncolor}{\textbf{Q. Were the individuals in question notified about the data collection?
} If so, please describe (or show with screenshots or other information) how
notice was provided, and provide a link or other access point to, or
otherwise reproduce, the exact language of the notification itself.
} 

The individuals were notified about the data collection and their participation is voluntary. The data collection was staged at an event called \textit{Meet the Chairs} at ACM MM 2019. The ConfLab web page (\url{https://conflab.ewi.tudelft.nl/}) served to communicate the aim of the event, what was being recorded, and how participants could sign up. This allowed us to embed the informed consent into this framework so we could keep track of sign ups. See Figure \ref{fig:conflabwebpage} for screenshots. This event website was also shared by the conference organizers and chairs (\url{https://2019.acmmm.org/conflab-meet-the-chairs/index.html}).

\textcolor{\questioncolor}{\textbf{Q. Did the individuals in question consent to the collection and use of their
data?
}
If so, please describe (or show with screenshots or other information) how
consent was requested and provided, and provide a link or other access
point to, or otherwise reproduce, the exact language to which the
individuals consented.
} 

All the individuals who participated in the data collection gave their consent by signing a consent form. A copy of the form is attached below in Figure \ref{fig:consent_form}. 

\textcolor{\questioncolor}{\textbf{Q. If consent was obtained, were the consenting individuals provided with a
mechanism to revoke their consent in the future or for certain uses?
}
 If so, please provide a description, as well as a link or other access
 point to the mechanism (if appropriate)
} 

Yes, the consenting individuals were informed about the possibility of revoking access to their data within a period of $3$ months after the data collection experiment, and not after that. The description is included in the consent form.

\textcolor{\questioncolor}{\textbf{Q. Has an analysis of the potential impact of the dataset and its use on data
subjects (e.g., a data protection impact analysis) been conducted?
}
}\\
No. 

\textcolor{\questioncolor}{\textbf{Q. Any other comments?
}}\\
None. 
\newpage
\begin{dsheetsec}
PREPROCESSING / CLEANING / LABELING
\end{dsheetsec} 

\begin{figure}[t!]
\includegraphics[width=\columnwidth,keepaspectratio=true, scale = 0.75]{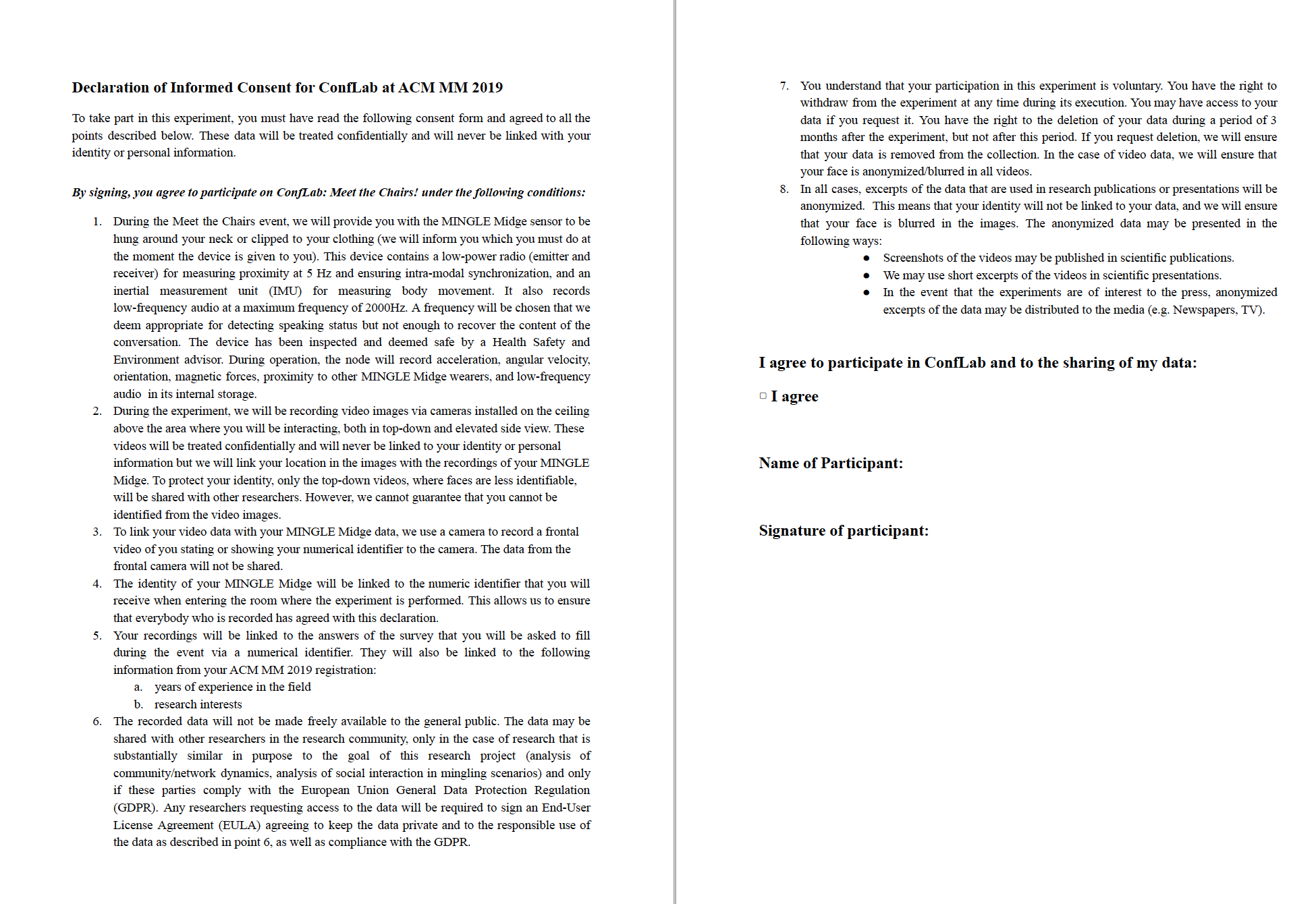}
\caption{Consent form signed by each participant in the data collection.}
\label{fig:consent_form}
\end{figure}

\textcolor{\questioncolor}{\textbf{Q. Was any preprocessing/cleaning/labeling of the data
done(e.g.,discretization or bucketing, tokenization, part-of-speech
tagging, SIFT feature extraction, removal of instances, processing of
missing values)?
}
If so, please provide a description. If not, you may skip the remainder of
the questions in this section.
} 

We did not pre-process the signals obtained from the wearable devices or cameras. The only exception is the audio data. Due to a hardware malfunction  (this is resolved for the Midges by using different SD cards), the audio needed to be post-processed in order to synchronize it with the other modalities. The synchronization against other modalities was manually checked.

Labeling of the dataset was done as explained in the \textit{Composition} section.

\textcolor{\questioncolor}{\textbf{Q. Was the “raw” data saved in addition to the preprocessed/cleaned/labeled
data (e.g., to support unanticipated future uses)?
}
} 

The dataset is separated into raw data and the post processed data. 
For the audio, the original raw data is not suitable for most use cases due to the mentioned synchronization issue. So we share the synchronized version in the raw part of the repository.

\textcolor{\questioncolor}{\textbf{Q. Is the software used to preprocess/clean/label the instances available?
}
If so, please provide a link or other access point.
} 

The processing / fixing of the audio files did not require special software.

The annotation of keypoints and speaking status was done by making use of the Covfee framework: \url{https://josedvq.github.io/covfee/}

\textcolor{\questioncolor}{\textbf{Q. Any other comments?
}} 

None. \\

\begin{dsheetsec}
USES
\end{dsheetsec}

\textcolor{\questioncolor}{\textbf{Q. Has the dataset been used for any tasks already?
}
If so, please provide a description.
} 

In the main paper, we have benchmarked three baseline tasks: person and keypoints detection, speaking status detection, and F-formation detection. The first task is a fundamental building block for automatically analyzing human social behaviors. The other two demonstrate how learned body keypoints can be used in the behavior analysis pipeline for inferring more socially related phenomena. We chose these benchmarking tasks since they have been studied on other in-the-wild behavior datasets.  

\textcolor{\questioncolor}{\textbf{Q. Is there a repository that links to any or all papers or systems that use the dataset?
}
} 

None at the time of writing of the paper.

\textcolor{\questioncolor}{\textbf{Q. What (other) tasks could the dataset be used for?
}
} 

Given the richness and the unscripted open-ended nature of the social interactions, ConfLab can be used for many other tasks.
\paragraph{Forecasting, causal relationship discovery} Recently, tasks pertaining to the forecasting low-level social cues in conversations have been receiving increased attention from the community \cite{ramanSocialProcessesSelfSupervised2021, barquero2022didn}. The real-life nature of ConfLab along with the increased data and annotation fidelity can prove a valuable resource for such tasks. Similarly, ConfLab can also be used for efforts towards discovering causal relationships between social behaviors \cite{raman2022did}. 
\paragraph{Data Association.} A crucial assumption made in many former multimodal datasets\cite{MnM2021,alameda2015salsa,Joo_2017_TPAMI} is that the association of video data to the wearable modality can be manually performed. Few works \cite{cabrera2016matching,cabrera2018hierarchical} have tried to address this issue but using movement cues alone to associate the modalities is challenging as conversing individuals are mostly stationary. This remains a significant and open question for future large scale deployable multimodal systems. One solution may be to annotate more social actions as a form of top-down supervision. However, detecting pose and actions robustly from overhead cameras remains to be solved.
\paragraph{Conversation floor and  F-formation estimation}
Prior analysis on the MatchNMingle dataset has demonstrated that F-formations can contain multiple simultaneous conversations when the F-formations contain a least 4 people \cite{ramanAutomaticEstimationConversation2019}. If this is the case for the ConfLab dataset, this may drastically change how F-formations should be labelled (e.g. returning to being a more subjective task \cite{hung2011detecting}) as more time-precise labelling could enable a more nuanced take on F-formation and conversation floor membership over time.
\paragraph{Multi-class social action estimation}
More annotations resources were focused on speaker status, F-formation, and keypoint estimation. However, there are a wealth of other social actions in the data that could be interesting to combine into a more complex multi-class social action estimation task. Example social actions include drinking, mobile phone use, hand and head gesture types \cite{MnM2021,hung2013classifying}.
\paragraph{Estimation and analysis of socially-related phenomena}
Beyond the modeling of human behavior which is of interest to the Computer Vision and Machine Learning communities, our benchmarked tasks form the basis for further explorations into downstream prediction of socially-related  constructs which is of interest to the Social Science and Social Psychology communities. Such constructs include conversation quality \cite{raj2020defining, raman2022perceived}, dominance \cite{5549893}, rapport \cite{Muller2018a}, and influence \cite{dong2007using}. 

\paragraph{Investigation of novel crossmodal fusion strategies }
The baseline tasks in our paper rely only on a late fusion strategy. However, ConfLab's sub-second expected cross modal latency of $\sim13$~ms along with higher sampling rate of features ($60$~fps video, $56$~Hz IMU) opens the gateway for the in-the-wild study of nuanced time-sensitive social behaviors like mimicry and synchrony (for predicting e.g. attraction \cite{quiros2021individual}) which need tolerances as low as $40$~ms \citep[Sec.3.2]{raman2020modular}. Prior works coped with lower tolerances by computing summary statistics over input windows \cite{Gedik2017a, Cabrera-Quiros2018b, Quiros2019}. ConfLab enables for the first time, the exploration of Multimodal machine learning approaches for social behaviour analysis in these highly dynamic in-the-wild settings \cite{MultimodalML2019}. Through the provided annotations Conflab also enables research in the topic of usage of mobile phones in small-group social interactions in-the-wild.
\paragraph{Person attribute estimation} Estimating individuals that are newcomers/old timers from the dataset may be possible based on their networking strategies.


\textcolor{\questioncolor}{\textbf{Q. Is there anything about the composition of the dataset or the way it was
collected and preprocessed/cleaned/labeled that might impact future uses?
}
For example, is there anything that a future user might need to know to
avoid uses that could result in unfair treatment of individuals or groups
(e.g., stereotyping, quality of service issues) or other undesirable harms
(e.g., financial harms, legal risks) If so, please provide a description.
Is there anything a future user could do to mitigate these undesirable
harms?
} 

Although ConfLab's long-term vision is towards developing technology to assist individuals in navigating social interactions, the data could also affect a community in unintended ways: for instance, cause worsened social satisfaction, a lack of agency, stereotype newcomers and veterans, or benefit only those members of the community who make use of resulting applications at the expense of the rest. More nefarious uses involve exploiting the data for developing methods that harmfully surveil or profile people. Researchers must consider such inadvertent effects must while developing downstream applications. Finally, since we recorded the dataset at a scientific conference and required voluntary participation, there is an implicit selection bias in the population represented in the data. Consequently, researchers using the data should be aware that resulting insights may not generalize to the general population.

\textcolor{\questioncolor}{\textbf{Q. Are there tasks for which the dataset should not be used?
}
If so, please provide a description.
} 

Beyond the cautionary discussion in the previous question, tasks involving the re-identifying the subjects is strictly against the End User License Agreement under which we share the dataset. 

\textcolor{\questioncolor}{\textbf{Q. Any other comments?
}} 

None. 

\begin{dsheetsec}
DISTRIBUTION
\end{dsheetsec}

\textcolor{\questioncolor}{\textbf{Q. Will the dataset be distributed to third parties outside of the entity
(e.g., company, institution, organization) on behalf of which the dataset
was created?
}
If so, please provide a description.
} 

The dataset is available for third parties outside of Delft University of Technology to use for academic research purposes subject signing and approval of our End User License Agreement. The dataset will be hosted by 4TU.ResearchData (see the Maintenance section for description of the 4TU entity).

\textcolor{\questioncolor}{\textbf{Q. How will the dataset will be distributed (e.g., tarball on website, API,
GitHub)?
}
Does the dataset have a digital object identifier (DOI)?
} 

The dataset will be distributed via the 4TU.ResearchData user interface where the data can be downloaded. The dataset has a DOI: \url{https://doi.org/10.4121/c.6034313} 

\textcolor{\questioncolor}{\textbf{Q. When will the dataset be distributed?
}
} 

The dataset has been available since June 9, 2022. 

\textcolor{\questioncolor}{\textbf{Q. Will the dataset be distributed under a copyright or other intellectual
property (IP) license, and/or under applicable terms of use (ToU)?
}
If so, please describe this license and/or ToU, and provide a link or other
access point to, or otherwise reproduce, any relevant licensing terms or
ToU, as well as any fees associated with these restrictions.
} 

The dataset will be distributed under a restricted copyleft license, specified  within our End User License Agreement, accessible through the 4TU.ResearchData dataset website. No fees are associated with the license. 

\textcolor{\questioncolor}{\textbf{Q. Have any third parties imposed IP-based or other restrictions on the data
associated with the instances?
}
} 

No. 

\textcolor{\questioncolor}{\textbf{Q. Do any export controls or other regulatory restrictions apply to the
dataset or to individual instances?
}
If so, please describe these restrictions, and provide a link or other
access point to, or otherwise reproduce, any supporting documentation.
} 

The terms of our EULA and the European General Data Protection Regulations (GDPR) apply.

\textcolor{\questioncolor}{\textbf{Any other comments?
}} \\
None.

\begin{dsheetsec}
MAINTENANCE
\end{dsheetsec}

\textcolor{\questioncolor}{\textbf{Q. Who is supporting/hosting/maintaining the dataset?
}
} 

The dataset is hosted by 4TU.ResearchData (\url{https://www.4tu.nl/en/about_4tu/}), and supported and maintained by The Socially Perceptive Computing Lab at TUDelft. 

\textcolor{\questioncolor}{\textbf{Q. How can the owner/curator/manager of the dataset be contacted (e.g., email address)?
}
} 

Via email: SPCLabDatasets-insy@tudelft.nl.

\textcolor{\questioncolor}{\textbf{Q. Is there an erratum?
}
} 

No.

\textcolor{\questioncolor}{\textbf{Q. Will the dataset be updated (e.g., to correct labeling errors, add new instances, delete instances)?
}
If so, please describe how often, by whom, and how updates will be
communicated to users (e.g., mailing list, GitHub)?
} 

Updates will be done as needed as opposed to periodically. Instances could be deleted, added, or corrected. The updates will be posted on the 4TU.ResearchData dataset website. 

\textcolor{\questioncolor}{\textbf{Q. If the dataset relates to people, are there applicable limits on the retention of the data associated with the instances (e.g., were individuals in question told that their data would be retained for a fixed period of time and then deleted)?
}
} 

No limits were communicated to our data participants.

\textcolor{\questioncolor}{\textbf{Q. Will older versions of the dataset continue to be supported/hosted/maintained?
}
If so, please describe how. If not, please describe how its obsolescence
will be communicated to users.
} 

Only the latest version of the dataset will be maintained. If applicable, we will also host older versions of the data, accessible through the 4TU.ResearchData website.

\textcolor{\questioncolor}{\textbf{Q. If others want to extend/augment/build on/contribute to the dataset, is there a mechanism for them to do so?
}
If so, please provide a description. Will these contributions be validated/verified? If so, please describe how. If not, why not? Is there a process for communicating/distributing these contributions to other users?
If so, please provide a description.
} 

We are open to contributions to the dataset. In accordance with our End User License Agreement, contributions should be made available, indicating if there are any restrictions on their contribution. We encourage the potential contributors to contact us to discuss how they wish to be attributed (e.g. citation of a paper or repository related to code/annotations). After finalizing the attribution discussion, we can add the attribution as an update following the same process explained above.


\clearpage
\FloatBarrier
\section{Sample Participant Report} \label{app:participant_reports}
\begin{figure}[!h]
    \centering
    \includegraphics[trim={2cm 3cm 2cm 3cm}, clip=True, width=0.49\textwidth]{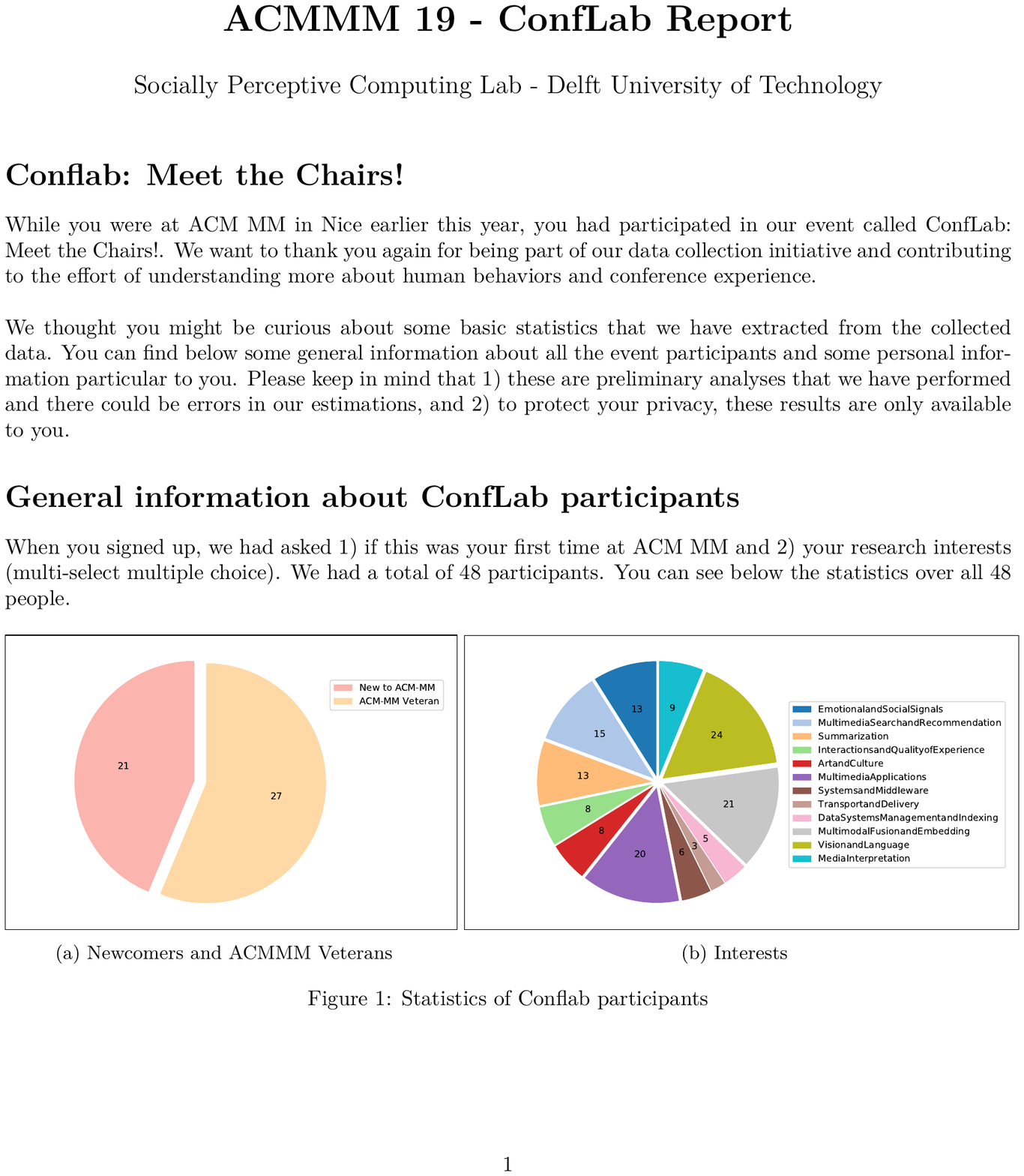}\hfill
    \includegraphics[trim={2cm 3cm 2cm 3cm}, clip=True, width=0.49\textwidth]{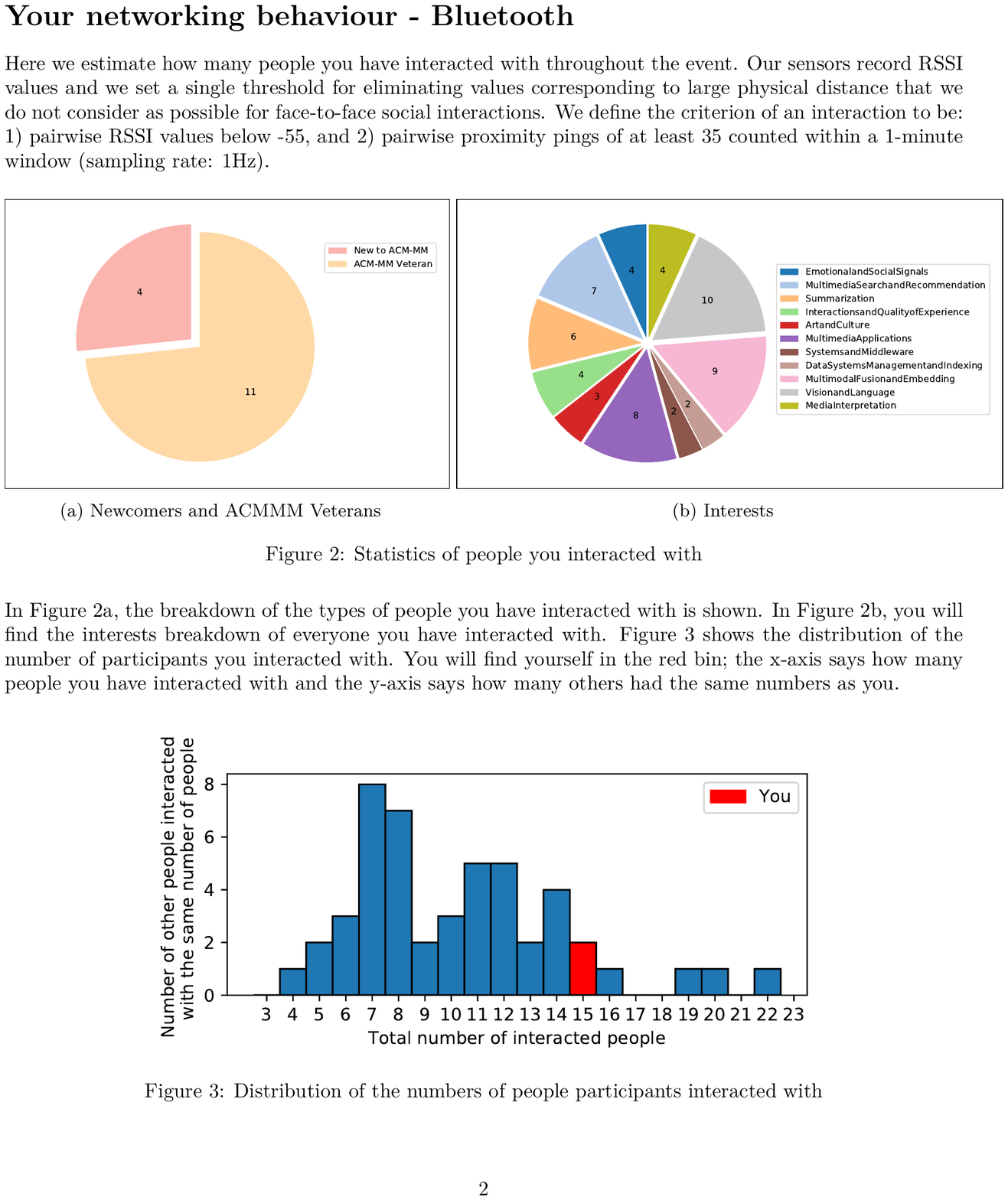}
    
    \vspace{20pt}
    \includegraphics[trim={2cm 3cm 2cm 3cm}, clip=True, width=0.49\textwidth]{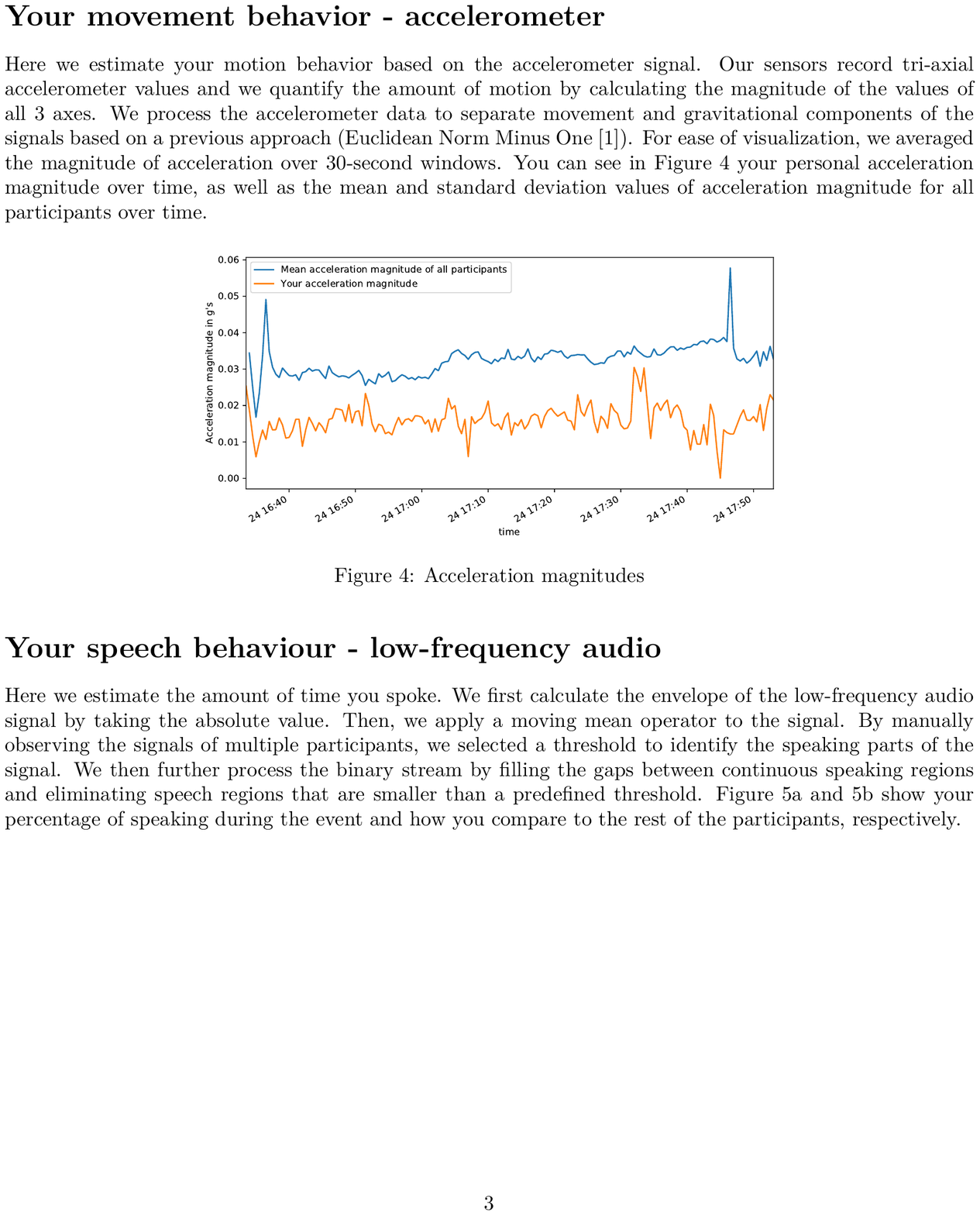}\hfill
    \includegraphics[trim={2cm 3cm 2cm 3cm}, clip=True, width=0.49\textwidth]{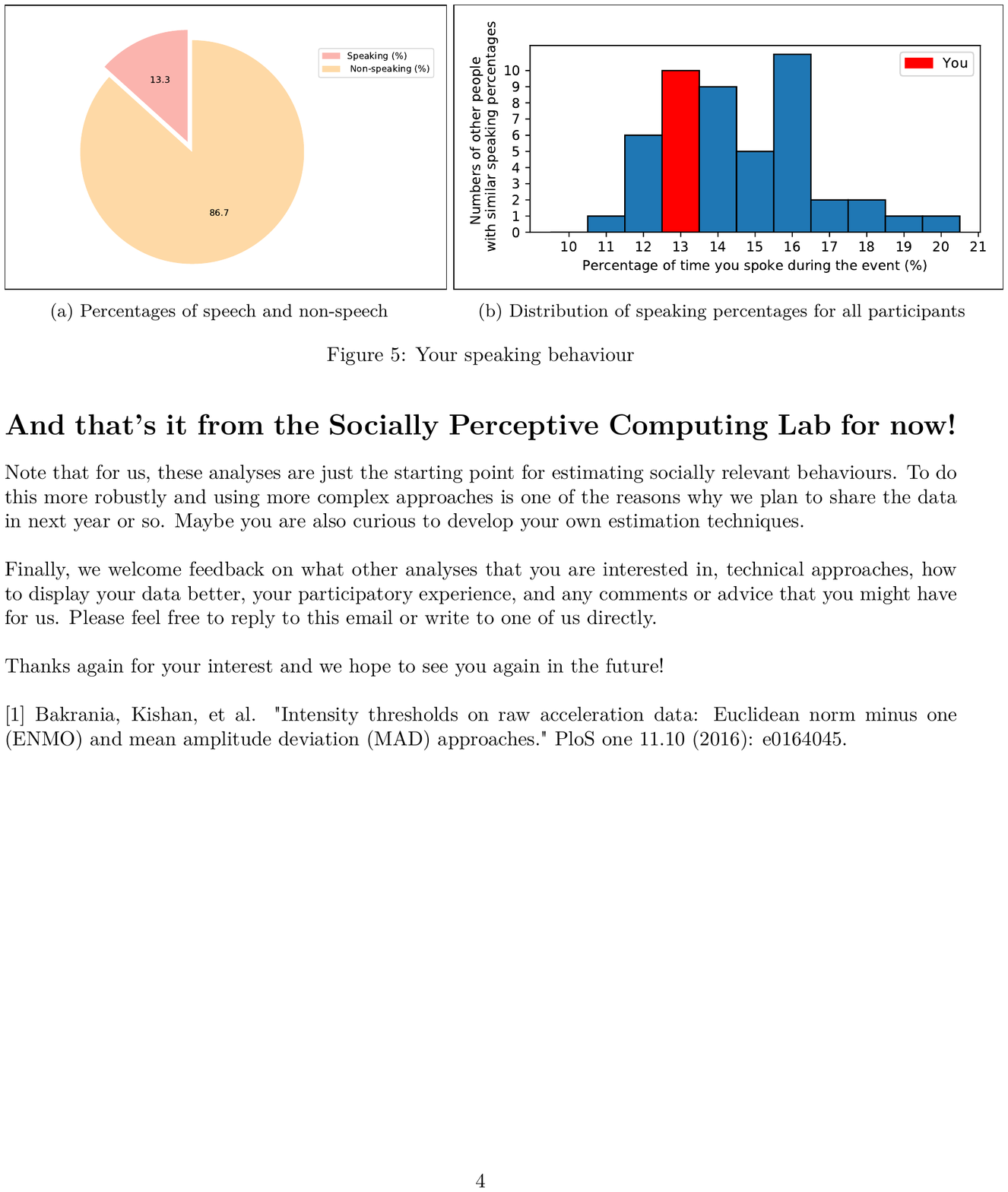}
    \caption{Sample post-hoc report sent to each participant of ConfLab. The report contains insights into the participant's networking behavior from the collected wearable-sensors data. This insight served as an additional incentive to participate in ConfLab, beyond interacting with the Chairs and contributing to a community-driven data endeavor (see main paper Section~\ref{sec:data-acquisition}).}
    \label{fig:participant-report}
\end{figure}

\clearpage
\section{Data Capture Setup Details}
\label{app:data-capture}

\paragraph{The Midge}\label{para:midge} We improved upon the Rhythm Badge in three ways towards enabling more fine-grained and flexible data capture: (i) enabling full audio recording with a frequency up to $48$~KHz, with an on-board switch to allow physical selection between high and low frequency capture directly at acquisition; (ii) 
adding a 9-axis Inertial Measurement Unit (IMU) with an on-board Digital Motion Processor (DMP) to record orientation; and (iii) an on-board SD card to directly store raw data, avoiding issues related to packet loss during wireless data transfer required by the Rhythm Badge. IMUs combine three tri-axial sensors: an accelerometer, a gyroscope, and a magnetometer. These measure acceleration, orientation, and angular rates respectively. These sensor measurements are combined on-chip by a Digital Motion Processor. Rough proximity estimation is performed by measuring the Received Signal Strength Indicator (RSSI) for Bluetooth packets broadcast every second ($1$~Hz) by every Midge. During the event, IMUs were set to record at $50$~Hz. We recorded audio at $1250$~Hz to mitigate extraction of verbal content while still ensuring robustness to cocktail-party noise.

\paragraph{Wireless Synchronization at Acquisition} The central idea for our syncrhonization approach involves using a common Network Time Protocol (NTP) signal as reference for the camera and wearables sub-networks. The set-up achieved a cross-modal latency of 13~ms at worst, which is well below the 40~ms latency tolerance suitable for behavior research in our setting \cite[Sec.~3.3]{raman2020modular}. Additionally, our synchronization approach allowed for dynamic addition of sensors to the network while still obtaining synchronized data streams. This is crucial in extreme in-the-wild events where some participants might arrive late. 

\paragraph{Sensor Calibration}\label{para:calibration}
For computing the camera extrinsics, we marked a grid of $1$~m $\times$ $1$~m squares in tape across the interaction area floor. We ensured line alignment and right angles using a laser level tool (STANLEY Cross90). For computing the camera intrinsics, we used the OpenCV asymmetric circles grid pattern \cite{2015opencv}. The calibration was performed using the Idiap multi camera calibration suite \cite{idiap-multicam}. All wearable sensors include one TDK InvenSense ICM-20948 IMU \cite{imu} unit that provides run time calibration. To establish a correspondence with the camera frame of reference, the sensors were lined up against a common reference-line visible in the cameras to acquire an alignment so that the camera data can offer drift and bias correction for the wearable sensors. 

\section{Implementation Details}
\label{app:implementation-details}
\subsection{Person and Keypoint Detection Models} \label{app-subsec:kp-implementation}

\paragraph{Data Cleaning} A few frames contained some incorrectly labeled keypoints, a product of annotation errors like mis-assignment of participant IDs. We removed these using a threshold on the proximity to other keypoints of the same person. Further, in some cases, a person might be partially outside a camera’s field of view. For the person detection task, we compute the bounding box from the keypoint ground-truth annotations. If more than half the body (50\% keypoints) is missing in the frame so that e.g. only their legs are visible (see top of  \autoref{fig:covfee}), we don’t consider the person for that frame in the person detection experiments. Note that due to the significant overlap between the camera views, the person would be considered for the corresponding frame in the next camera. If they move back into the original view, we again take them into consideration for the original camera for the corresponding frame. Moreover, if there are more than 10\% missing keypoints across all people in an image, we also discard that image from the experiment. This preprocessing resulted in a training set with 112k frames (1809k person instances) and a test set with 7k frames (158k person instances).

\paragraph{Training} We resized the images to $960\times540$, and augmented the data by randomizing brightness and horizontal flips. The learning rate was set to $0.02$ and batch size to $4$. We trained the models for $50$~k iterations, using the COCO-pretrained weights for initialization. All hyper-parameters were chosen based on the performance on a separate hold-out camera chosen as validation set. During training, any missing ground-truth keypoints (resulting from the person being partially outside the camera's view for instance) are ignored during back-propagation.

\subsection{F-formation Detection} \label{app-subsec:ff-implementation}
\paragraph{Data Cleaning} Because keypoint annotations of the subjects are based on camera view and that the F-formation clustering methods cannot group subjects that do not exist under one camera view (e.g., when there are more identities than in associated ground truths), we processed the ground truth also based on camera number. This filtering pre-processing was decided based on the best camera view of the F-formations. 

\paragraph{Feature Extraction} The required features of GCFF and GTCG include location and orientation of the subjects. We used the X and Y position of subjects' head (as it is the most visible from the top-down view) for location, and extracted orientations for head, shoulders and hips. The orientations are calculated based on corresponding vectors determined by head and nose keypoints, left and right shoulder keypoints, and left and right hip keypoints, respectively.

\paragraph{Training} We used pre-trained parameters for field of view (FoV) and frustum aperture (GTCG) and minimum description length (GCFF), provided in these models trained on the Cocktail Party. FOV and aperture are related to human eye gaze and head anatomical constraints reported by \cite{BaAndOdobez}, and hence not dataset specific. The minimum description length is an initialized prior dictated by the same form of the Akaike Information Criterion, and becomes part of the optimization formulation.  We tuned parameters such as frustum length (GTCG) and stride (GCFF) to account for average interpersonal distance in ConfLab based on Camera 6, as they vary across different datasets.

\section{Additional Results}\label{app:extra_results}

\subsection{Person and Keypoints Detection}\label{app:kp_results}

\begin{figure}[!t]
    \centering
    \includegraphics[width=0.3\linewidth]{imgs/keypoint/td/vid2-seg8-scaled-denoised_000082.jpg}
    \includegraphics[width=0.3\linewidth]{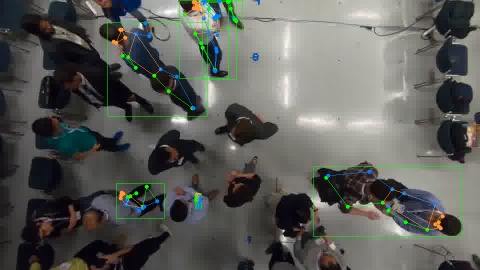}
    \includegraphics[width=0.3\linewidth]{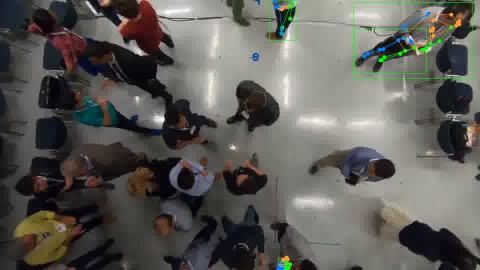}
    
    \includegraphics[width=0.3\linewidth]{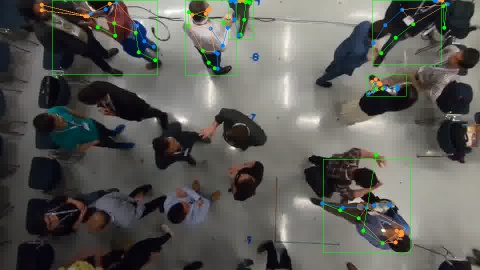}
    \includegraphics[width=0.3\linewidth]{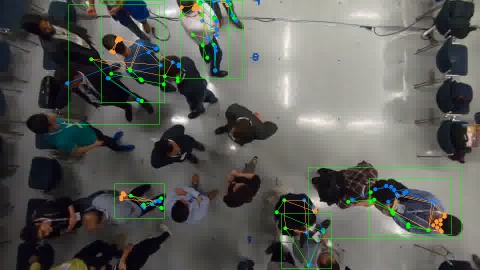}
    \includegraphics[width=0.3\linewidth]{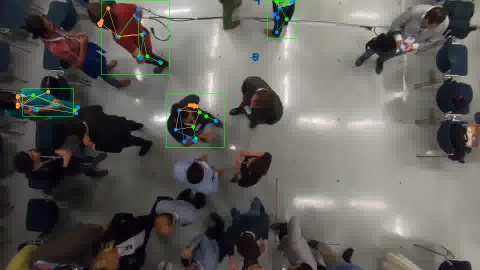}
    
    \includegraphics[width=0.3\linewidth]{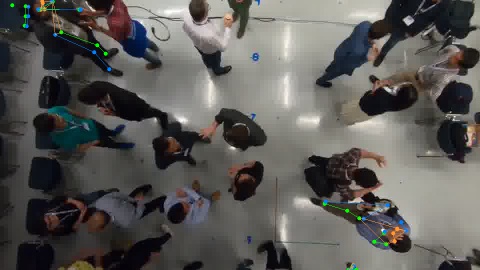}
    \includegraphics[width=0.3\linewidth]{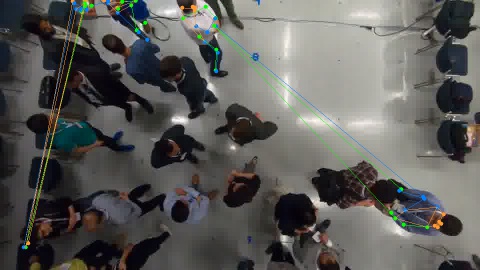}
    \includegraphics[width=0.3\linewidth]{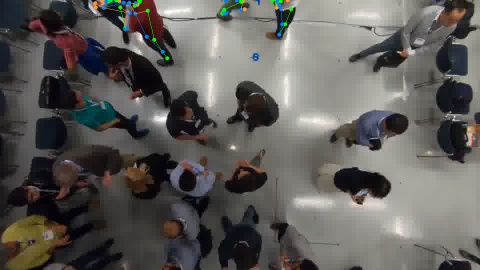}
    
    \includegraphics[width=0.3\linewidth]{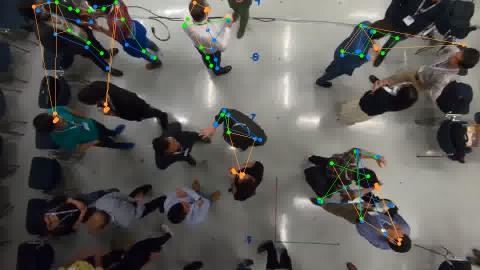}
    \includegraphics[width=0.3\linewidth]{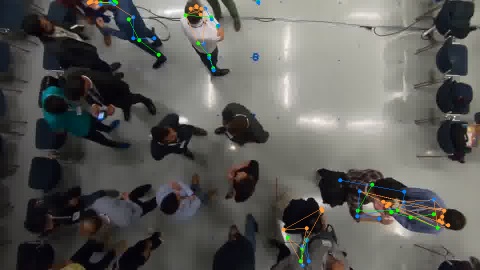}
    \includegraphics[width=0.3\linewidth]{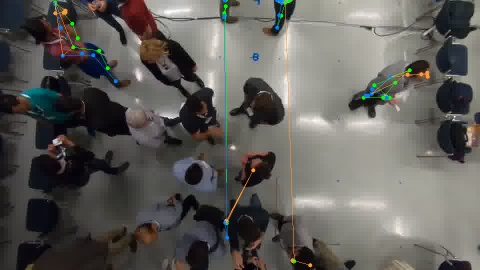}
    
    \caption{Results from Pretrained keypoint detection models. From top to bottom - predictions from RSN \cite{cai2020learningRSN}, MSPN\cite{li2019rethinkingmspn}, HigherHRNet \cite{Cheng_2020_CVPRhigherhrnet}, and HourglassAENet \cite{newell2017associative_posehglass}. Results show that \emph{SOTA 2D body keypoint detection models fail to capture the body keypoints in the ConfLab dataset}.}
    \label{fig:pretrained_pred}
\vspace{5pt}
\end{figure}

\paragraph{Predictions from pretrained SOTA models}  Figure \ref{fig:pretrained_pred} shows predictions from SOTA human keypoint estimation models, namely, RSN \cite{cai2020learningRSN}, MSPN\cite{li2019rethinkingmspn}, HigherHRNet \cite{Cheng_2020_CVPRhigherhrnet}, and HourglassAENet \cite{newell2017associative_posehglass}, for the testing images of the Conflab dataset. Note that RSN and MSPN are top-down networks, i.e., they require person bounding boxes to predict the keypoints in each bounding box. We use COCO pretrained faster-RCNN network for bounding box estimation. HigherHRNet and HourglassAENet are bottom-up models, i.e., they directly predict keypoints from the full image. We use publicly available COCO pretrained checkpoints for prediction. The results show that the \emph{state-of-the-arts 2D body keypoint detection models fail to capture the body keypoints in the Conflab dataset}. We infer that training on the dataset (e.g., COCO) that contains mostly side-view images does not work well in top-view images, for which Conflab dataset is important to the community.   

\begin{figure}[!t]
    \centering
    \includegraphics[width=0.3\linewidth]{imgs/keypoint/det2_pretrained/vid2-seg8-scaled-denoised_006802.jpg}
    \includegraphics[width=0.3\linewidth]{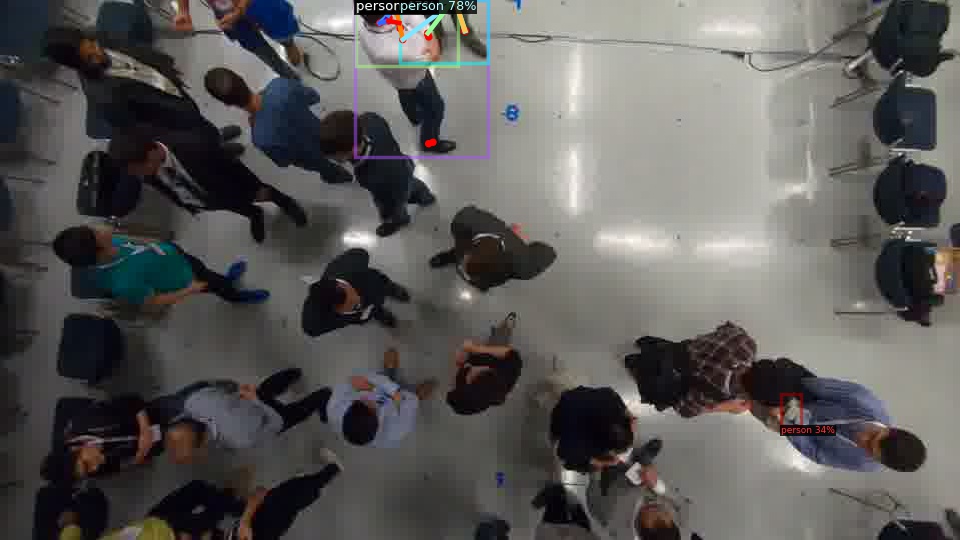}
    \includegraphics[width=0.3\linewidth]{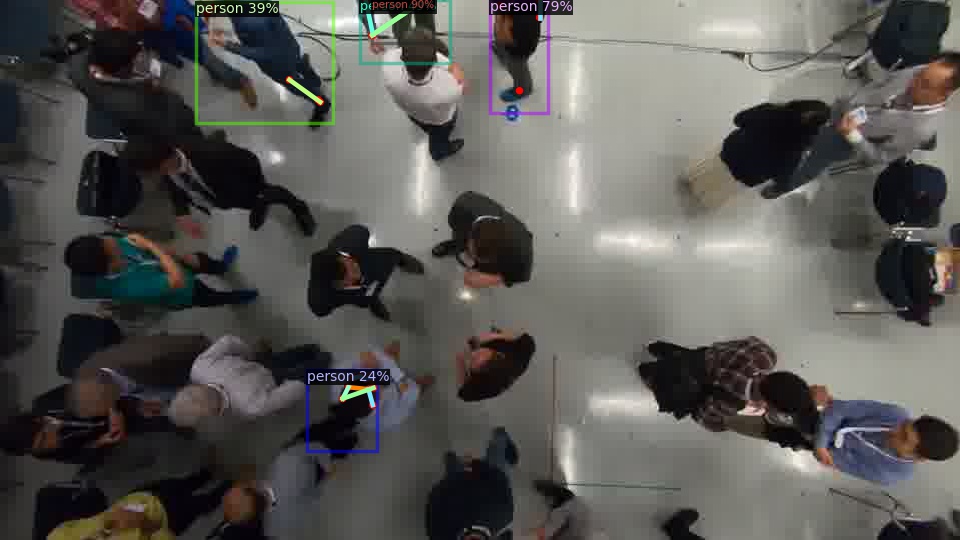}
    
    \includegraphics[width=0.3\linewidth]{imgs/keypoint/det2/vid2-seg8-scaled-denoised_006802.jpg}
    \includegraphics[width=0.3\linewidth]{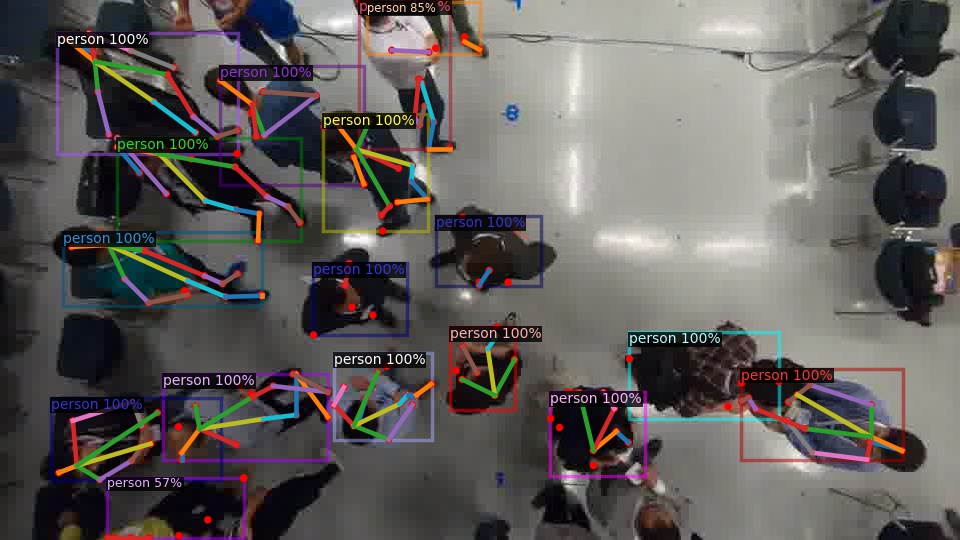}
    \includegraphics[width=0.3\linewidth]{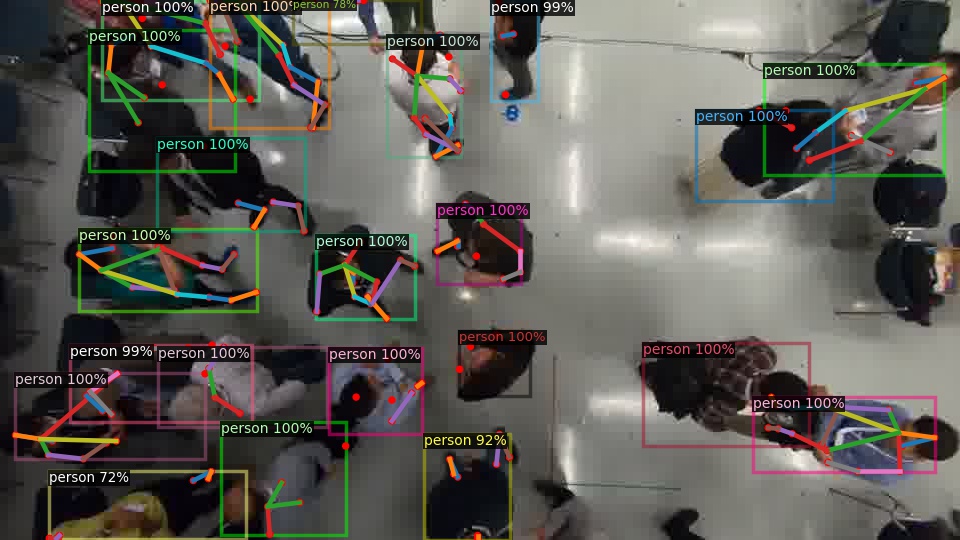}
    \caption{Results from (top) COCO pretrained Mask-RCNN model, (bottom) our ConfLab finetuned Mask-RCNN model.}
    \label{fig:det2pred}
\vspace{15pt}
\end{figure}

\begin{figure}[!h]
\begin{minipage}[t]{0.45\textwidth}
\captionsetup{type=table}
\caption{Effect of varying \% frames from each camera at training on keypoint estimation.}     \label{tab:abl_n}
\setlength{\tabcolsep}{2pt}
    \centering
    \begin{tabular}{@{}lc@{}}
    \toprule
       {\% of training samples}  & AP$^\text{OKS}_\text{50}$ \\
    \midrule
    1.6\% & 29.0 \\
    3.2\% & 35.9 \\
    8\% & 39.0 \\
    16\% & 44.5 \\
    100\% & 45.3 \\
    \bottomrule
    \end{tabular}
\end{minipage}\hfill
\begin{minipage}[t]{0.52\textwidth}
\captionsetup{type=table}
\caption{Effect of adding all frames from individual cameras to the training set on keypoint estimation.} \label{tab:abl_data_amount}
\setlength{\tabcolsep}{2pt}
    \centering
    \begin{tabular}{@{}lcc@{}}
    \toprule
      Train Camera & \#(training samples) & AP$^\text{OKS}_\text{50}$ \\
    \midrule
    cam 2 & 34k & 8.6 \\
    cam 2 + cam 4 & 69k & 31.1 \\
    cam 2 + cam 4 + cam 8 & 112k & 45.3 \\
    \bottomrule
    \end{tabular}
\end{minipage}
\end{figure}

\begin{figure}[t]
\begin{minipage}[b]{0.45\textwidth}
\ra{1.2}
\captionsetup{type=table}
\caption{Keypoint estimation ablation with keypoints from different body sections: head and shoulders (5), + torso (9), + hips (11), + knees and feet (full 17).}\label{tab:num_kp}
\centering
\begin{adjustbox}{max width=\linewidth}
\begin{tabular}{@{}lccc@{}} 
\toprule
\#Keypoints &
 AP$^\text{OKS}_\text{50}$ & AP$^\text{OKS}$ & AP$^\text{OKS}_\text{75}$ \\
\midrule
5 & 26.6 & 7.1 & 1.4 \\
9 & 26.5 & 6.9 & 2.0\\
11 & 35.8 & 9.5 & 2.2 \\
17 & 45.3 & 13.5 & 3.3\\
\bottomrule
\end{tabular}
\end{adjustbox}
\end{minipage}\hfill
\begin{minipage}[b]{0.52\textwidth}
\centering
\captionsetup{type=table}
\caption{ROC AUC and accuracy for different sensor modalities from out 9-dof IMU in speaking status detection using the Minirocket classifier \cite{minirocket}. The number of channels in the corresponding modality is indicated in parentheses.}
\label{tab:ss_modalities}
\begin{tabular}{@{}lcc@{}}
\toprule
Input Modality &   AUC & Accuracy \\
\midrule
Acceleration (3) & 0.813 &    0.768 \\
Gyroscope (3) & 0.765 &    0.716 \\
Magnetometer (3)  & 0.610 &    0.656 \\
Rotation vector (4)  & 0.726 &    0.696 \\
All (13)  & 0.774 &    0.739 \\
\bottomrule
\end{tabular}
\end{minipage}
\end{figure}

\paragraph{Qualitative Results from ResNet-50 Finetuning}
\figurename~\ref{fig:det2pred} illustrates more qualitative results from our finetuning experiments. We find that finetuning on our non-invasive top-down camera perspective significantly improves the keypoint estimation performance.

\paragraph{Ablations}
Tables~\ref{tab:abl_n} and \ref{tab:abl_data_amount} include the results of our experiments investigating the effect of varying the training data size on keypoint detection performance (see main paper Section~\ref{subsec:kp-experiments}).
In Table~\ref{tab:num_kp}, we show keypoint detection scores for experiments with different number of keypoints. We first focus on the five upper body keypoints: \{head, nose, neck, rightShoulder, leftShoulder\}. We then additionally considered the torso region keypoints for a total of nine: \{rightElbow, rightWrist, leftElbow, leftWrist\}. Finally, we add the hip keypoints \{rightHip, leftHip\} to the set. The experiments in the main paper are performed with all 17 keypoints. The results show that performance drops slightly when adding the arms keypoints ($5\to9$, AP$^\text{OKS}_\text{50}$ and AP$^\text{OKS}$), and that the relative gain when adding the hip keypoints ($9\to11$) is lower than when adding the lower body keypoints ($11\to17$, especially AP$^\text{OKS}_\text{75}$). We believe this is largely due to the lower body being more static relative to the arms that move a lot to execute gestures during conversations.  

\subsection{Speaking Status Detection}\label{app:ss_extra} 

\paragraph{Experiments with different sensor modalities}

Table \ref{tab:ss_modalities} displays the results from experiments using specific modalities from our IMUs for the task of speaking status detection. We used the best performing classifier (Minirocket \cite{minirocket}) among the ones tested in Table~\ref{tab:ss-baseline}. The experiment setup is the same as detailed in Section \ref{sec:speaking-status}, and the model is not changed between runs, except for the fact that different modalities may have a different number of input channels.

\section{Reproducibility Checklist}\label{app:reproducibility} 

\subsection{Person and Keypoints Detection}

\begin{itemize}
    \item Source code link: \url{https://github.com/TUDelft-SPC-Lab/conflab}
    \item Data used for training: $112$k frames ($1809$k person instances).
    \item Pre-processing: See Section~\ref{sec:annotation}, Appendix~\ref{app-subsec:kp-implementation}.
    \item How samples were allocated for train/val/test: cameras $2$, $4$, and $8$ are selected for training. For hyperparameter tuning, camera $8$ are held out for validation.
    \item Hyperpatameter consideration: We considered learning rates $(0.001/0.005/0.05/0.01)$, number of epochs $(10/20/50/100)$, detection backbone (R50-FPN/R50-C4). Also see Appendix~\ref{app-subsec:kp-implementation}
    \item Number of evaluation runs: $5$
    \item How experiments were ran: See Section~\ref{subsec:kp-experiments}.
    \item Evaluation metrics: Average precision at different thresholds.
    \item Results: See Section~\ref{subsec:kp-experiments} and Appendix~\ref{app:kp_results}.
    \item Computing infrastructure used: All baseline experiments were ran on Nvidia V100 GPU ($16$GB) with IBM POWER9 Processor.
\end{itemize}

\subsection{Speaking Status Detection}

\begin{itemize}
    \item Source code link: \url{https://github.com/TUDelft-SPC-Lab/conflab}
    \item Data used for training: $42884$ windows ($3$~seconds), extracted from $48$ participants’ wearable data and speaking status annotations
    \item Pre-processing: Data was windowed into $3$-second segments (see Section \ref{sec:speaking-status}). The source code includes this pre-processing step. 
    \item How samples were allocated for train/val/test: 10-fold cross-validation at the subject level ($48$ subjects) to test generalization to unseen data subjects. The splits can be reproduced exactly using the source code.
    \item Hyperparameter considerations: For acceleration-based methods, we used default network hyper-parameters and architectures from their tsai implementation \cite{tsai}. For the MS-G3D baseline \cite{Gupta2020}, we used default hyperparameters from the authors' implementation. For both, we determined the early stoppage point using a small subset ($10\%$) of the training set.
    \item Number of evaluation runs: 1 run of 10-fold cross-validation
    \item How experiments were ran: For each fold, the early stoppage point was first determined using $10\%$ of the training data as validation set and AUC as performance metric. The model at this stoppage point was then applied to the test set for evaluation.
    \item Evaluation metrics: Area under the ROC curve (AUC)
    \item Results: See Section \ref{sec:speaking-status} 
    \item Computing infrastructure used: Experiments were ran on a personal computer with GPU acceleration (NVidia RTX3080).
\end{itemize}

\subsection{F-formation Detection}
\begin{itemize}
    \item Source code link: \url{https://github.com/TUDelft-SPC-Lab/conflab}
    \item Data used for training: Camera $6$
    \item Pre-processing: See Section~\ref{app-subsec:ff-implementation} for data cleaning and feature extraction.
    \item How samples were allocated for train/val/test: samples from Camera $6$ were used to select the best model parameters. The rest are for test (evaluation). However, we note that Table \ref{tab:ff-baseline} shows averaged performance on all cameras to provide a holistic view of the F-formation detection performance on ConfLab.   
    \item (Hyper)parameter considerations: Both baseline methods are not deep-learning based and model parameters are interpretable. For GTCG, the parameters are frustum length ($275$), frustum aperture ($160$), frustum samples ($2000$), and sigma for affinity matrix ($0.6$). For GCFF, the parameters are minimum description length ($30000$) and stride ($70$).
    \item Number of evaluation runs: 1 
    \item How experiments were ran: A total of eight experiments were run for choosing the best parameters, and three for evaluation (for camera $2$, $4$, and $8$). The parameters were chosen based on grid-search. For optimizing frustum length in GTCG, we searched over $[170,195,220, 245, 275]$ with $275$ being averaged interpersonal distance based on Camera $6$. For optimizing stride $D$ in GCFF, we searched over $[30,50,70]$. 
    \item Evaluation metrics: F1
    \item Results: See Section \ref{sec: fformation} 
    \item Computing infrastructure used: The experiments were run on Linux-based cluster instances on CPU with Matlab 2018a.
\end{itemize}

\end{document}